\documentclass[a4paper]{article}
\let\ACMmaketitle=\maketitle
\renewcommand{\maketitle}{\begingroup\let\footnote=\thanks \ACMmaketitle\endgroup}

\usepackage{amsmath}
\usepackage{amssymb}
\usepackage{cancel}

\usepackage{graphicx} 

\usepackage{multirow}
\usepackage{mathtools}
\usepackage{amsfonts}
\usepackage{verbatim}
\usepackage{subcaption}
\usepackage{wasysym}
\usepackage{color}
\usepackage{tikz}
\usetikzlibrary{shapes}
\usepackage{scrextend}

\usepackage{geometry}
\usepackage{fancyhdr}

\definecolor{blue}{rgb}{0,0,1}
\definecolor{green}{rgb}{0,0.5,0}
\definecolor{red}{rgb}{1,0,0}
\definecolor{yellow}{rgb}{0.75,0.75,0}
\definecolor{magenta}{rgb}{0.75, 0.0, 0.75}
\definecolor{orange}{rgb}{1.0, 0.27058823529411763, 0.0}
\definecolor{black}{rgb}{0.0,0.0,0.0}

\newcommand{\bluelineStar}{\raisebox{2pt}{\tikz{\node[star,blue,fill=blue, scale=0.5] at (2.5mm,0.0mm) {};
\draw[-,blue,solid,line width = 1.0pt](0.,0.0mm) -- (5mm,0.0mm)
}}}
\newcommand{\greenlineRectangle}{\raisebox{2pt}{\tikz{\node[rectangle,green,fill=green, scale=0.8] at (2.5mm,0.0mm) {};
\draw[-,green,solid,line width = 1.0pt](0.,0.0mm) -- (5mm,0.0mm)
}}}
\newcommand{\redlineDiamond}{\raisebox{2pt}{\tikz{\node[diamond,red,fill=red, scale=0.45] at (2.5mm,0.0mm) {};
\draw[-,red,solid,line width = 1.0pt](0.,0.0mm) -- (5mm,0.0mm)
}}}
\newcommand{\bluelineRectangle}{\raisebox{1pt}{\tikz{\node[rectangle,blue,fill=blue, scale=0.8] at (2.75mm,0.0mm) {};
\draw[-,blue,solid,line width = 1.1pt](0.,0.0mm) -- (5.5mm,0.0mm)
}}}
\newcommand{\greenlineDottedRectangle}{\raisebox{1pt}{\tikz{\node[rectangle,green,fill=green, scale=0.8] at (2.75mm,0.0mm) {};\draw[-,green,dotted,line width = 1.1pt](0.,0.0mm) -- (5.5mm,0.0mm)
}}}

\newcommand{\blueline}{\raisebox{3pt}{\tikz{\draw[-,blue,solid,line width = 0.9pt](0,0mm) -- (5mm,0mm);}}}
\newcommand{\bluelineDashed}{\raisebox{3pt}{\tikz{\draw[-,blue,dashed,line width = 0.9pt](0,0mm) -- (5mm,0mm);}}}
\newcommand{\greenlineDotted}{\raisebox{3pt}{\tikz{\draw[-,green,dotted,line width = 0.9pt](0,0) -- (5mm,0);}}}

\newcommand{\yellowlineDashed}{\raisebox{3pt}{\tikz{\draw[-,yellow,dashed,line width = 0.9pt](0,0) -- (5mm,0);}}}
\newcommand{\blacklineDashed}{\raisebox{3pt}{\tikz{\draw[-,black,dashed,line width = 0.9pt](0,0) -- (5mm,0);}}}
\newcommand{\blacklineDashedStar}{\raisebox{2pt}{\tikz{\node[star,black,fill=black, scale=0.5] at (2.5mm,0.0mm) {};
\draw[-,black,dashed,line width = 1.0pt](0.,0.0mm) -- (5mm,0.0mm)
}}}

\newcommand{\blacklineStar}{\raisebox{2pt}{\tikz{\node[star,black,fill=black, scale=0.5] at (2.5mm,0.0mm) {};
\draw[-,black,solid,line width = 1.0pt](0.,0.0mm) -- (5mm,0.0mm)
}}}

\newcommand{\orangelineDiamond}{\raisebox{2pt}{\tikz{
\definecolor{orange}{rgb} {1.0, 0.64, 0.0}
\node[diamond,orange,fill=orange, scale=0.45] at (2.5mm,0.0mm) {};
\draw[-,orange,solid,line width = 1.0pt](0.,0.0mm) -- (5mm,0.0mm)
}}}

\newcommand{\orangelineDashedDiamond}{\raisebox{2pt}{\tikz{
\definecolor{orange}{rgb} {1.0, 0.64, 0.0}
\node[diamond,orange,fill=orange, scale=0.45] at (2.5mm,0.0mm) {};
\draw[-,orange,dashed,line width = 1.0pt](0.,0.0mm) -- (5mm,0.0mm)
}}}

\newcommand{\redlineRectangle}{\raisebox{1pt}{\tikz{
\definecolor{red}{rgb} {1.0, 0.0, 0.0}
\node[rectangle,blue,fill=red, scale=0.8] at (2.75mm,0.0mm) {};
\draw[-,red,solid,line width = 1.1pt](0.,0.0mm) -- (5.5mm,0.0mm)
}}}

\newcommand{\redlineDashedRectangle}{\raisebox{1pt}{\tikz{
\definecolor{red}{rgb} {1.0, 0.0, 0.0}
\node[rectangle,blue,fill=red, scale=0.8] at (2.75mm,0.0mm) {};
\draw[-,red,dashed,line width = 1.1pt](0.,0.0mm) -- (5.5mm,0.0mm)
}}}

\newcommand{\magentalineDashed}{\raisebox{3pt}{\tikz{\draw[-,magenta,dashed,line width = 0.9pt](0,0) -- (5mm,0);}}}
\newcommand{\redlineDashed}{\raisebox{3pt}{\tikz{\draw[-,red,dashed,line width = 0.9pt](0,0) -- (5mm,0);}}}

\newcommand{\redlineDotted}{\raisebox{3pt}{\tikz{\draw[-,red,dotted,line width = 0.9pt](0,0) -- (5mm,0);}}}
\newcommand{\greenlineDashed}{\raisebox{3pt}{\tikz{\draw[-,green,dashed,line width = 0.9pt](0,0) -- (5mm,0);}}}

\newcommand{\greenline}{\raisebox{2pt}{\tikz{\draw[-,green,solid,line width = 0.9pt](0,0) -- (5mm,0);}}}
\newcommand{\redline}{\raisebox{2pt}{\tikz{\draw[-,red,solid,line width = 0.9pt](0,0) -- (5mm,0);}}}

\newcommand{\R}{\mathbb{R}}
\newcommand{\h}{\mathbf{h}}
\renewcommand{\i}{\mathbf{i}}
\renewcommand{\o}{\mathbf{o}}
\newcommand{\W}{\mathbf{W}}

\title{Data-Driven Forecasting of High-Dimensional Chaotic Systems with Long-Short Term Memory Networks}

\author{
Pantelis R. Vlachas\footnote{Chair of Computational Science, ETH Zurich, Clausiusstrasse 33, Zurich, CH-8092, Switzerland} , Wonmin Byeon${}^{*}$, Zhong Y. Wan\footnote{Department of Mechanical Engineering, Massachussetts Institute of Technology, 77 Massachusetts Ave., Cambridge, MA 02139, United States} , \\ Themistoklis P. Sapsis${}^{\dagger}$, Petros Koumoutsakos${}^{*}$\footnote{Corresponding author. Email: petros@ethz.ch}}
\date{11 July 2018}

\begin{document}
\maketitle
\thispagestyle{fancy}

\begin{abstract}
We introduce a data-driven forecasting method for high dimensional, chaotic systems using Long Short-Term Memory (LSTM) recurrent neural networks. The proposed LSTM neural networks perform inference of high dimensional dynamical systems in their reduced order space and are shown to be an effective set of non-linear approximators of their attractor. We demonstrate the forecasting performance of the LSTM and compare it  with Gaussian processes (GPs) in time series obtained from  the Lorenz 96 system, the Kuramoto-Sivashinsky equation and a prototype climate model. The LSTM networks outperform the GPs in short-term forecasting accuracy in all applications considered. A hybrid architecture, extending the LSTM with a mean stochastic model (MSM-LSTM), is proposed to ensure convergence to the invariant measure. This novel hybrid method is fully data-driven and extends the forecasting capabilities of LSTM networks.
\end{abstract}

\section{Introduction}

Natural systems, ranging from climate and ocean circulation to organisms and cells, involve complex dynamics extending over multiple spatio-temporal scales. Centuries old efforts to comprehend and forecast the dynamics of such systems have spurred developments in large scale simulations, dimensionality reduction techniques and a multitude of forecasting methods. The goals of understanding and prediction have been complementing each other but have been hindered by the high dimensionality and chaotic behavior of these systems. In recent years we observe a convergence of these approaches due to advances in computing power, algorithmic innovations and the ample availability of data. A major beneficiary of this convergence are data-driven dimensionality reduction methods \cite{Rowley2005,Williams2015,Tu2014,Kutz2016,
Arbabi2016,Kerschen2005,Sapsis2013}, model identification procedures \cite{Krischer1993,Milano2002, Brunton2016,Duriez2016,Majda2014,Schaeffer2017,Farazmand2016,Bongard2007} and forecasting techniques \cite{
Einicke1999, Julier1999, Lee2016, Comeau2017, Tatsis2017, Quade2016, Mirmomeni2010, Gholipour2007, Mirmomeni2011a, Mirmomeni2011b, Mirmomeni2006, Abdollahzade2015, Ye2011, Cousins2014, Cousins2016} that aim to provide precise short-term predictions while capturing the long-term statistics of these systems. Successful forecasting methods address the highly non-linear energy transfer mechanisms between modes not captured effectively by the dimensionality reduction methods.  

The pioneering technique of analog forecasting proposed in \cite{Lorenz1969} inspired a widespread research in non-parametric prediction approaches.  Two dynamical system states are called analogues if they resemble one another on the basis of a specific criterion. This class of methods uses a training set of historical observations of the system. The system evolution is predicted using the evolution of the closest analogue from the training set corrected by an error term. This approach has led to promising results in practice \cite{Prince2007} but the selection of the resemblance criterion to pick the optimal analogue is far from straightforward. Moreover, the geometrical association between the current state and the training set is not exploited. More recently \cite{Zhao2016}, analog forecasting is performed using a weighted combination of data-points based on a localized kernel that quantifies the similarity of the new point and the weighted combination. This technique exploits the  local geometry  instead of selecting a single optimal analogue.
Similar kernel-based methods, \cite{Chiavazzo2014} use diffusion maps to globally parametrize a low dimensional manifold capturing the slower time scales. Moreover, non-trivial interpolation schemes are investigated in order to encode the system dynamics in this reduced order space as well as map them to the full space (lifting). Although the geometrical structure of the data is taken into account, the solution of an eigen-system with a size proportional to the training data is required, rendering the approach computationally expensive. In addition, the inherent uncertainty due to sparse observations in certain regions of the attractor introduces prediction errors which cannot be modeled in a deterministic context. In \cite{Zhong2017} a method based on Gaussian process regression (GPR) \cite{Rasmussen2006} was proposed for prediction and uncertainty quantification in the reduced order space. The technique is based on a training set that sparsely samples the attractor. Stochastic predictions exploit the geometrical relationship between the current state and the training set, assuming a Gaussian prior over the modeled latent variables. A key advantage of GPR is that uncertainty bounds can be analytically derived from the hyper-parameters of the framework. Moreover, in \cite{Zhong2017} a Mean Stochastic Model (MSM) is used for under-sampled regions of the attractor to ensure accurate modeling of the steady state in the long-term regime. However the resulting  inference and training have a quadratic cost in terms of the number of data samples $O(N^2)$.

Some of the earlier approaches to capture the evolution of time series in chaotic systems using recurrent neural networks were developed during the inception of the Long Short-Term Memory networks (LSTM) \cite{Hochreiter1997}. However, to the best of our knowledge, these methods have been used only on low-dimensional chaotic systems \cite{Gers2002}. Similarly, other machine learning techniques based on Multi Layer Perceptrons (MLP) \cite{Rico1992}, Echo State Networks (ESNs) \cite{Jaeger2004,Chatzis2011} or radial basis functions \cite{Broomhead1988,Kim2000ControlOC} have been successful, albeit only for low order dynamical systems. Recent work in \cite{Pathak2018, Pathak2017} demonstrated promising results of ESNs for high dimensional chaotic systems.

In this paper, we propose LSTM based methods that exploit information of the recent history of the reduced order state to predict the high-dimensional dynamics. Time-series data are used to train the model while no knowledge of the underlying system equations is required. Inspired by Taken's theorem \cite{Takens1981} an embedding space is constructed using time delayed versions of the reduced order variable. The proposed method tries to identify an approximate forecasting rule globally for the reduced order space. In contrast to GPR \cite{Zhong2017}, the method has a deterministic output while its training cost scales linearly with the number of training samples and it exhibits an $\mathcal{O}(1)$  inference computational cost. Moreover, following \cite{Zhong2017}, LSTM is combined with a MSM, to cope with attractor regions that are not captured in the training set. In attractor regions, under-represented in the training set, the MSM is used to guarantee convergence to the invariant measure and avoid an exponential growth of the prediction error. The effectiveness of the proposed hybrid method in accurate short-term prediction and capturing the long-term behavior is shown in the Lorenz 96 system and the Kuramoto-Sivashinsky system. Finally the method is also tested on predictions of a prototypical climate model.

The structure of the paper is as follows: In Section \ref{sec:lstm} we explain how the LSTM can be employed for modeling and prediction of a reference dynamical system and a blended LSTM-MSM technique is introduced. In Section \ref{sec:benchmark} three other state of the art methods, GPR, MSM and the hybrid GPR-MSM scheme are presented and two comparison metrics are defined. The proposed LSTM technique and its LSTM-MSM extension are benchmarked against GPR and GPR-MSM in three complex chaotic systems in Section \ref{sec:applications}. In Section \ref{sec:complexity} we discuss the  computational complexity of training and inference in LSTM. Finally, Section \ref{sec:conclusion} offers a summary and discusses future research directions.

\section{Long Short-Term Memory (LSTM) Recurrent Neural Networks}
\label{sec:lstm}

\begin{figure}[!httb]
\centering
\begin{subfigure}{.25\textwidth}
\centering
\includegraphics[height=3cm]{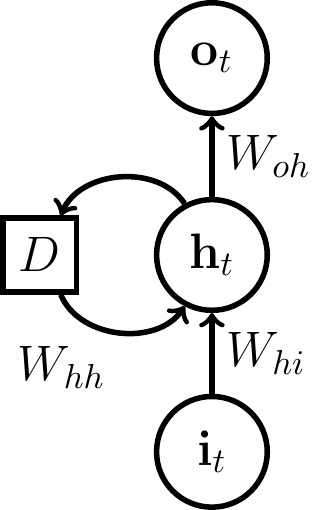}
\caption{}
\label{fig:RNN_1}
\end{subfigure}
\begin{subfigure}{.7\textwidth}
\centering
\includegraphics[height=3cm]{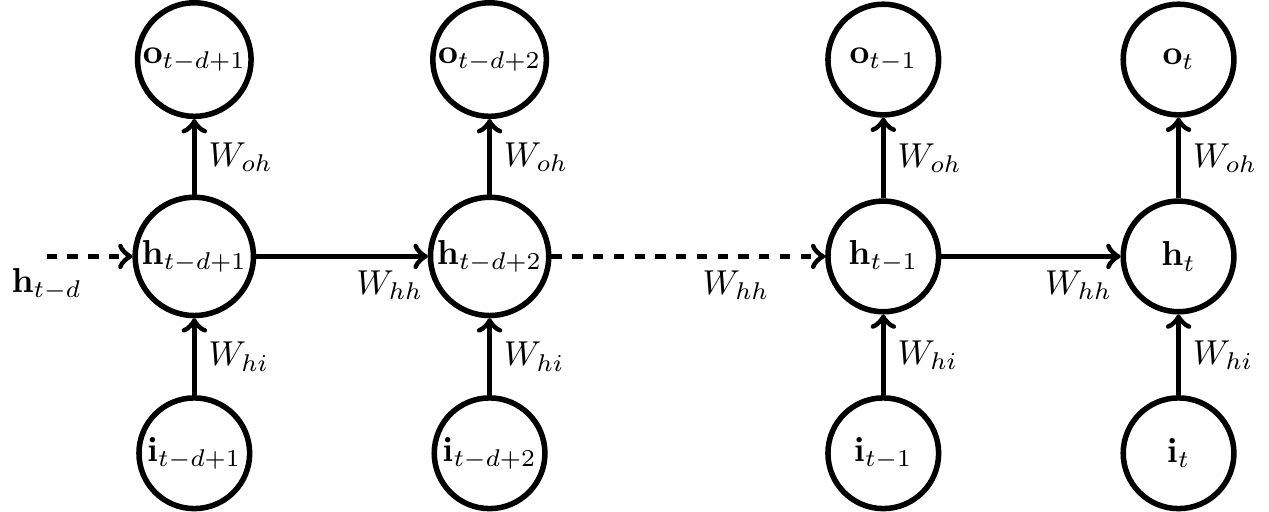}
\caption{}
\label{fig:RNN_2}
\end{subfigure}
\caption{\textbf{a)} A recurrent neural network cell, where $D$ denotes a delay. The hidden cell state $\h_t$ depends on the input $\i_t$ and its previous value $\h_{t-1}$. The output $\o_t$ depends on the hidden state. The weight matrices are parameters of the cell. \textbf{b)} A recurrent neural network unfolded in time (unfolding the delay). The same weights are used at each time step to compute the output $\o_t$ that depends on the current input $\i_t$ and short-term history (recursively) encoded in $h_{t-1}$.}
\label{fig:RNN}
\end{figure}

The LSTM was introduced in order to regularize the training of recurrent neural networks (RNNs)\cite{Hochreiter1997}. RNNs contain loops that allow information to be passed between consecutive temporal steps (see Figure \ref{fig:RNN_1}) and can be expressed as: 
\begin{gather}
\h_t = \sigma_{h} \big( \W_{hi} \i_{t} + \W_{hh} \h_{t-1} + b_{h}  \big), \\
\o_t = \sigma_{o} \big( \W_{oh} \h_t + b_{o}  \big) = f^{w}(\i_t, \h_{t-1})
\label{eq:rnnequations}
\end{gather}
where $\i_t \in \mathbb{R}^{d_i}$, $\o_t \in  \mathbb{R}^{d_o}$ and $\h_t\in  \mathbb{R}^{d_h}$ are the input, the output and the hidden state of the RNN at time step $t$. The weight matrices are $\W_{hi} \in \R^{d_h \times d_i}$ (input-to-hidden), $\W_{hh} \in \R^{d_h \times d_h}$ (hidden-to-hidden), $\W_{oh} \in \R^{d_o \times d_h}$ (hidden-to-output), $b_h $ and $b_o $. Moreover, $\sigma_{h}$ and $\sigma_{o}$ are the hidden and output activation functions, while $b_{h}\in \R^{d_h}$ and $b_{o}\in \R^{d_o}$ are the respective biases. Temporal dependencies are captured by the hidden-to-hidden weight matrix $\W_{hh}$, which couples two consecutive hidden states together. A schematic of the RNN architecture is given in Figure \ref{fig:RNN}.

In many practical applications, RNNs  suffer from the vanishing (or exploding) gradient problem and have failed to  capture long-term dependencies \cite{Hochreiter1991, Bengio1994}. Today the RNNs owe their renaissance largely  to the LSTM, that copes effectively with the aforementioned problem using \textit{gates}. The LSTM has been successfully applied in sequence modeling \cite{Wierstra2005}, speech recognition \cite{Graves2013}, hand-writing recognition \cite{Graves2007} and language translation \cite{Wu2016}.  

The equations of the LSTM are
\begin{equation}
\begin{aligned}
g^f_t &= \sigma_f \big(\W_f [\h_{t-1}, \i_t ] + b_f\big) 
&& g^{i}_t = \sigma_i \big( \W_i [\h_{t-1}, \i_t ] +b_i \big) \\
\tilde{C}_t &= tanh \big( \W_C [\h_{t-1}, \i_t ] +b_C \big) 
&& C_t = g^f_t C_{t-1} + g^{i}_t \tilde{C}_t   \\
g^o_t &= \sigma_h \big( \W_h [\h_{t-1}, \i_t ] + b_h \big) 
&& \h_t =  g^o_t \, tanh(C_t),
\end{aligned}
\label{eq:lstmequations}
\end{equation}
where $g^f_t, g^{i}_t, g^{o}_t \in \R^{d_h \times (d_h+d_i)}$, are the gate signals (forget, input and output gates), $\i_{t} \in \R^{d_i}$ is the input, $\h_{t} \in \R^{d_h}$ is the hidden state, $\mathbf{C}_{t}\in \R^{d_h}$ is the cell state, while $\W_f$, $\W_i$, $\W_C, \W_h$ $\in \R^{d_h \times (d_h+d_i)}$, are weight matrices and $b_f, b_i, b_C, b_h \in \R^{d_h}$  biases. The activation functions $\sigma_f$, $\sigma_i$ and $\sigma_h$ are sigmoids. For a more detailed explanation on the LSTM architecture refer to \cite{Hochreiter1997}. In the following we refer to the LSTM hidden and cell states ($\h_t$ and $C_t$) jointly as \textit{LSTM states}. The dimension of these states is called the number of hidden units $h=d_{h}$ and it controls the capability of the cell to encode history information. In practice we want the output to have a specific dimension $d_o$. For this reason, a fully connected final layer without activation function $\W_{oh} \in \mathbb{R}^{d_o \times h}$ is added
\begin{gather}
\o_t = \W_{oh} \h_t = f^w(\i_t, \h_{t-1}, C_{t-1}),
\label{eq:outputlayerequations}
\end{gather}
where all parameters (weights and biases) are encoded in $w = \{\W_f$, $\W_i$, $\W_C, \W_h, b_f, b_i, b_C, b_h \}$ and $f^{w}$ is the LSTM cell function that maps the previous \textit{LSTM States} $\h_{t-1}, C_{t-1}$  and current input $\i_t$ to the output. By unfolding the LSTM $d$ time-steps in the past and ignoring dependencies longer that $d$ we get 
\begin{gather}
\o_t =  f^w(z_t, \h_{t-1}, C_{t-1}) = \mathcal{F}^w(\underbrace{z_t, z_{t-1}, \cdots, z_{t-d+1}}_{z_{t:t-d+1}}, \color{red}{\cancelto{0}{\h_{t-d}}}
 , \cancelto{0}{C_{t-d}}
 \color{black}{
 ),
 }
\label{eq:iterativelstm}
\end{gather}
where $\mathcal{F}^w$ represents the iterative application of $f^w$ and computation of the \textit{LSTM states} for $d$ time steps. For a more detailed explanation of the formula for $\mathcal{F}^w$, and a Figure of the neural network architecture refer to the Appendix.

In this work, we consider the reduced order problem where the state of a dynamical system is projected in the reduced order space. The system is considered to be autonomous, while $\dot{z}_t = \frac{dz_t}{dt}$ is the system state derivative at time step $t$. Following \cite{Gers2002}, The LSTM is trained using time series data from the system in the reduced order space $D=\{z_{1:T},  \dot{z}_{1:T}\}$ to predict the reduced state derivative $\dot{z}_t$ from a short history of the reduced order state $\{z_{t}, z_{t-1}, \dots, z_{t-d+1} \}$ consisting of $d$ past temporally consecutive states. In this work, we approximated the derivative from the original time series using first order forward differences. The loss that has to be minimized is defined as
\begin{equation}
\mathcal{L}(D, w) = \frac{1}{T-d+1} \sum_{t=d}^{T} ||\underbrace{\mathcal{F}^w(z_{t:t-d+1})}_{\o_t} - \dot{z}_t||^2.
\end{equation}
The short-term history for the states before $z_d$ is not available, that is why in total we have $T-d+1$ training samples from a time series with $T$ samples. During training the weights of the LSTM are optimized according to $w^{\star} = \underset{w}{\operatorname{argmin}} \, \mathcal{L}(D, w)$. The parameter $d$ is denoted as truncation layer and time dependencies longer than $d$ are not explicitly captured in the loss function.

Training of this model is performed using Back-propagation through time, truncated at layer $d$ and mini-batch optimization with the Adam method \cite{Kingma2017} with an adaptive learning rate (initial learning rate $\eta=0.0001$). The LSTM weights are initialized using the method of Xavier \cite{Xavier2010}. Training is stopped when convergence of the training error is detected or the maximum of $1000$ epochs is reached. During training the loss of the model is evaluated on a separate validation dataset to avoid overfitting. The training procedure is explained in detail in the Appendix.

\begin{figure}[!httb]
\centering
\includegraphics[height=4.4cm]{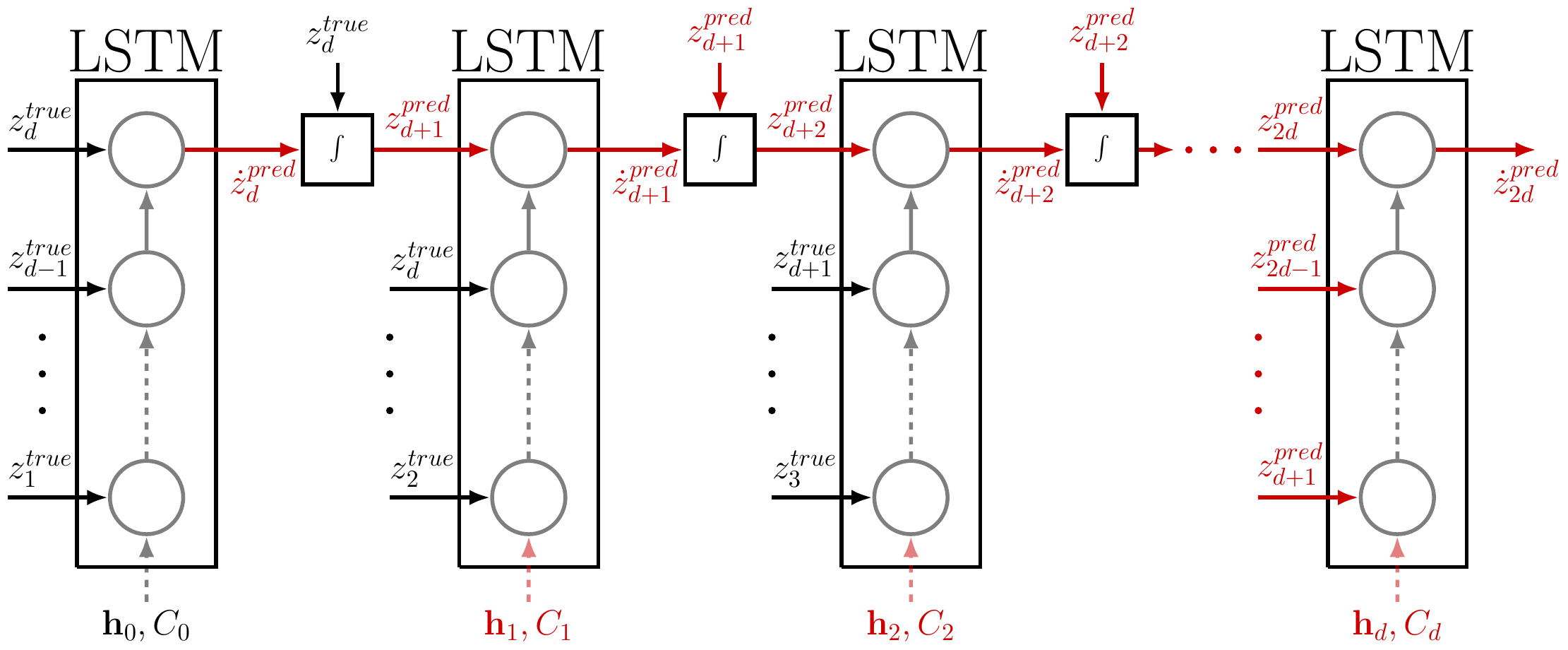}
\caption{Iterative prediction using the trained LSTM model. A short-term history of the system, i.e. $z_1^{true}, \dots, z_d^{true}$, is assumed to be known. Initial \textit{LSTM states} are $h_0, C_0$. The trained LSTM is used predict the derivative $\dot{ z}_d^{pred}=\mathcal{F}^w(z_{d:1}^{true}, \h_{0}, C_{0})$. The state prediction $z_{d+1}^{pred}$ is obtained by integrating this derivative. This value is used for the next prediction in an iterative fashion. After $d$ time-steps only predicted values are fed in the input. The short-term history is illustrated with black arrows, while predicted values with red. In \textit{stateless} LSTM, $\h$ and $C$ are initialized to zero before every prediction.
}
\label{fig:lstm_iterative_prediction}
\end{figure}

An important issue is how to select the hidden state dimension $h$ and how to initialize the \textit{LSTM states} $\h_{t-d}, C_{t-d}$ at the truncation layer $d$. A small $h$ reduces the expressive capabilities of the LSTM and deteriorates inference performance. On the other hand, a big $h$ is more sensitive to overfitting and the computational cost of training rises. For this reason, $h$ has to be tuned depending on the observed training behavior. In this work, we performed a grid search and selected the optimal $h$ for each application. For the truncation layer $d$, there are two alternatives, namely \textit{stateless} and \textit{stateful} LSTM. In \textit{stateless} LSTM the \textit{LSTM states} at layer $d$ are initialized to zero as in equation (\ref{eq:iterativelstm}). As a consequence, the LSTM can only capture dependencies up to $d$ previous time steps. In the second variant, the \textit{stateful} LSTM, the state is always propagated for $p$ time steps in the future and then reinitialized to zero, to help the LSTM capture longer dependencies. In this work, the systems considered exhibit chaotic behavior and the dependencies are inherently short-term, as the states in two time steps that differ significantly can be considered statistically independent. For this reason, the short temporal dependencies can be captured without propagating the hidden state for a long horizon. As a consequence, we consider only the  \textit{stateless} variant $p=0$. We also applied \textit{stateful} LSTM without any significant improvement so we omit the results for brevity. The trained LSTM model can be used to iterative predict the system dynamics as illustrated in Figure \ref{fig:lstm_iterative_prediction}. This is a solely data-driven approach and no explicit information regarding the form of the underlying equations is required.

\subsection{Mean Stochastic Model (MSM) and Hybrid LSTM-MSM}
\label{sec:hybrid}

The MSM is a powerful data-driven method used to quantify uncertainty and perform forecasts in turbulent systems  with high intrinsic attractor dimensionality \cite{Majda2012, Zhong2017}. It is parameterized a priori to capture global statistical information of the attractor by design, while its computationally complexity is very low compared to LSTM or GPR. The concept behind MSM is to model each component of the state $z^i$ independently with an Ornstein-Uhlenbeck (OU) process that captures the energy spectrum and the damping time scales of the statistical equilibrium. The process takes the following form
\begin{gather}
dz^i=c_i z^i dt + \xi_i dW_i,
\end{gather}
where $c_i, \xi_i$ are parameters fitted to the centered training data and $W_i$ is a Wiener process. In the statistical steady state the mean, energy and damping time scale of the process are given by
\begin{gather}
\mu_i = \mathbb{E}[z^i]=0, \quad
E_i = \mathbb{E}[z^i (z^i)^{*}]=-\frac{\xi^2}{2 c_i}, \quad T_i=-\frac{1}{c_i},
\label{eq:msm}
\end{gather}
where $(z^i)^*$ denotes the complex conjugate of $z^i$. In order to fit the model parameters $c_i, \xi_i$ we directly estimate the variance $\mathbb{E}[z^i (z^i)^*]$ from the time series training data and the decorrelation time using
\begin{gather}
T_i = \frac{1}{\mathbb{E}[z^i (z^i)^*]} \int_0^{\infty} \mathbb{E} \big[z^i(t) \big(z^i(t+\tau) \big)^* \big] d\tau.
\end{gather}
After computing these two quantities we replace in (\ref{eq:msm}) and solve with respect to $c_i$ and $\xi_i$. Since the MSM is modeled a priori to mimic the global statistical behavior of the attractor, forecasts made with MSM can never escape. This is not the case with LSTM and GPR, as prediction errors accumulate and iterative forecasts escape the attractor due to the chaotic dynamics, although short-term predictions are accurate. This problem has been addressed with respect to GPR in \cite{Zhong2017}. In order to cope effectively with this problem we introduce a hybrid LSTM-MSM technique that prevents forecasts from diverging from the attractor.

The state dependent decision rule for forecasting in LSTM-MSM is given by
\begin{gather}
\dot{z}_t =
\begin{cases}
 (\dot{z}_t)_{LSTM}, \,\,\, \text{if} \,\,\, p^{train}(z_t)=\prod p^{train}_i(z_t^i) >\delta \\
 (\dot{z}_t)_{MSM}, \,\,\, \text{otherwise} 
\end{cases}
\end{gather}
where $p^{train}(z_t)$ is an approximation of the probability density function of the training dataset and $\delta \approx 0.01$ a constant threshold tuned based on $p^{train}(z_t)$. We approximate $p^{train}(z_t)$ using a mixture of Gaussian kernels. This hybrid architecture exploits the advantages of LSTM and MSM. In case there is a high probability that the state $z_i$ lies close to the training dataset (interpolation) the LSTM having memorized the local dynamics is used to perform inference. This ensures accurate LSTM short-term predictions. On the other hand, close to the boundaries the attractor is only sparsely sampled $p^{train}(z_i) <\delta$ and errors from LSTM predictions would lead to divergence. In this case, MSM guarantees that forecasting trajectories remain close to the attractor, and that we converge to the statistical invariant measure in the long-term.

\section{Benchmark and Performance Measures}
\label{sec:benchmark}

The performance of the proposed LSTM based prediction mechanism is benchmarked against the following state-of-the-art methods:
\begin{itemize}
\item Mean Stochastic Model (MSM)
\item Gaussian Process Regression (GPR) 
\item Mixed Model (GPR-MSM)
\end{itemize}

In order to guarantee that the prediction performance is independent of the initial condition selected, for all applications and all performance measures considered the average value of each measure for a number of different initial conditions sampled independently and uniformly from the attractor is reported. The ground truth trajectory is obtained by integrating the discretized reference equation starting from each initial condition, and projecting the states to the reduced order space. The reference equation and the projection method are of course application dependent. 

From each initial condition, we generate an empirical Gaussian ensemble of dimension $N_{en}$ around the initial condition with a small variance $\sigma_{en}$. This noise represents the uncertainty in the knowledge of the initial system state. We forecast the evolution of the ensemble by iteratively predicting the derivatives and integrating (deterministically for each ensemble member for the LSTM, stochastically for GPR) and we keep track of the mean. We select an ensemble size $N_{en}=50$, which is the usual choice in environmental science, e.g. weather prediction and short-term climate prediction \cite{Majda2005}.

The ground truth trajectory at each time instant $z$ is then compared with the predicted ensemble mean $\widetilde{z}$. As a comparison measure we use the root mean square error (RMSE) defined as
$RMSE(z_k) = \sqrt{1/V \, \sum_{i=1}^{V} \Big( z_k^{i} - \widetilde{z}_k^{\, i} \Big)^2},
$
where index $k$ denotes the $k$\textsuperscript{th} component of the reduced order state $z$, $i$ is the initial condition, and $V$ is the total number of initial conditions. The RMSE is computed at each time instant for each component $k$ of the reduced order state, resulting in error curves that describe the evolution of error with time.  Moreover, we use the standard deviation $\sigma$ of the attractor samples in each dimension as a relative comparison measure. Assuming that the attractor consists of samples $\{z_1,z_2,\dots, z_N \}$, with $z_j \in \R^{d_i}$, the attractor standard deviation in dimension $i \in \{1,\dots,d_i \}$ is defined as $\sigma_i = \sqrt{ \mathbb{E}[(z^i-\overline{z}^i)^2]) }$, where $\overline{z}^i$ is the mean of the samples in this dimension. If the prediction error is bigger than this standard deviation, then a trivial mean predictor performs better.

Moreover, we use the mean Anomaly Correlation Coefficient (ACC) \cite{Allgaier2012} over $V$ initial conditions to quantify the pattern correlation of the predicted trajectories with the ground-truth. The ACC is defined as
\begin{equation}
ACC =  \frac{1}{V} \sum_{i=1}^{V} \frac{ \sum_{k=1}^{r_{dim}} w_k \Big( z_{k}^{i} - \overline{z}_{k} \Big)   \Big( \widetilde{z}_{k}^{\, i} - \overline{z}_{k} \Big)}{ \sqrt{ \sum_{k=1}^{r_{dim}} w_k \Big( z_{k}^{i} - \overline{z}_{k} \Big)^2   \sum_{k=1}^{r_{dim}} w_k \Big(
\widetilde{z}_{k}^{\, i} - \overline{z}_{k} \Big) ^2  }},
\label{eq:acc}
\end{equation}
where $k$ refers to the mode number, $i$ refers to the initial condition, $w_k$ are mode weights selected according to the energies of the modes after dimensionality reduction and $\overline{z}_k$ is the time average of the respective mode, considered as reference. This score ranges from $-1.0$ to $1.0$. If the forecast is perfect, the score equals to $1.0$. The ACC coefficient is a widely used forecasting accuracy score in the meteorological community \cite{Basnarkov2012}.


\section{Applications}
\label{sec:applications}

In this section, the effectiveness of the proposed method is demonstrated with respect to three chaotic dynamical systems, exhibiting different levels of chaos, from weakly chaotic to fully turbulent, i.e. the Lorenz 96 system, the Kuramoto-Sivashinsky equation and a prototypical barotropic climate model.


\subsection{The Lorenz 96 System}
\label{sec:lorenz96}

In \cite{Lorenz96} a model of the large-scale behavior of the mid-latitude atmosphere is introduced. This model describes the time evolution of the components $X_j$ for $j \in \{0,1,\dots,J-1 \}$ of a spatially discretized (over a single latitude circle) atmospheric variable. In the following we refer to this model as the Lorenz 96. The Lorenz 96 is usually used (\cite{Basnarkov2012, Zhong2017} and references therein) as a toy problem to benchmark methods for weather prediction.

\begin{figure}[!ht]
\centering
\includegraphics[width=0.8\textwidth, trim={0 1.5cm 0 0}, clip]{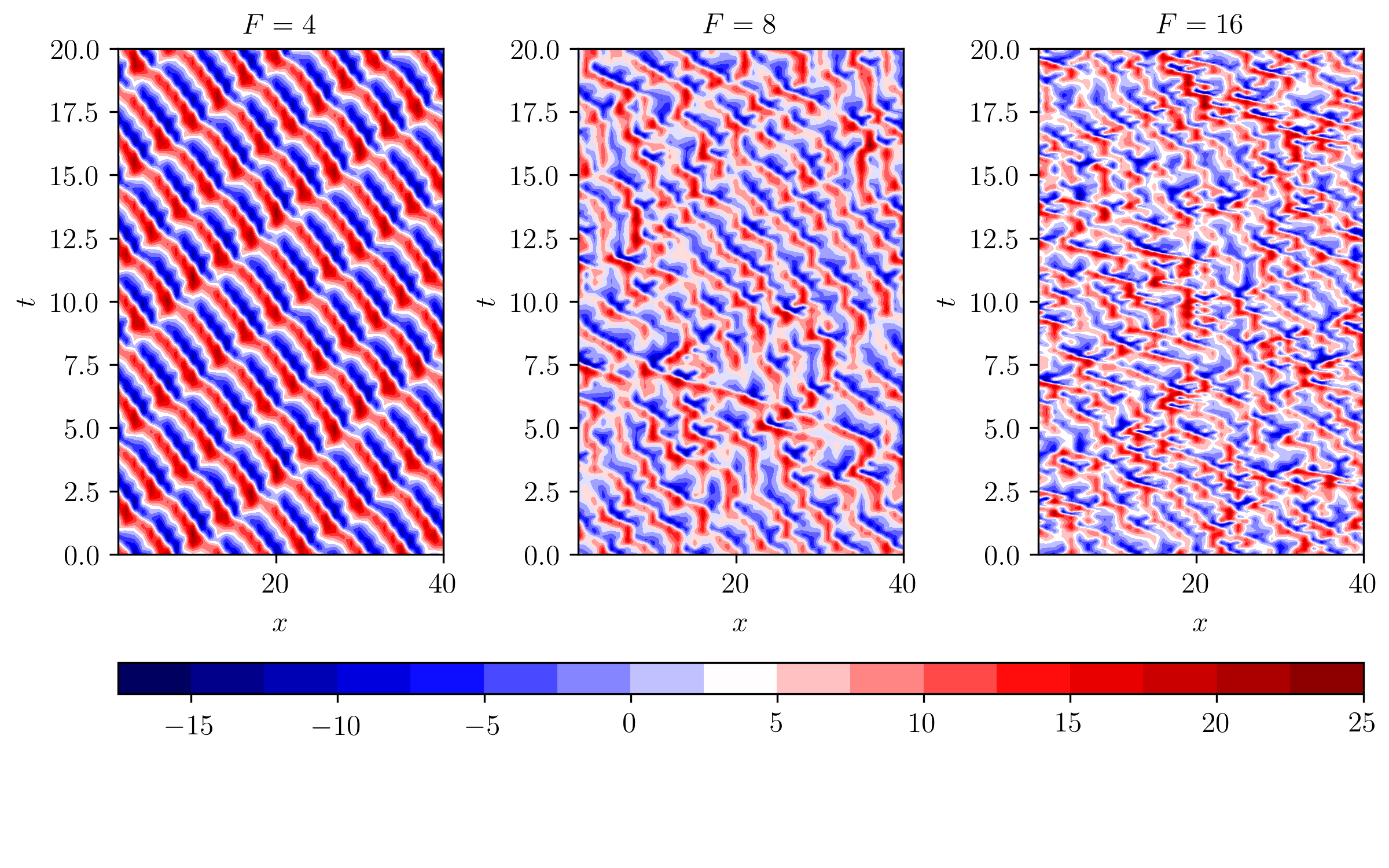}
\caption{Lorenz 96 contour plots for different forcing regimes $F$. Chaoticity rises with bigger values of $F$.}
\label{fig:Plot_U_Lorenz}
\end{figure}

The system of differential equations that governs the Lorenz 96 is defined as
\begin{equation}
\frac{dX_j}{dt} = (X_{j+1}-X_{j-2})X_{j-1}-X_j+F, 
\label{eq:Lorenz96}
\end{equation}
for $j \in \{0,1,\dots,J-1 \}$, where by definition $X_{-1} = X_{J},X_{-2} = X_{J-1}$. In our analysis $J = 40$. The right-hand side of (\ref{eq:Lorenz96}) consists of a non-liner adjective term $(X_{j+1}-X_{j-2})X_{j-1}-X_j$, a linear advection (dissipative) term $-X_j$ and a positive external forcing term $F$. The discrete energy of the system remains constant throughout time and the Lorenz 96 states $X_j$ remain bounded. By increasing the external forcing parameter F the behavior that the system exhibits changes from periodic $F < 1$ to weakly chaotic ($F = 4$) to end up in fully turbulent regimes ($F = 16$). These regimes can be observed in Figures \ref{fig:Plot_U_Lorenz}.

Following \cite{Zhong2017, Majda2005} we apply a shifting and scaling to standardize the Lorenz 96 states $X_j$. The discrete or Dirichlet energy is given by $E = \frac{1}{2} \sum_{j=1}^J X_j^2$. In order for the scaled Lorenz 96 states to have zero mean and unit energy we transform them using 
\begin{equation}
\tilde{X}_j = \frac{X_j - \overline{X}}{\sqrt{E_p}}, \quad d\tilde{t}=\sqrt{E_p}dt, \quad E_p = \frac{1}{2T} \sum_{j=0}^{J-1} \int_{T_0}^{T_0+T} (X_j-\overline{X})^2 dt,
\label{eq:transformation}
\end{equation}
where $E_p$ is the average energy fluctuation. In this way the scaled energy is $\tilde{E}=\frac{1}{2} \sum_{j=0}^{J-1} \tilde{X}_j^2=1$ and the scaled variables have zero mean
$\overline{\tilde{X}} =\frac{1}{J} \sum_{j=0}^{J-1} \tilde{X}_j=0,$ with $\overline{X}$ the mean state. The scaled Lorenz 96 states $\tilde{X}_j$ obey the following differential equation
\begin{equation}
\begin{split}
\frac{d \tilde{X}_j}{d \tilde{t}} = \frac{F-\overline{X}}{E_p}+ \frac{(\tilde{X}_{j+1}-\tilde{X}_{j-2})\overline{X} - \tilde{X}_{j}}{\sqrt{E_p}} + (\tilde{X}_{j+1} - \tilde{X}_{j-2} )\tilde{X}_{j-1}
\end{split}
\end{equation}

\subsubsection{Dimensionality Reduction: Discrete Fourier Transform}
\label{sec:lorenzdimred}

Firstly, the Discrete Fourier Transform (DFT) is applied to the energy standardized Lorenz 96 states $\tilde{X}_j$. The Fourier coefficients $\hat{X}_k \in \mathbb{C}$ and the inverse DFT to recover the Lorenz 96 states are given by
\begin{equation}
\hat{X}_k = \frac{1}{J} \sum_{j=0}^{J-1}\tilde{X}_j e^{-2 \pi i k j /J}, \quad \tilde{X}_j = \sum_{k=0}^{J-1} \hat{X}_k e^{2 \pi i k j /J}.
\end{equation}

\begin{figure}[!httb]
\centering
\begin{subfigure}{.46\textwidth}
\centering
\includegraphics[width=1\textwidth]{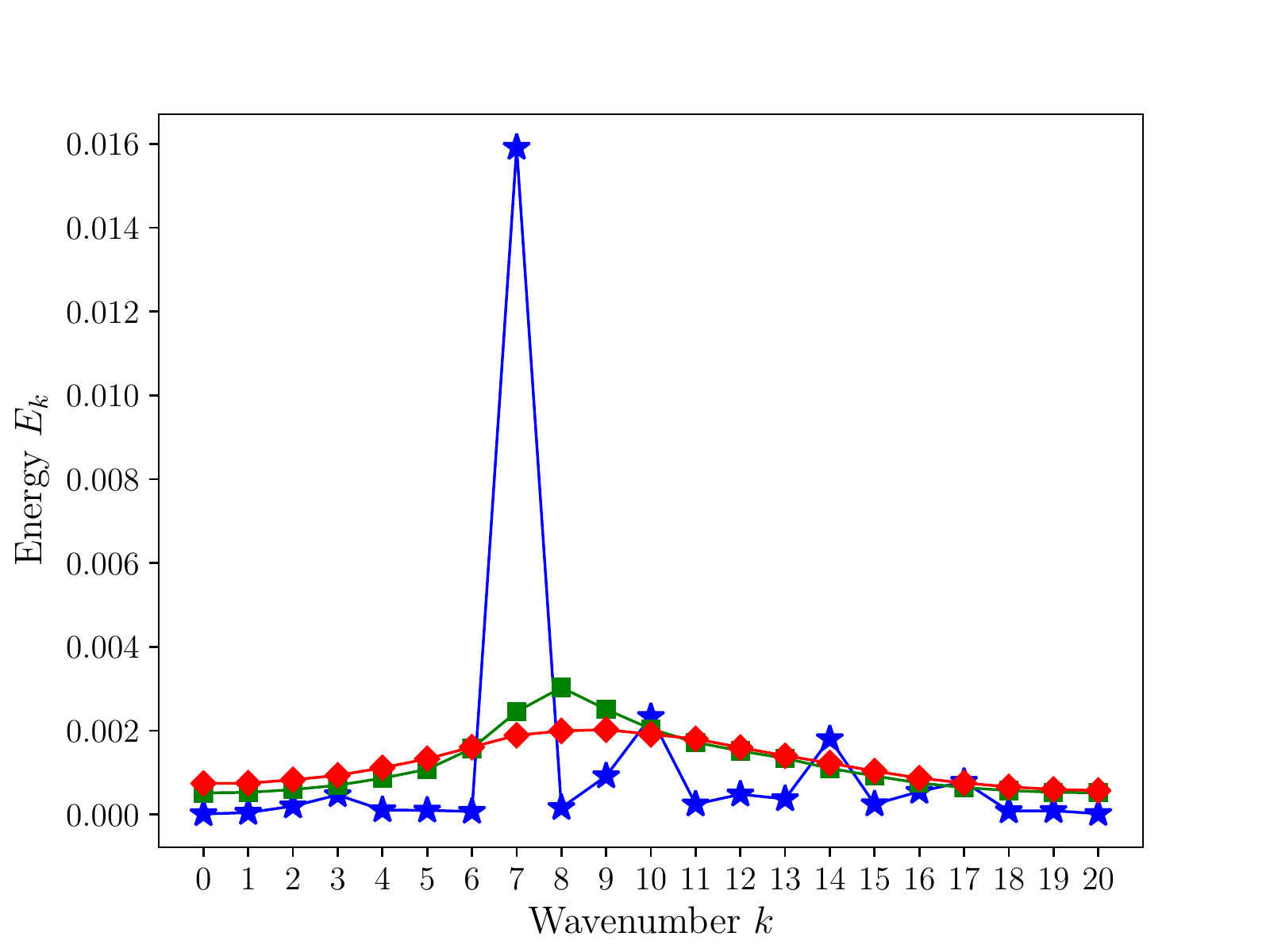}
\end{subfigure} 
\hspace{0.6cm}
\begin{subfigure}{.46\linewidth}
\includegraphics[width=1\textwidth]{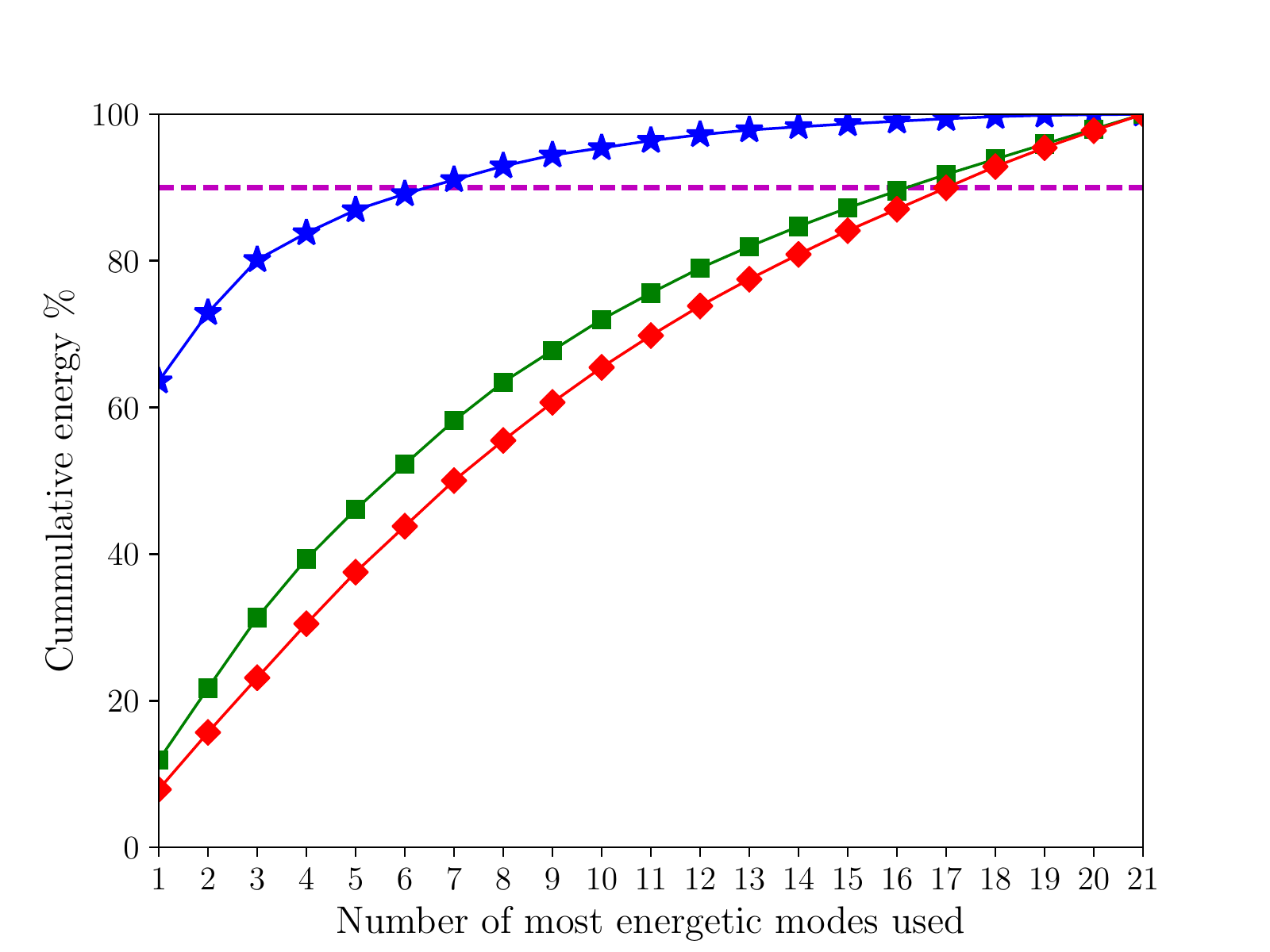}
\end{subfigure}
\caption{Energy spectrum $E_k$ and cumulative energy with respect to the number of most energetic modes used for different forcing regimes of Lorenz 96 system. As the forcing increases, more chaoticity is introduced to the system. \\ $F=4$ \protect\bluelineStar ; $F=8$\protect \greenlineRectangle  ; $F=16$ \protect \redlineDiamond ; $90\%$ of the total energy \protect \magentalineDashed}
\label{fig:Plot_E_lorenz}
\end{figure}

After applying the DFT to the Lorenz 96 states we end up with a symmetric energy spectrum that can be uniquely characterized by $J/2+1$ ($J$ is considered to be an even number) coefficients $\hat{X}_k$ for $k \in K = \{0,1,\cdots,J/2\}$. In our case $J = 40$, thus we end up with $|K| = 21$ complex coefficients $\hat{X}_k \in  \mathbb{C}$. These coefficients are referred to as the Fourier modes or simply modes. The Fourier energy of each mode is defined as $E_k =Var(\hat{X}_k)=\mathbb{E} [ (\hat{X}_k(\tilde{t})-\overline{\hat{X}}_k)(\hat{X}_k(\tilde{t})-\overline{\hat{X}}_k)^{*} ]$.

The energy spectrum of the Lorenz 96 system is plotted in Figure \ref{fig:Plot_E_lorenz} for different values of the forcing term $F$. We take into account only the $r_{dim} = 6$ modes corresponding to the highest energies and the rest of the modes are truncated. For the different forcing regimes $F=4,8,16$, the six most energetic modes correspond to approximately $89\%$, $52\%$ and $43.8\%$ of the total energy respectively. The space where the reduced variables live in is referred to as the reduced order phase space and the most energetic modes are notated as $\hat{X}_{k}^{r}$ for $k\in \{1, \dots , r_{dim} \}$. As shown in \cite{Crommelin2004} the most energetic modes are not necessarily the ones that capture better the dynamics of the model. Including more modes, or designing a criterion to identify the most important modes in the reduced order space may boost prediction accuracy. However, in this work we are not interested in an optimal reduced space representation, but rather in the effectiveness of a prediction model given this space. The truncated modes are ignored for now. Nevertheless, their effect can be modeled stochastically as in \cite{Zhong2017}.

Since each Fourier mode $\hat{X}_{k}^{r}$ is a complex number, it consists of a real part and an imaginary part. By stacking these real and imaginary parts of the $r_{dim}$ truncated modes we end up with the $2 \, r_{dim}$ dimensional reduced model state
\begin{equation}
\mathbf{X} \equiv  [ Re(\hat{X}_{1}^{r}), \dots, Re(\hat{X}_{r_{dim}}^{r}), Im(\hat{X}_{1}^{r}), \dots, Im(\hat{X}_{r_{dim}}^{r}) ]^T
\label{eq:reducedmodelstate}
\end{equation}
Assuming that $X_j^t$ for $j \in \{0,1,\dots,J-1\}$ are the Lorenz 96 states at time instant $t$, the mapping $X_j^t$, $\forall j \to \mathbf{X} $ is unique and the reduced model state of the Lorenz 96 has a specific vector value.

\subsubsection{Training and Prediction in Lorenz 96}

The reduced Lorenz 96 system states $\mathbf{X}_{t}$ are considered as the true reference states $z_t$. The LSTM is trained to forecast the derivative of the reduced order state $\dot{z}_t$ as elaborated in Section \ref{sec:lstm}. We use a \textit{stateless LSTM} with $h=20$ hidden units and the back-propagation truncation horizon set to $d=10$.

In order to obtain training data for the LSTM, we integrate the Lorenz 96 system state Eg. (\ref{eq:Lorenz96}) starting from an initial condition $X_j^0$ for $j \in \{0,1,\dots,J-1\}$ using a Runge-Kutta 4th order method with a time step $dt=0.01$ up to $T=51$. In this way a time series $X_j^t, \, t \in \{0,1,\cdots\}$ is constructed. We obtain the reduced order state time series $\mathbf{X}_t, \, t \in \{0,1,\cdots \}$, using the DFT mapping $X_{t}^j \forall j \to \mathbf{X}_{t}$. From this time series we discard the first $10^4$ initial time steps as initial transients, ending up with a time series with $N^{train}=50000$ samples. A similar but independent process is repeated for the validation set.

\subsubsection{Results}
\label{sec:lorenzresults}

The trained LSTM models are used for prediction based on the iterative procedure explained in Section \ref{sec:lstm}. In this section, we demonstrate the forecasting capabilities of LSTM and compare it with GPs. $100$ different initial conditions uniformly sampled from the attractor are simulated. For each initial condition, an ensemble with size $N_{en}=50$ is considered by perturbing it with a normal noise with variance $\sigma_{en}=0.0001$.

\begin{figure}[!httb]
\centering
\begin{subfigure}{.31\textwidth}
\centering
\includegraphics[width=1\textwidth]{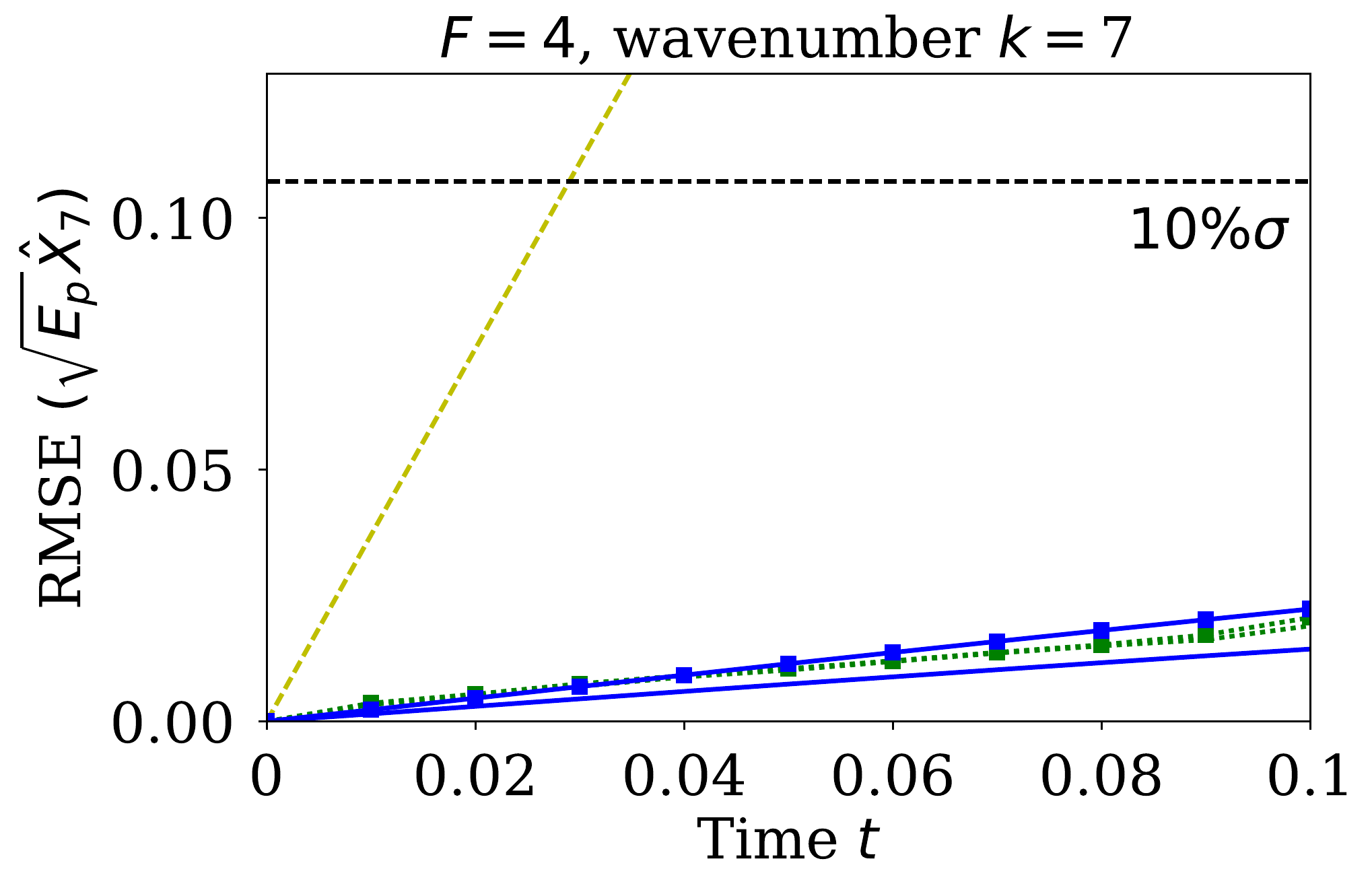}
\vspace{-0.75cm}
\caption{}
\label{results_lorenz:a}
\end{subfigure} 
\hspace{0.3cm}
\begin{subfigure}{.31\textwidth}
\centering
\includegraphics[width=1\textwidth]{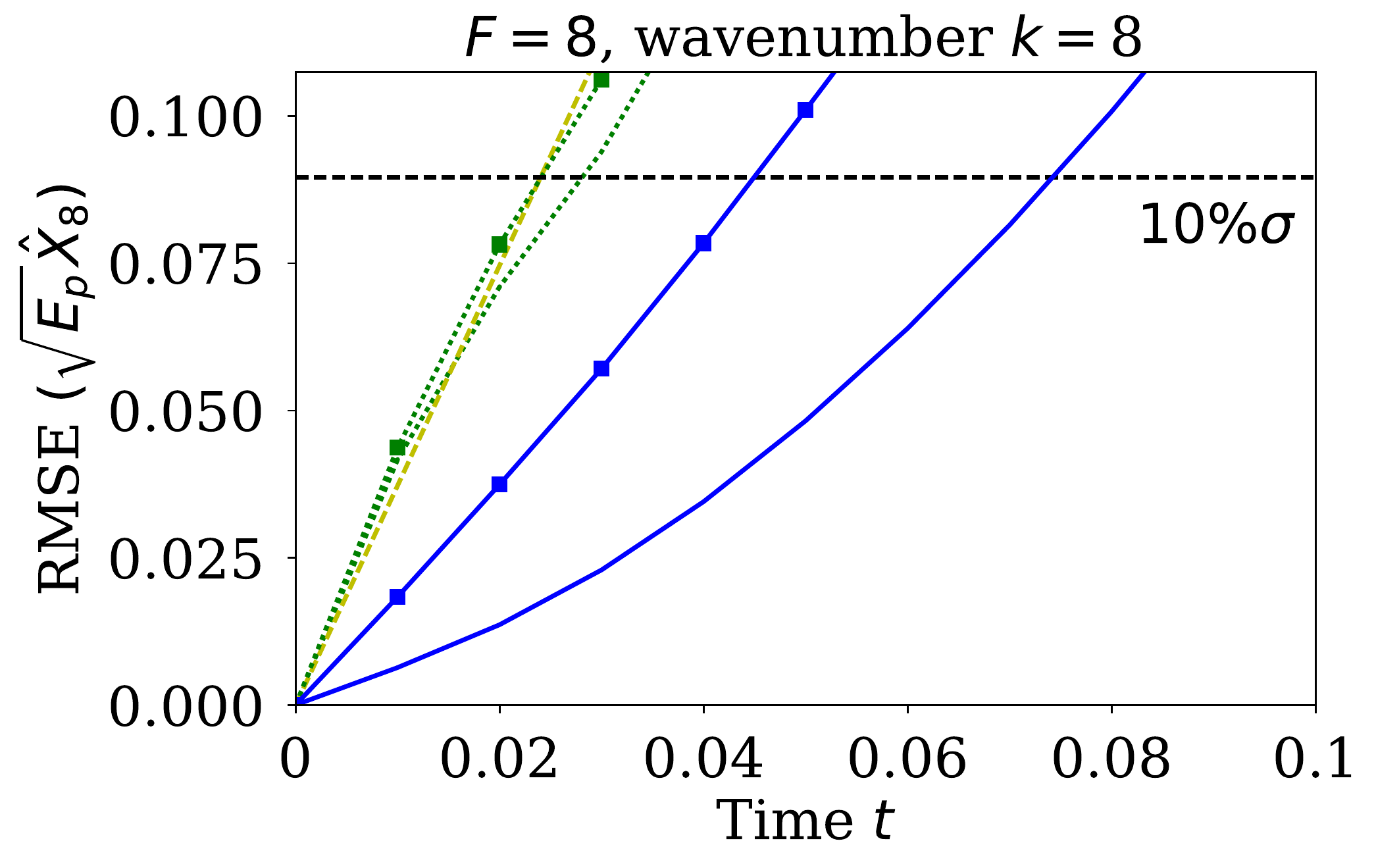}
\vspace{-0.75cm}
\caption{}
\label{results_lorenz:b}
\end{subfigure}
\centering
\begin{subfigure}{.31\textwidth}
\centering
\includegraphics[width=1\textwidth]{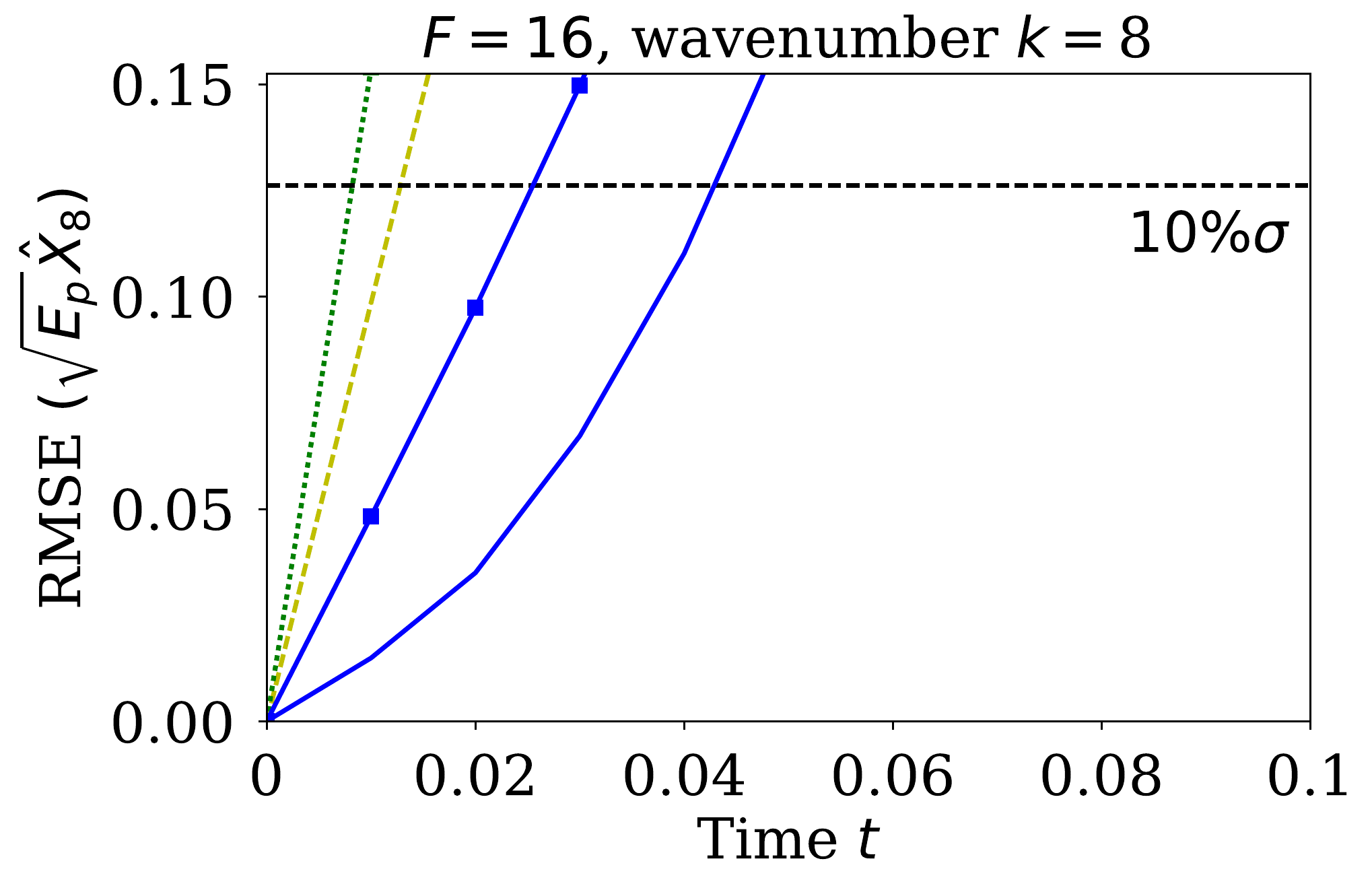}
\vspace{-0.75cm}
\caption{}
\label{results_lorenz:c}
\end{subfigure}
\begin{subfigure}{.31\textwidth}
\centering
\includegraphics[width=1\textwidth]{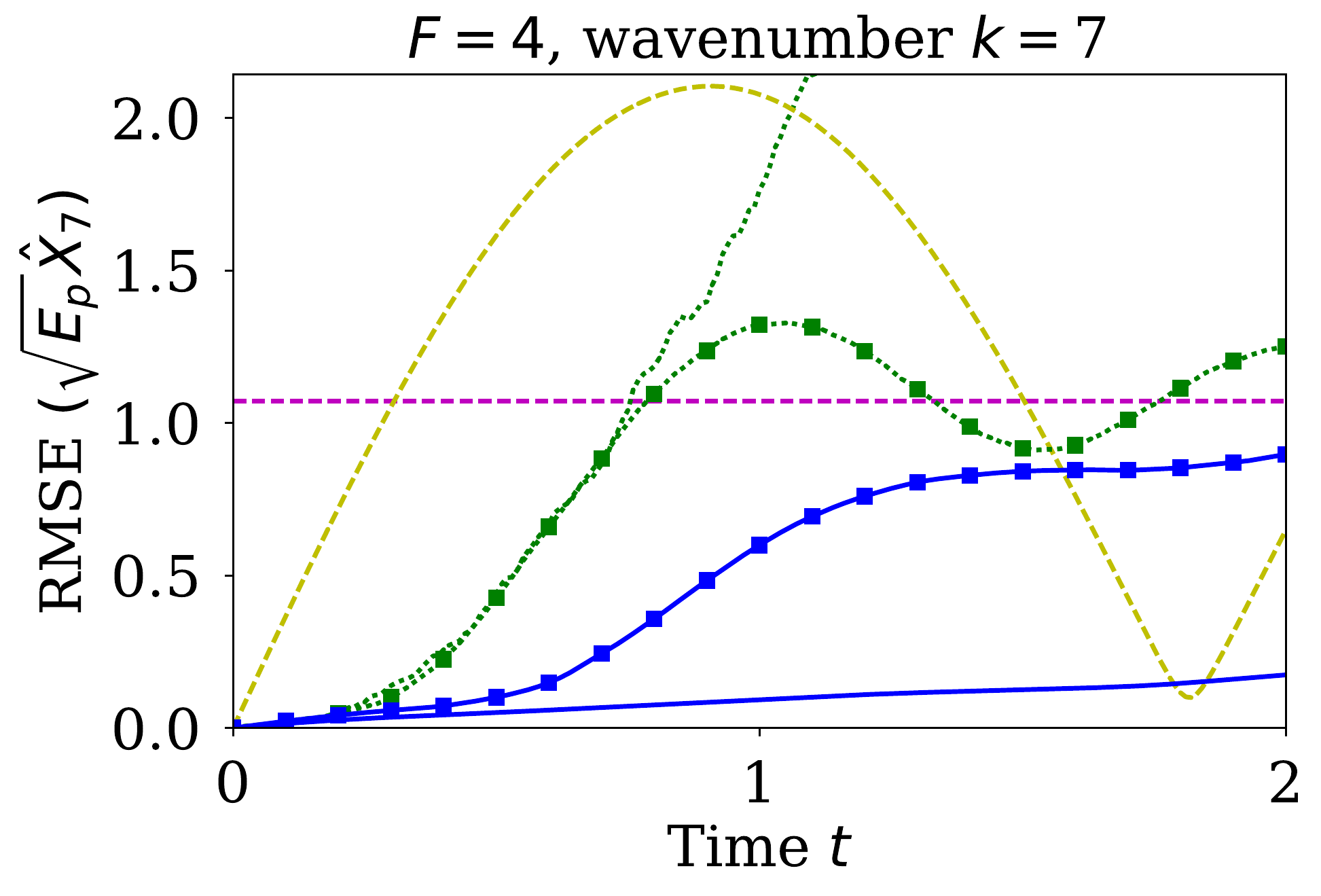}
\vspace{-0.75cm}
\caption{}
\label{results_lorenz:d}
\end{subfigure} 
\hspace{0.3cm}
\begin{subfigure}{.31\textwidth}
\centering
\includegraphics[width=1\textwidth]{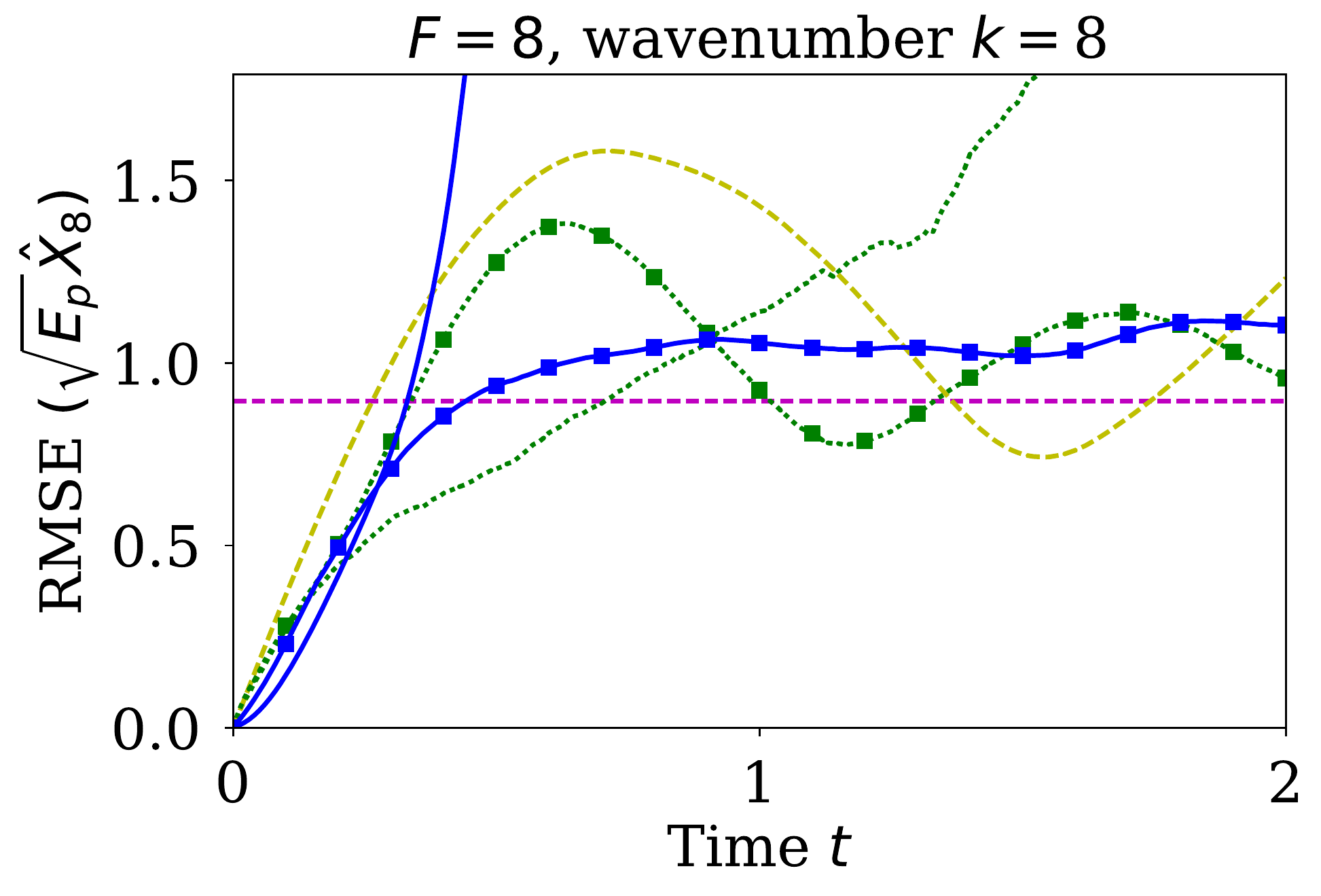}
\vspace{-0.75cm}
\caption{}
\label{results_lorenz:e}
\end{subfigure}
\centering
\begin{subfigure}{.31\textwidth}
\centering
\includegraphics[width=1\textwidth]{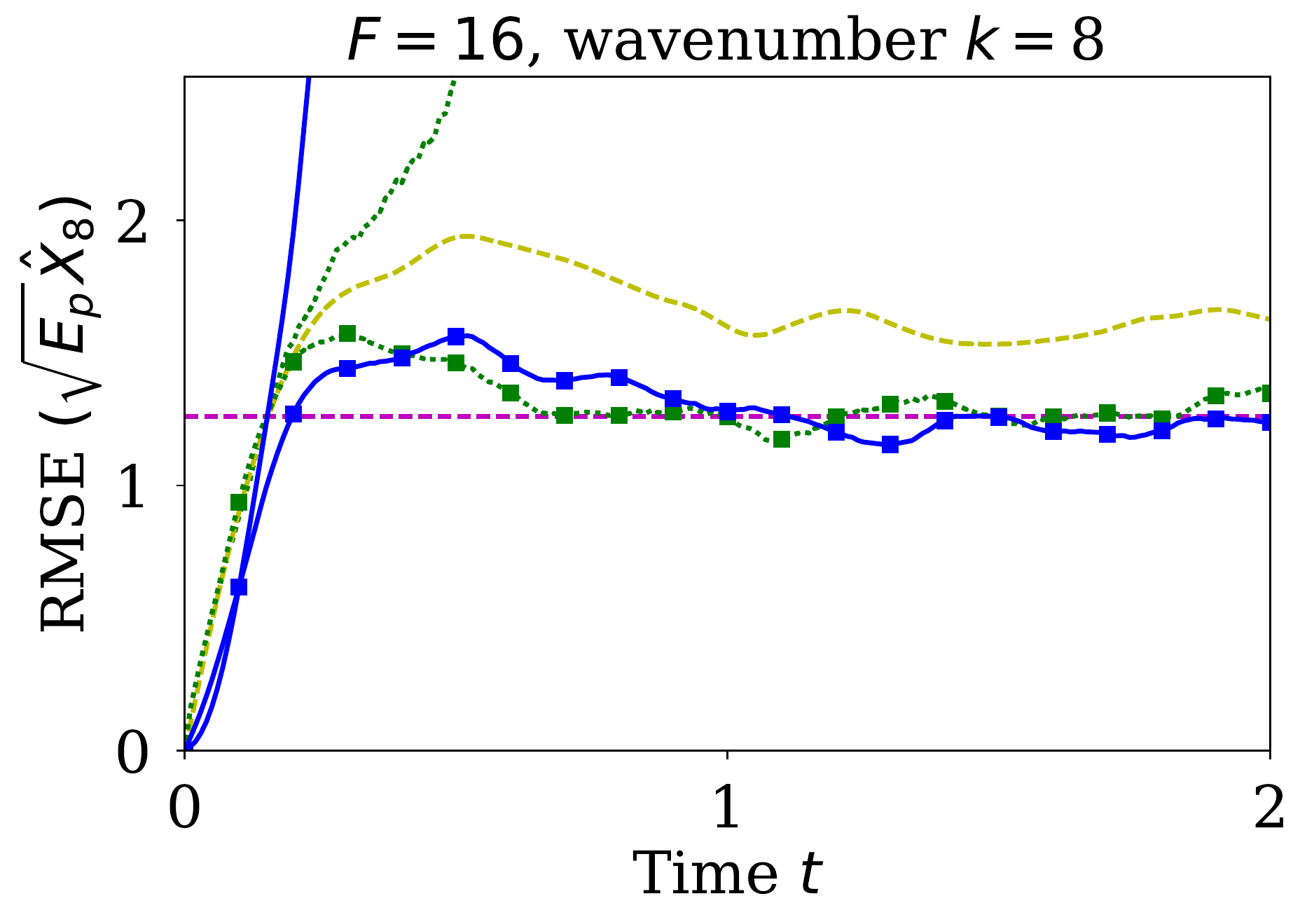}
\vspace{-0.75cm}
\caption{}
\label{results_lorenz:f}
\end{subfigure}
\begin{subfigure}{.31\textwidth}
\centering
\includegraphics[width=1\textwidth]{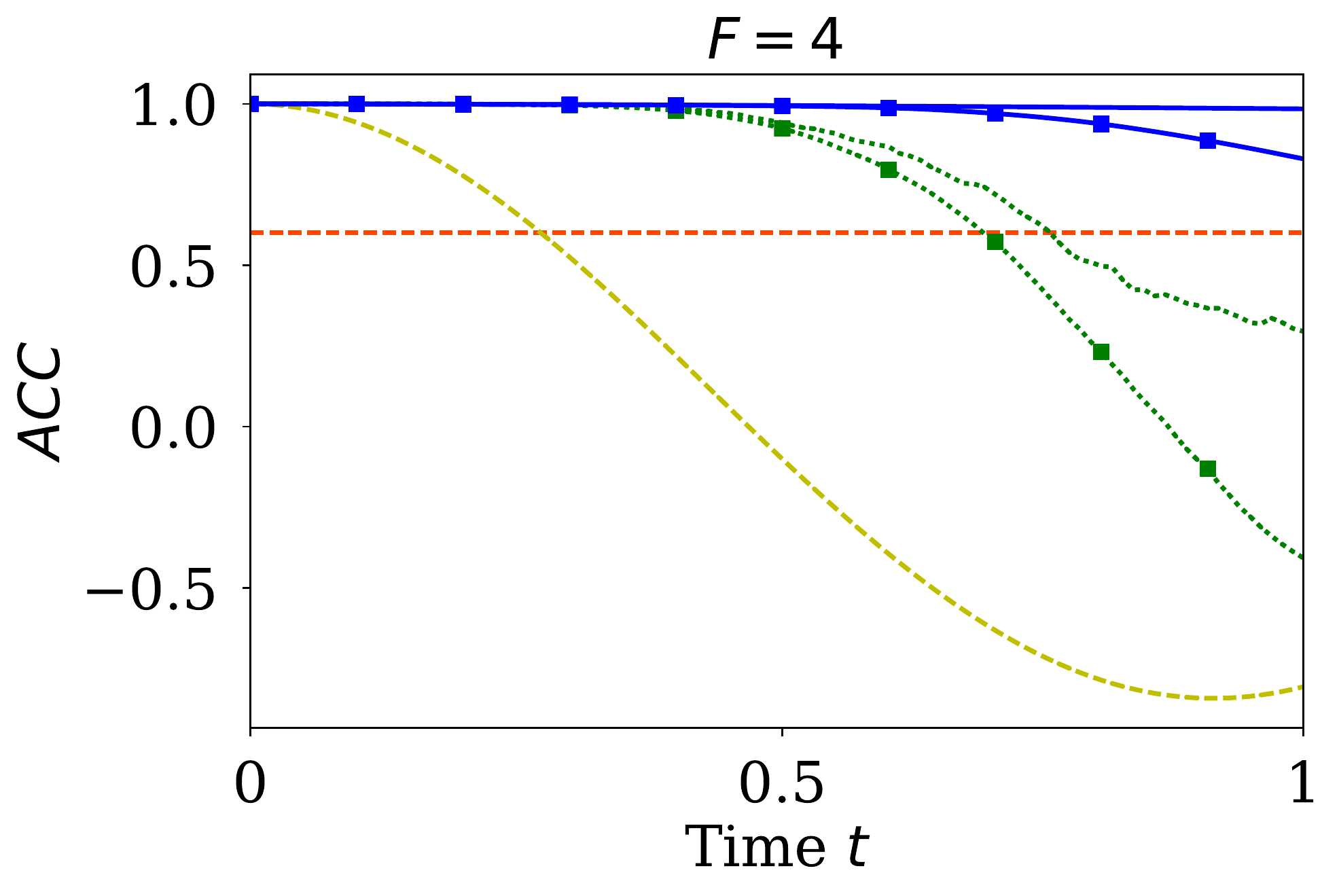}
\vspace{-0.75cm}
\caption{}
\label{results_lorenz:g}
\end{subfigure} 
\hspace{0.3cm}
\begin{subfigure}{.31\textwidth}
\centering
\includegraphics[width=1\textwidth]{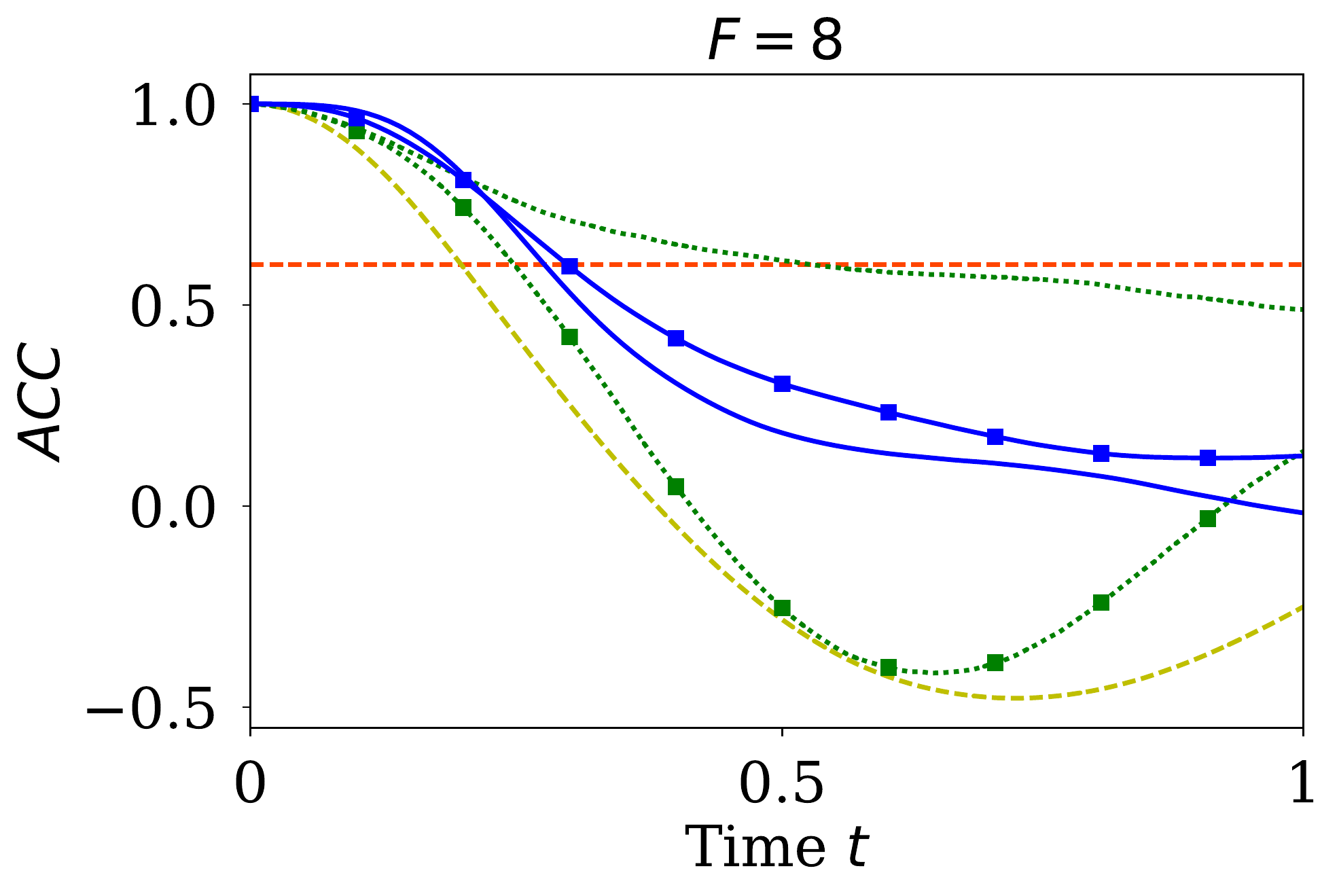}
\vspace{-0.75cm}
\caption{}
\label{results_lorenz:h}
\end{subfigure}
\centering
\begin{subfigure}{.31\textwidth}
\centering
\includegraphics[width=1\textwidth]{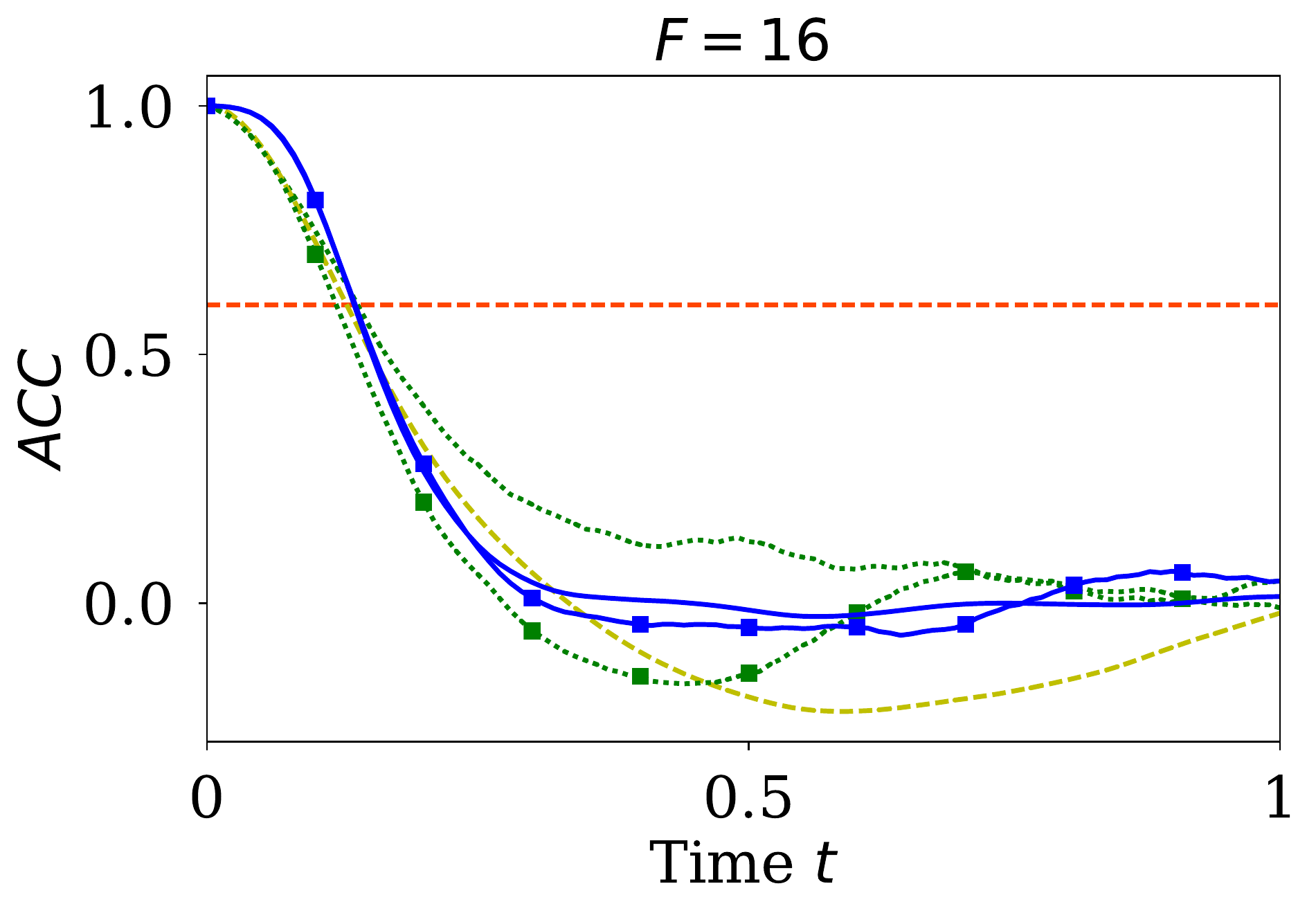}
\vspace{-0.75cm}
\caption{}
\label{results_lorenz:i}
\end{subfigure}
\caption{\textbf{(a), (b), (c)} Short-term RMSE evolution of the most energetic mode for forcing regimes $F=4,8,16$ respectively of the Lorenz 96 system. \textbf{(d), (e), (f)} Long-term RMSE evolution. \textbf{(g), (h), (i)} Evolution of the ACC coefficient.  (In all plots average over $1000$ initial conditions is reported). \\
$10\% \, \sigma_{attractor}$  \protect \blacklineDashed; $\sigma_{attractor}$ \protect\magentalineDashed; $ACC=0.6$ threshold \protect\redlineDashed; MSM\protect\yellowlineDashed; GPR\protect \greenlineDotted; GPR-MSM\protect\greenlineDottedRectangle; LSTM\protect\blueline; LSTM-MSM\protect\bluelineRectangle}
\label{fig:results_lorenz}
\end{figure}

In Figures \ref{results_lorenz:a}, \ref{results_lorenz:b}, and \ref{results_lorenz:c} we report the mean RMSE prediction error of the most energetic mode $\hat{X}_1^r \in \mathbb{C}$, scaled with  $\sqrt{E_p}$ for the forcing regimes $F \in \{ 6,8,16\}$ for the first $N=10$ time steps ($T=0.1$). In the RMSE the complex norm $||v||_2=vv^{*}$ is taken into account. The $10 \%$ of the standard deviation of the attractor is also plotted for reference ($10 \% \sigma$). As $F$ increases, the system becomes more chaotic and difficult to predict. As a consequence, the number of prediction steps that remain under the $10 \% \sigma$ threshold are decreased. The LSTM models extend this predictability horizon for all forcing regimes compared to GPR and MSM. However, when LSTM is combined with MSM the short-term prediction performance is compromised. Nevertheless, hybrid LSTM-MSM models outperform GPR methods in short-term prediction accuracy.

In Figures \ref{results_lorenz:d}, \ref{results_lorenz:e}, and \ref{results_lorenz:f}, the RMSE error for $T=2$ is plotted. The standard deviation from the attractor $\sigma$ is plotted for reference. We can observe that both GPR and LSTM diverge, while MSM and blended schemes remain close to the attractor in the long-term as expected.

In Figures \ref{results_lorenz:g}, \ref{results_lorenz:h}, and \ref{results_lorenz:i}, the mean ACC over 1000 initial conditions is given. The predictability threshold of $0.6$ is also plotted. After crossing this critical threshold, the methods do not predict better than a trivial mean predictor. For $F=4$ GPR methods show inferior performance compared to LSTM approaches as analyzed previously in the RMSE comparison. However, for $F=8$ LSTM models do not predict better than the mean after $T\approx 0.35$, while GPR shows better performance. In turn, when blended with MSM the compromise in the performance for GPR-MSM is much bigger compared to LSTM-MSM. The LSTM-MSM scheme shows slightly superior performance than GPR-MSM during the entire relevant time period ($ACC>0.6$). For the fully turbulent regime $F=16$, LSTM shows comparable performance with both GPR and MSM and all methods converge as chaoticity rises, since the intrinsic dimensionality of the system attractor increases and the system become inherently unpredictable. 

\begin{figure}[!httb]
\centering
\begin{subfigure}{.4\textwidth}
\centering
\includegraphics[width=1\textwidth]{Lorenz96/RMSE_ST_8_mode_0.pdf}
\vspace{-0.75cm}
\caption{}
\label{results_lorenz_F8:a}
\end{subfigure} 
\hspace{0.3cm}
\begin{subfigure}{.4\textwidth}
\centering
\includegraphics[width=1\textwidth]{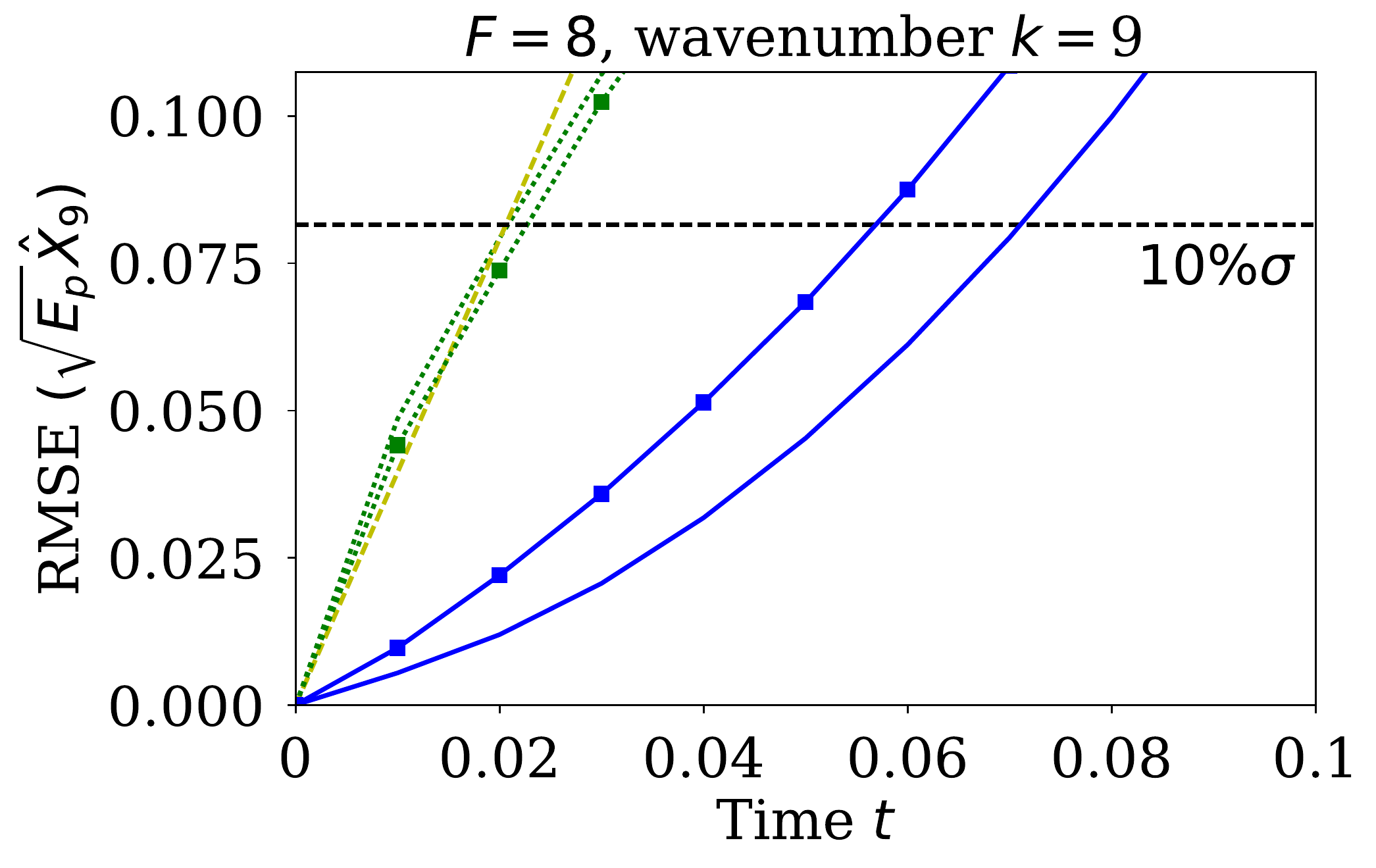}
\vspace{-0.75cm}
\caption{}
\label{results_lorenz_F8:b}
\end{subfigure}
\centering
\begin{subfigure}{.4\textwidth}
\centering
\includegraphics[width=1\textwidth]{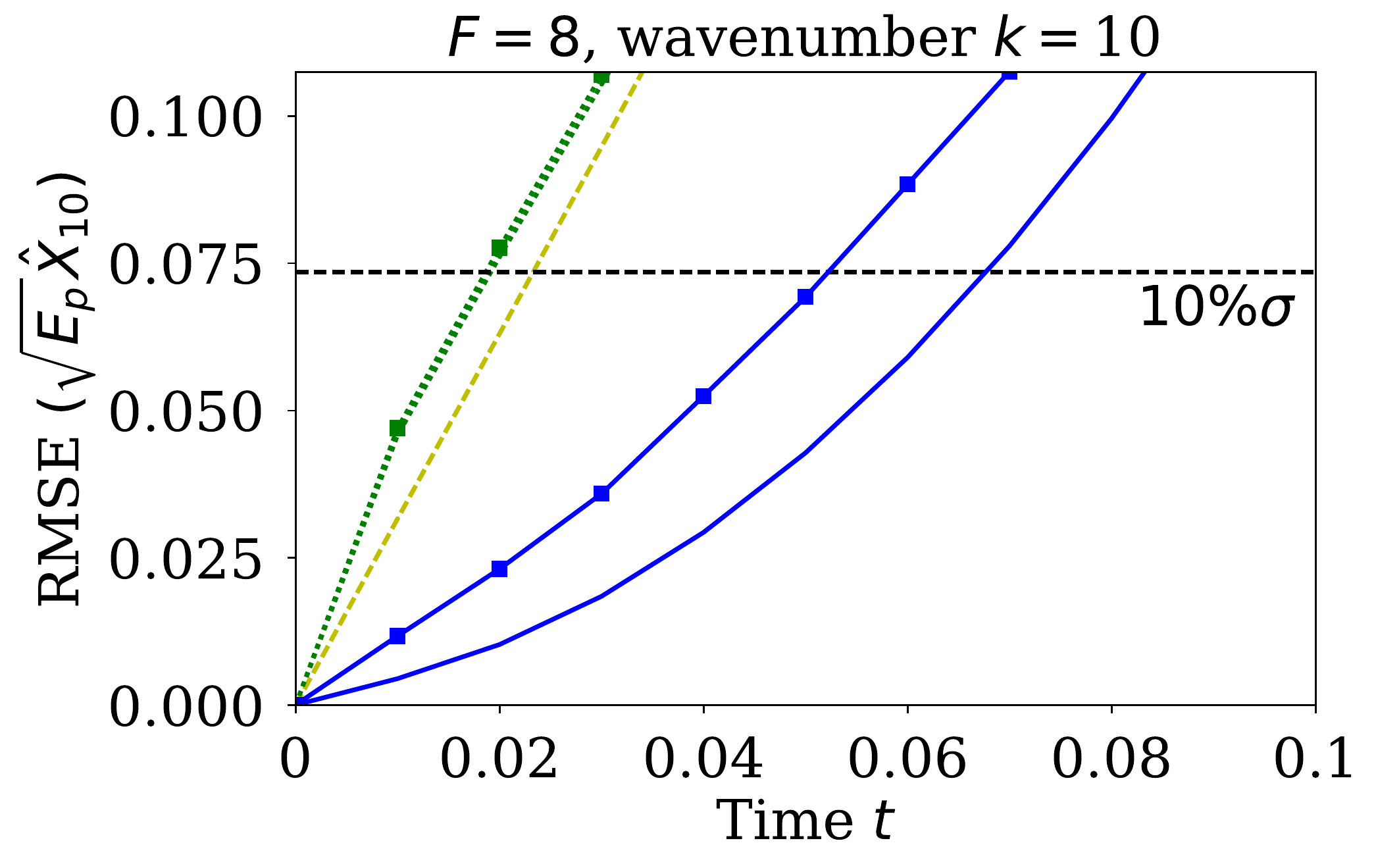}
\vspace{-0.75cm}
\caption{}
\label{results_lorenz_F8:c}
\end{subfigure} 
\hspace{0.3cm}
\begin{subfigure}{.4\textwidth}
\centering
\includegraphics[width=1\textwidth]{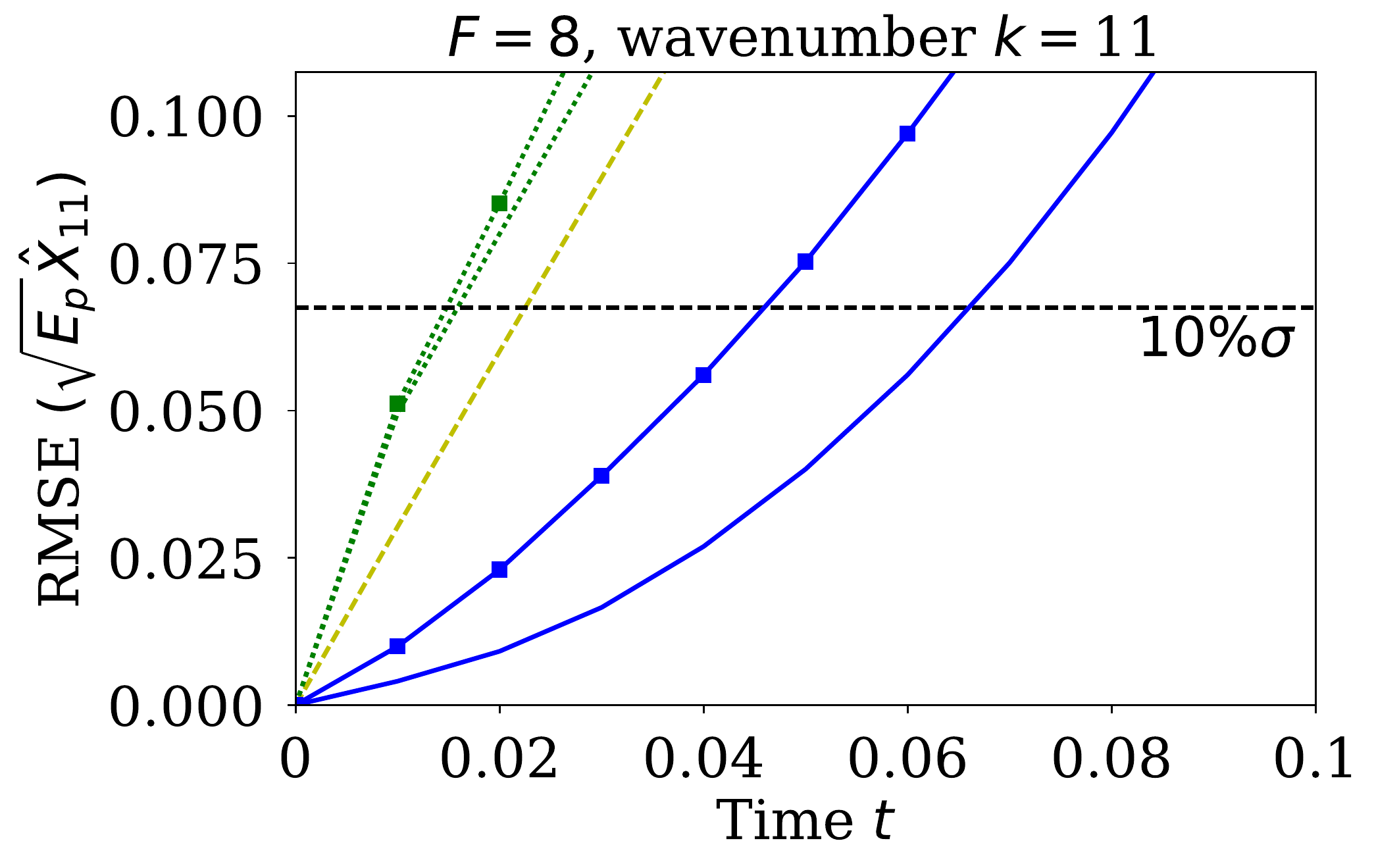}
\vspace{-0.75cm}
\caption{}
\label{results_lorenz_F8:d}
\end{subfigure}
\caption{RMSE prediction error evolution of four energetic modes for the Lorenz 96 system with forcing $F=8$. \textbf{(a)} Most energetic mode $k=8$. \textbf{(b)} Low energy mode $k=9$. \textbf{(c)} Low energy mode $k=10$. \textbf{(d)} Low energy mode $k=11$. (In all plots average over $1000$ initial conditions reported)\\
 $10\% \, \sigma_{attractor}$  \protect \blacklineDashed; MSM\protect\yellowlineDashed; GPR\protect\greenlineDotted; GPR-MSM\protect\greenlineDottedRectangle; LSTM\protect\blueline; LSTM-MSM\protect\bluelineRectangle}
\label{fig:results_lorenz_F8}
\end{figure}

In Figure \ref{fig:results_lorenz_F8}, the evolution of the mean RMSE over $1000$ initial conditions of the wavenumbers $k=8,9,10,11$ of the Lorenz 96 with forcing $F=8$ is plotted. In contrast to GPR, the RMSE error of LSTM is much lower in the moderate and low energy wavenumbers $k=9,10,11$ compared to the most energetic mode $k=8$. This difference among modes is not observed in GPR. This can be attributed to the highly non-linear energy transfer mechanisms between these lower energy modes as opposed to the Gaussian and locally linear energy transfers of the most energetic mode.

\begin{figure}[!httb]
\centering
\begin{subfigure}{.45\textwidth}
\centering
\includegraphics[width=0.95\textwidth]{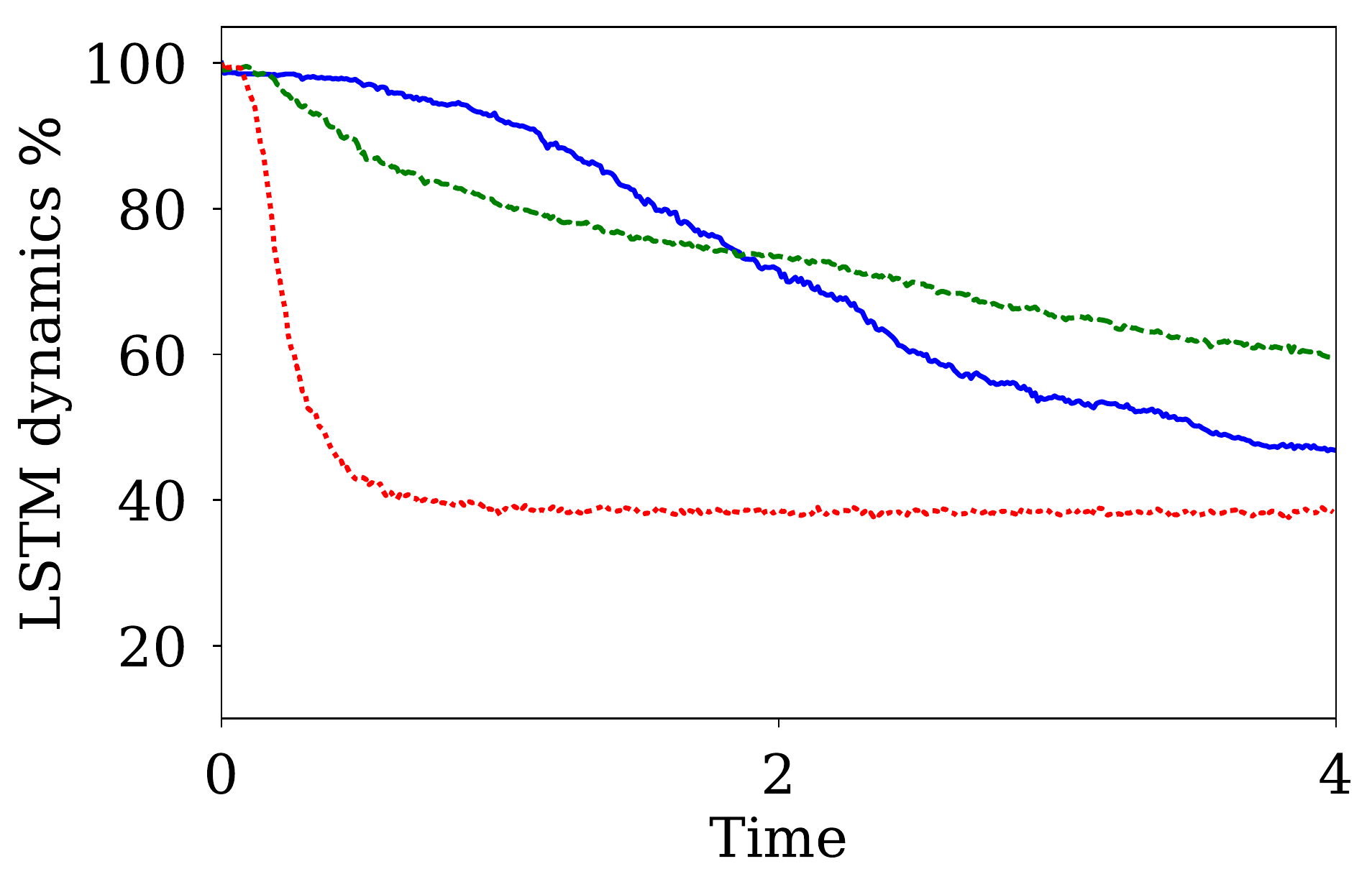}
\vspace{-0.5cm}
\caption{}
\label{Lorenz_Plot_MSM_Quotient_500:a}
\end{subfigure}
\begin{subfigure}{.45\textwidth}
\centering
\includegraphics[width=0.95\textwidth]{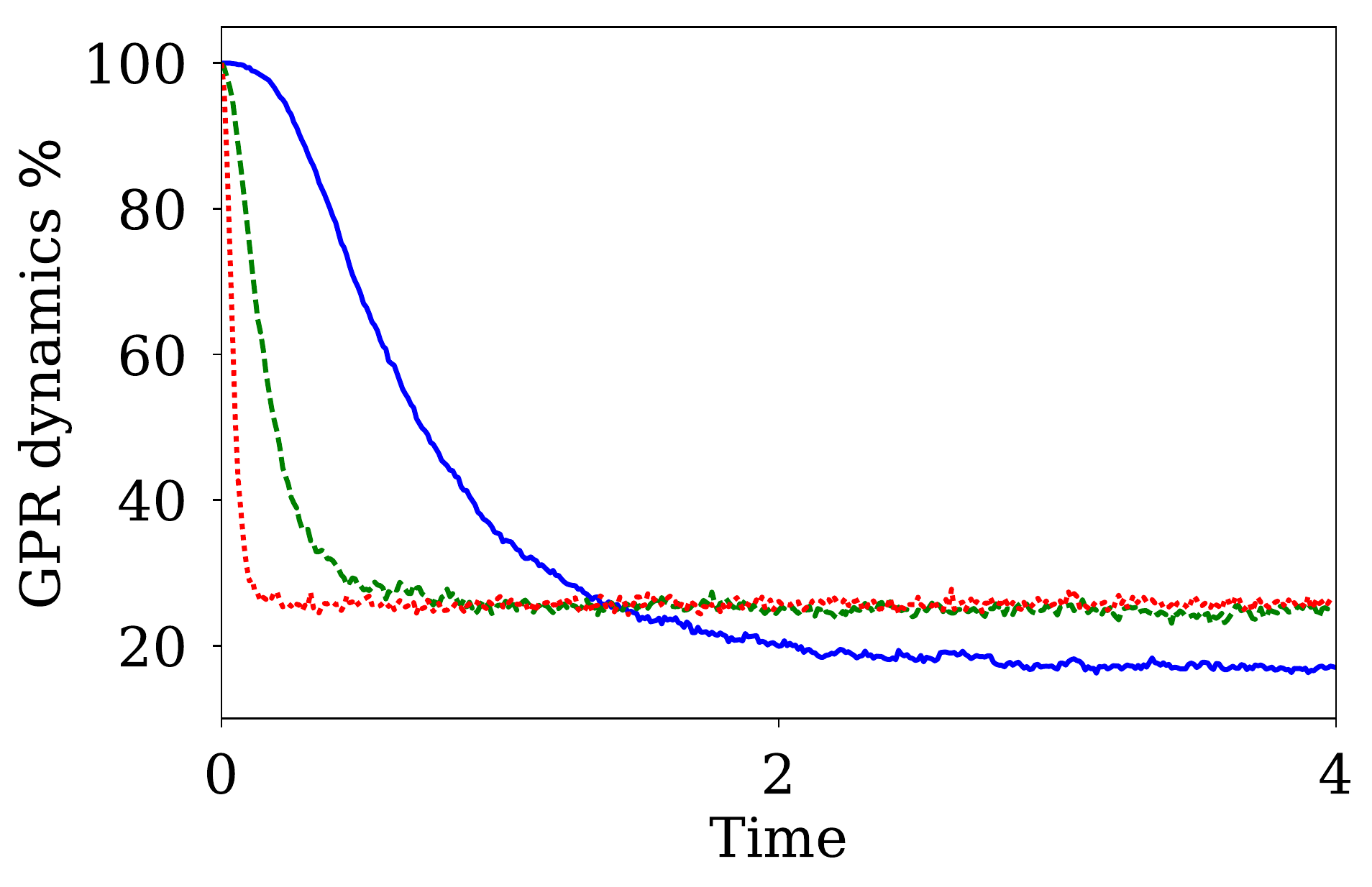}
\vspace{-0.5cm}
\caption{}
\label{Lorenz_Plot_MSM_Quotient_500:b}
\end{subfigure}
\caption{\textbf{(a)} Ratio of the ensemble members evaluated using the LSTM model over time for different Lorenz 96 forcing regimes in the hybrid LSTM-MSM method and  \textbf{(b)} the same for GPR in the hybrid GPR-MSM method. (average over 500 initial conditions). \\ $F=4$ \protect\blueline ; $F=8$\protect\greenlineDashed   ; $F=16$ \protect\redlineDotted}
\label{fig:Lorenz_Plot_MSM_Quotient_500}
\end{figure}

As illustrated before, the hybrid LSTM-MSM architecture effectively combines the accurate short-term prediction performance of LSTM with the long-term stability of MSM. The ratio of ensemble members modeled by LSTM in the hybrid scheme is plotted with respect to time in Figure \ref{Lorenz_Plot_MSM_Quotient_500:a}. Starting from the initial ensemble of size $50$, as the LSTM forecast might deviate from the attractor, the MSM is used to forecast in the hybrid scheme. As a consequence, the ratio of ensemble members modeled by LSTM decreases with time. In parallel with the GPR results presented in \cite{Zhong2017} and plotted in Figure \ref{Lorenz_Plot_MSM_Quotient_500:b}, the slope of this ratio curve increases with $F$ up to time $t\approx 1.5$. However, the LSTM ratio decreases slower compared to GPR.

\subsection{Kuramoto-Sivashinsky Equation}
\label{sec:kuramoto}

The Kuramoto-sivashinsky (K-S) system is extensively used in many scientific fields to model a multitude of chaotic physical phenomena. It was first derived by Kuramoto \cite{Kuramoto1976, Kuramoto1978} as a turbulence model of the phase gradient of a slowly varying amplitude in a reaction-diffusion type medium with negative viscosity coefficient. Later, Sivashinsky \cite{Sivashinsky1977} studied the spontaneous instabilities of the plane front of a laminar flame ending up with the K-S equation, while in \cite{Sivashinsky1980} the K-S equation is found to describe the surface behavior of viscous liquid in a vertical flow. 

For our study, we restrict ourselves to the one dimensional K-S equation with boundary and initial conditions given by
\begin{equation}
\begin{split}
&\frac{\partial u}{ \partial t} = - \nu \frac{\partial^4 u}{ \partial x ^4 } - \frac{\partial^2 u}{ \partial x ^2 } - u  \frac{\partial u}{ \partial x},\\
&u(0,t)=u(L,t)
= \frac{\partial u}{\partial x} \bigg|_{x=0}
= \frac{\partial u}{\partial x} \bigg|_{x=L} = 0, \\
& u(x,0)=u_0(x),
\end{split}
\label{eq:kuramoto}
\end{equation}
where $u(x,t)$ is the modeled quantity of interest depending on a spatial variable $x \in [0,L]$ and time $t \in [0,\infty)$. The negative viscosity is modeled by the parameter $\nu>0$. We impose Dirichlet and second-type boundary conditions to guarantee ergodicity \cite{Blonigan2014}. In order to spatially discretize (\ref{eq:kuramoto}) we use a grid size $\Delta x$ with $D=L/\Delta x + 1$ the number of nodes. Further, we denote with $u_i=u(i \Delta x)$ the value of $u$ at node $i \in \{0,\dots,D-1\}$. Discretization using a second order differences scheme yields
\begin{equation}
\begin{split}
\frac{d u_i}{ d t} = - \nu \frac{u_{i-2}-4u_{i-1}+6u_{i}-4u_{i+1}+u_{i+2}}{\Delta x ^4} - \frac{u_{i+1}-2u_{i}+u_{i-1}}{\Delta x^2} -
\frac{u_{i+1}^2-u_{i-1}^2}{4 \Delta x}.
\end{split}
\label{eq:kuramotodiscretized}
\end{equation}
Further, we impose $u_0=u_{D-1}=0$ and add ghost nodes $u_{-1}=u_1$, $u_{D}=u_{D-2}$ to account for the Dirichlet and second-order boundary conditions. In our analysis, the number of nodes is $D=513$. The Kuramoto-Sivashinsky equation exhibits different levels of chaos depending on the bifurcation parameter $\tilde{L}=L/2\pi \sqrt{\nu}$ \cite{Kevrekidis1990}. Higher values of $\tilde{L}$ lead to more chaotic systems \cite{Zhong2017}. 

\begin{figure}[!ht]
\centering
\begin{subfigure}{.7\textwidth}
\centering
\includegraphics[height=5cm, trim={0 1.5cm 0 0},clip]{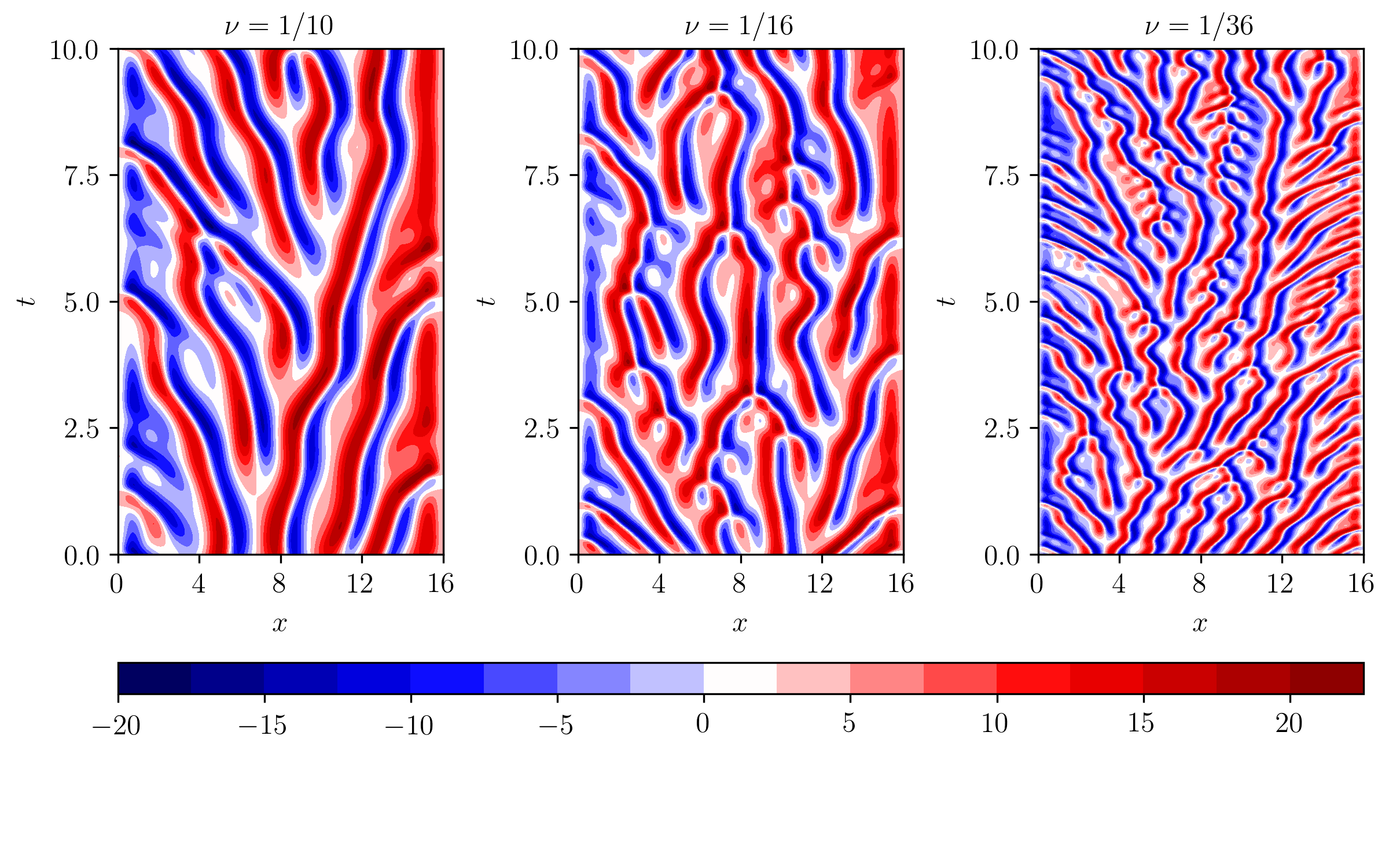}
\vspace{-0.5cm}
\caption{}
\label{fig:Plot_U_KS}
\end{subfigure} 
\begin{subfigure}{.28\textwidth}
\centering
\includegraphics[height=4.5cm]{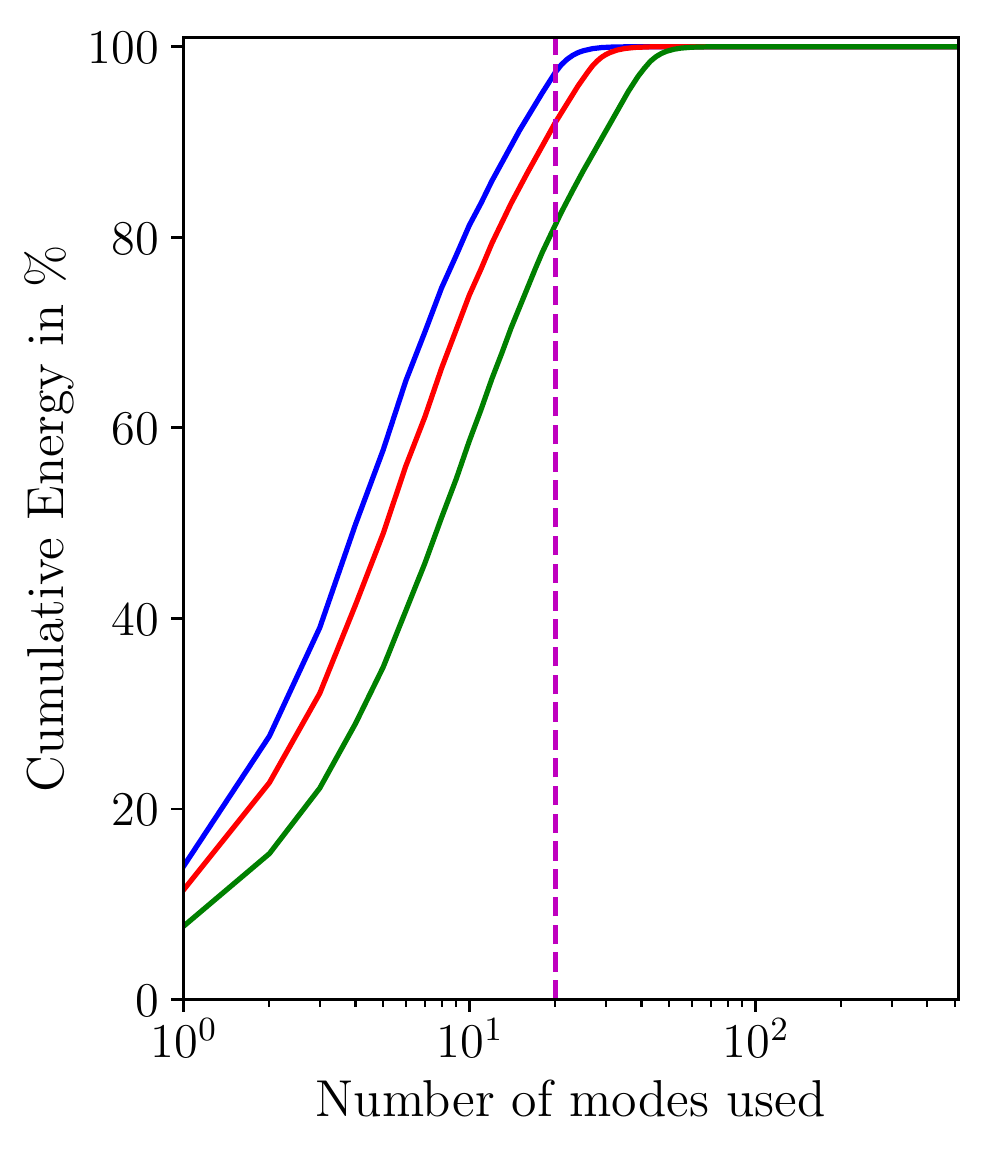}
\vspace{-0.5cm}
\caption{}
\label{fig:Plot_E_cum_20_paper}
\end{subfigure} 
\caption{\textbf{(a)} Contour plots of the solution $u(x,t)$ of the Kuramoto-Sivashinsky system for different values of $\nu$ in steady state. Chaoticity rises with smaller values of $\nu$. \textbf{(b)} Cumulative energy as a function of the number of the PCA modes for different values of $\nu$.\\ $1/\nu=10$ \protect\blueline ; $1/\nu=16$ \protect\redline ; $1/\nu=36$ \protect\greenline ; 20 modes \protect\magentalineDashed}
\end{figure}

In our analysis the spatial variable bound is held constant to $L=16$ and chaoticity level is controlled through the negative viscosity $\nu$, where a smaller value leads to a system with a higher level of chaos (see Figure \ref{fig:Plot_U_KS}). In our study, we consider two values, namely $\nu=1/10$ and $\nu=1/16$ to benchmark the prediction skills of the proposed method. The discretized equation (\ref{eq:kuramotodiscretized}) is integrated with a time interval $dt=0.02$ up to $T=11000$. The data points up to $T=1000$ are discarded as initial transients. Half of the remaining data ($N=250000$ samples) are used for training and the other half for validation.

\subsubsection{Dimensionality Reduction: Singular Value Decomposition}

The dimensionality of the problem is reduced using Singular Value Decomposition (SVD). By subtracting the temporal mean $\overline{\mathbf{u}}$ and stacking the data, we end up with the data matrix $\mathbf{U}\in \mathbb{R}^{N\times 513}$, where $N$ is the number of data samples ($N=500000$ in our case). Performing SVD on  $\mathbf{U}$ leads to 
\begin{equation}
\mathbf{U} = \mathbf{M} \mathbf{\Sigma} \mathbf{V}^T, \quad \mathbf{M} \in \mathbb{R}^{N\times N}, \quad \mathbf{\Sigma} \in \mathbb{R}^{N\times 513}, \quad \mathbf{V} \in \mathbb{R}^{513 \times 513},
\label{eq:svd}
\end{equation}
with $\Sigma$ diagonal, with descending diagonal elements. The right singular vectors corresponding to the $r_{dim}$ largest singular values are the first columns of $\mathbf{V}=[\mathbf{V_r}, \mathbf{V_{-r}}]$. Stacking these singular vectors yields $\mathbf{V_r} \in \mathbb{R}^{513 \times r_{dim}}$. Assuming that $\mathbf{u}_t \in \mathbb{R}^{513}$ is a vector of the discretized values of $u(x,t)$ in time $t$, in order to get a reduced order representation $\mathbf{c} \equiv  [c_1,\dots,c_{r_{dim}}]^T$ corresponding to the $r_{dim}$ components with the highest energies (singular values) we multiply
\begin{equation}
\mathbf{c} = \mathbf{V_r}^T \mathbf{u}  , \quad  \mathbf{c}  \in \mathbb{R}^{r_{dim}}.
\label{eq:svd_2}
\end{equation}
The percentage of cumulative energy w.r.t. to the number of PCA modes considered is plotted in Figure \ref{fig:Plot_E_cum_20_paper}. In our study, we pick $r_{dim}=20$ (out of $513$) most energetic modes, as they explain approximately $90\%$ of the total energy.

\subsubsection{Results}

\begin{figure}[!httb]
\centering
\begin{subfigure}{.45\textwidth}
\centering
\includegraphics[width=1\textwidth]{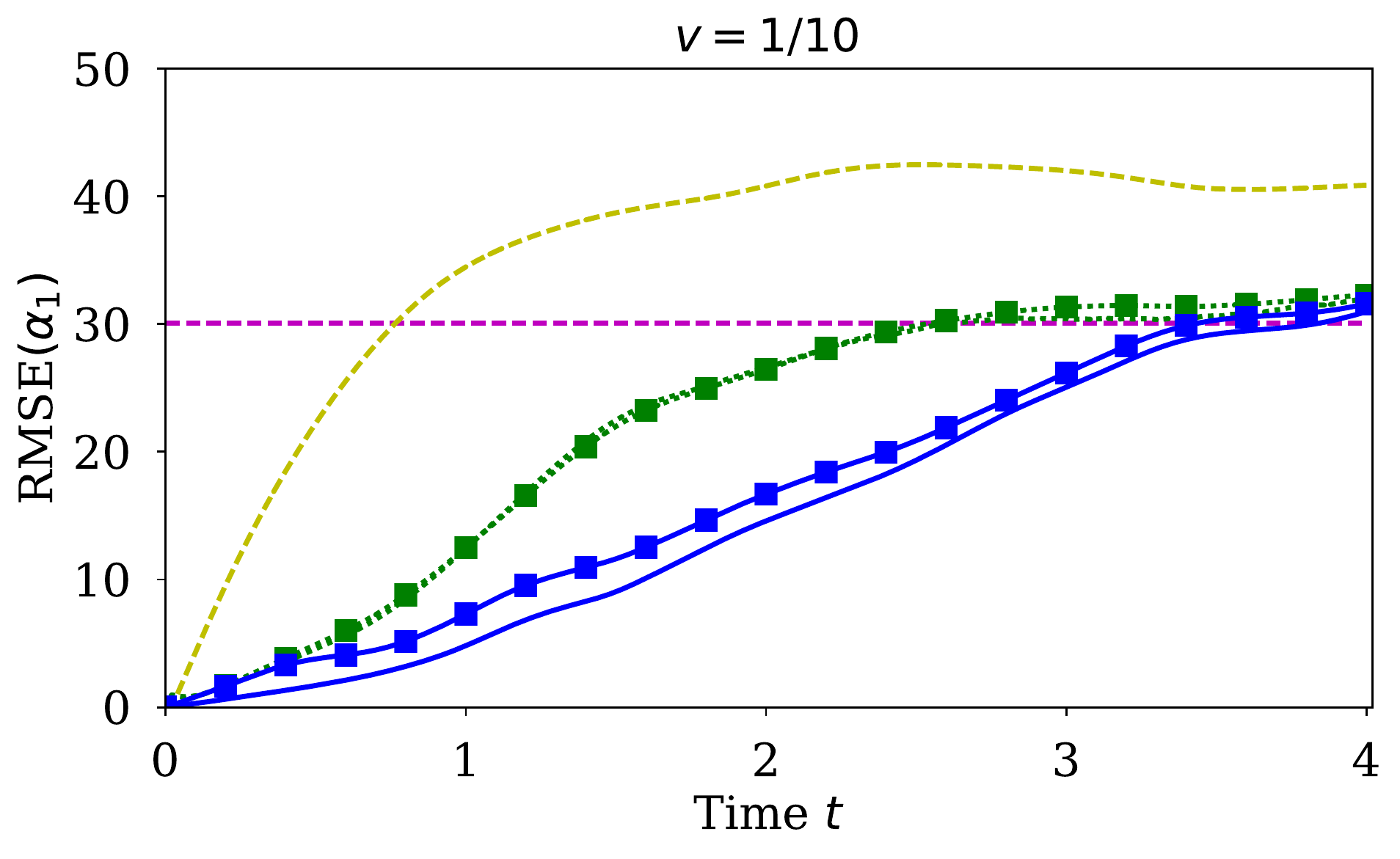}
\vspace{-0.75cm}
\caption{}
\label{results_kuramoto:a}
\end{subfigure} 
\hspace{0.3cm}
\begin{subfigure}{.45\textwidth}
\centering
\includegraphics[width=1\textwidth]{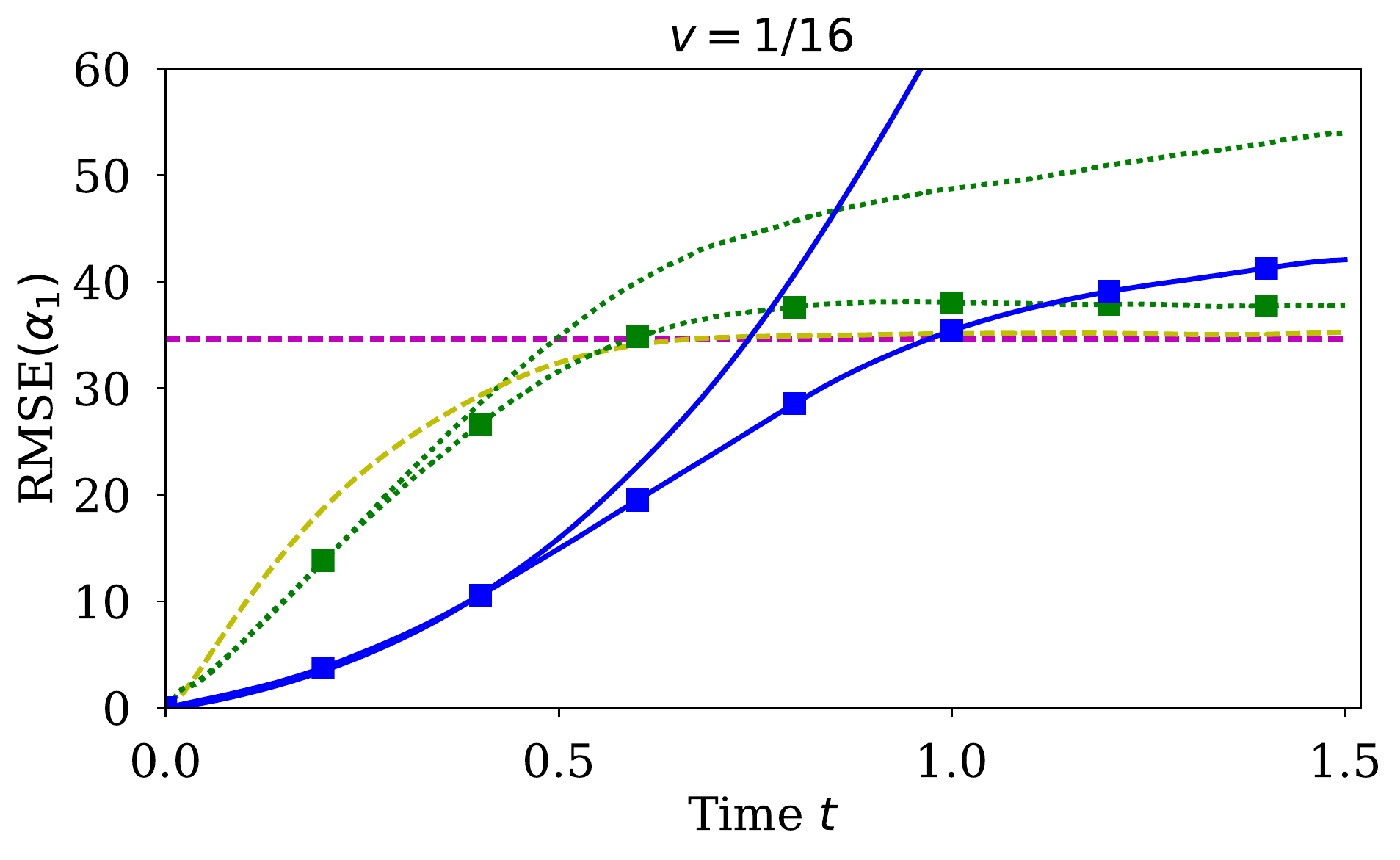}
\vspace{-0.75cm}
\caption{}
\label{results_kuramoto:b}
\end{subfigure}
\centering
\begin{subfigure}{.45\textwidth}
\centering
\includegraphics[width=1\textwidth]{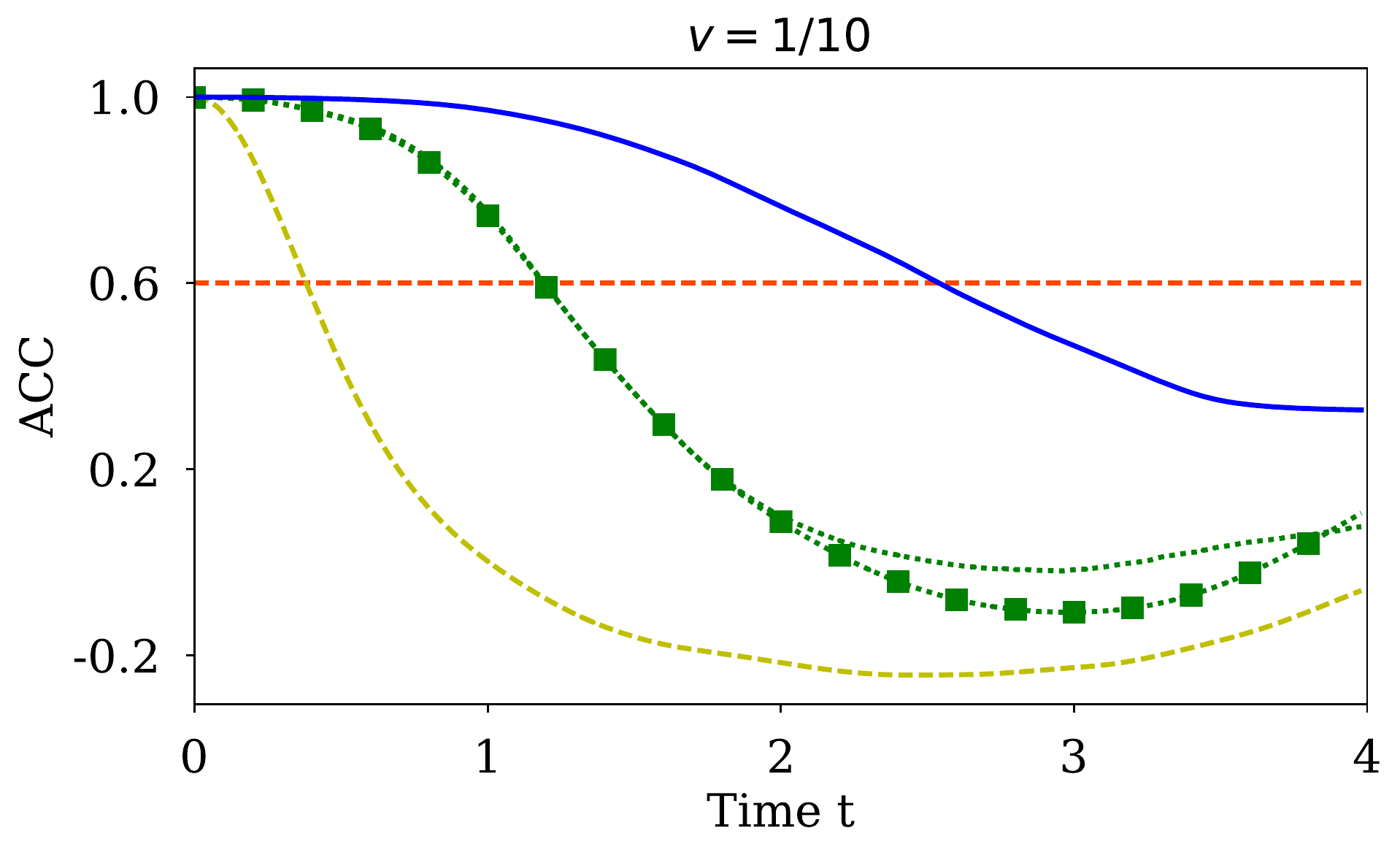}
\vspace{-0.75cm}
\caption{}
\label{results_kuramoto:c}
\end{subfigure} 
\hspace{0.3cm}
\begin{subfigure}{.45\textwidth}
\centering
\includegraphics[width=1\textwidth]{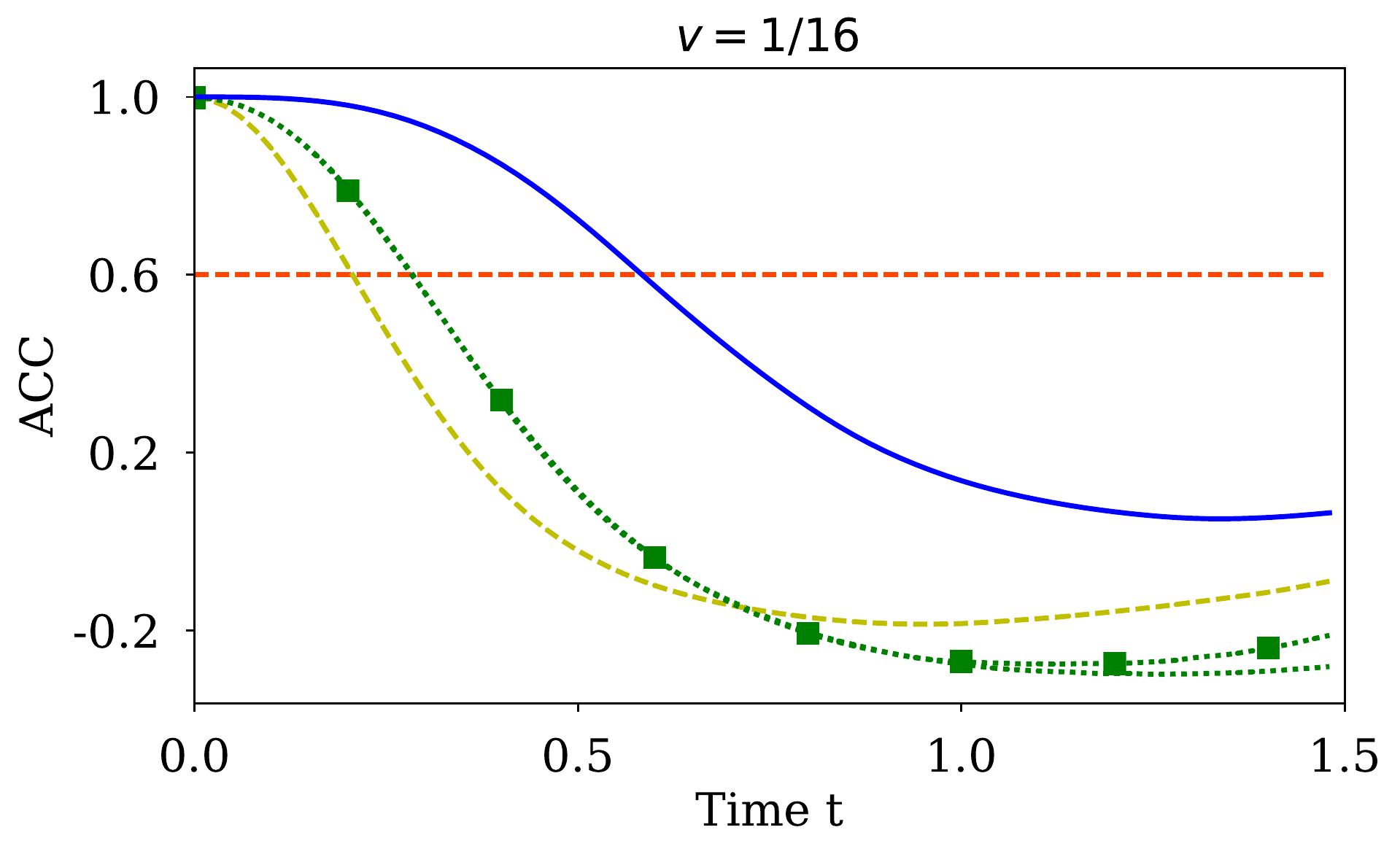}
\vspace{-0.75cm}
\caption{}
\label{results_kuramoto:d}
\end{subfigure}
\caption{
\textbf{(a)}, \textbf{(b)}  RMSE evolution of the most energetic mode of the K-S equation with $1/\nu=10$ and $1/\nu=16$.
\textbf{(c)}, \textbf{(d)}  ACC evolution of the most energetic mode of the K-S equation with $1/\nu=10$ and $1/\nu=16$.
(In all plots, average value over $1000$ initial conditions is reported) \\
$\sigma_{attractor}$\protect\magentalineDashed; $ACC=0.6$ threshold\protect\redlineDashed; MSM\protect\yellowlineDashed; GPR\protect\greenlineDotted; GPR-MSM\protect\greenlineDottedRectangle; LSTM\protect\blueline; LSTM-MSM\protect\bluelineRectangle
}
\label{fig:results_kuramoto}
\end{figure}

We train \textit{stateless} LSTM models with $h=100$ and $d=50$. For testing, starting from $1000$ initial conditions uniformly sampled from the attractor, we generate a Gaussian ensemble of dimension $N_{en}=50$ centered around the initial condition in the original space with standard deviation of $\sigma=0.1$. This ensemble is propagated using the LSTM prediction models, and GPR, MSM and GPR-MSM models trained as in \cite{Zhong2017}. The root mean square error between the predicted ensemble mean and the ground-truth is plotted in Figures \ref{results_kuramoto:a}, \ref{results_kuramoto:b} for different values of the parameter $\nu$. All methods reach the invariant measure much faster for $1/\nu = 16$ compared to the less chaotic regime $1/\nu = 10$ (note the different integration times $T=4$ for $1/\nu = 10$, while $T=1.5$ for $1/\nu = 16$).

In both chaotic regimes $1/\nu = 10$ and $1/\nu = 16$, the reduced order LSTM outperforms all other methods in the short-term before escaping the attractor. However, in the long-term, LSTM does not stabilize and will eventually diverge faster than GPR (see Figure \ref{results_kuramoto:b}). Blending LSTM with MSM alleviates the problem and both accurate short-term predictions and long-term stability is attained. Moreover, the hybrid LSTM-MSM has better forecasting capabilities compared to GPR. 

The need for blending LSTM with MSM in the KS equation is less imperative as the system is less chaotic than the Lorenz 96 and LSTM methods diverge much slower, while they sufficiently capture the complex nonlinear dynamics. As the intrinsic dimensionality of the attractor rises LSTM diverges faster.

The mean ACC (\ref{eq:acc}) is plotted with respect to time in Figures \ref{results_kuramoto:c} and \ref{results_kuramoto:d} for $\nu=10$ and $16$ respectively. The evolution of the ACC justifies the aforementioned analysis. The mean ACC of the trajectory predicted with LSTM remains above the predictability threshold of $0.6$ for a highest time duration compared to other methods. This predictability horizon is approximately $2.5$ for $\nu=1/10$ and $0.6$ for $\nu=1/16$, since the chaoticity of the system rises and accurate predictions become more challenging. For a plot of the time evolution of the ratio of the ensemble members that are modeled with LSTM dynamics in the hybrid LSTM-MSM refer to the Appendix.

\subsection{A Barotropic Climate Model}
\label{sec:barotropic}

In this section, we examine a standard barotropic climate model \cite{Selten1995} originating from a realistic winter circulation. The model equations are given by
\begin{equation}
\frac{ \partial \zeta}{\partial t}=-\mathcal{J}(\psi, \zeta+f+h) + k_1 \zeta + k_2 \delta^3 \zeta + \zeta^{*},
\label{eq:barotropic}
\end{equation}
where $\psi$ is the stream function, $\zeta=\delta \psi$ the relative vorticity, $f$ the Coriolis parameter, $\zeta^{*}$ a constant vorticity forcing, while $k_1$ and $k_2$ are the Ekman damping and the scale-selective damping coefficient. $\mathcal{J}$ is the Jacobi operator given by
\begin{equation}
\mathcal{J}(a,b) = \Big( \frac{\partial a}{\partial \lambda} \frac{\partial B}{\partial \mu}-\frac{\partial a}{\partial \mu} \frac{\partial B}{\partial \lambda} \Big),
\end{equation}
where $\mu$ and $\lambda$ denote the sine of the geographical latitude and longitude respectively. The equation of the barotropic model (\ref{eq:barotropic}) is non-dimensionalized using the radius of the earth as unit length and the inverse of the earth angular velocity as time unit. The non-dimensional orography $h$ is related to the real Northern Hemisphere orography $h^{'}$ by $h=2sin(\phi_0)A_0h^{'}/H$, where $phi_0$ is a fixed amplitude of $45^{\circ} N$, $A_0$ is a factor expressing the surface wind strength blowing across the orography, and $H$ a scale height \cite{Selten1995}. The stream-function $\psi$ is expanded into a spherical harmonics series and truncated at wavenumber 21, while modes with an even total wavenumber are excluded, avoiding currents across the equator and ending up with a hemispheric model with $231$ degrees of freedom.

\begin{figure}[!httb]
\centering
\begin{subfigure}{0.3\textwidth}
\centering
\includegraphics[height=3.6cm]{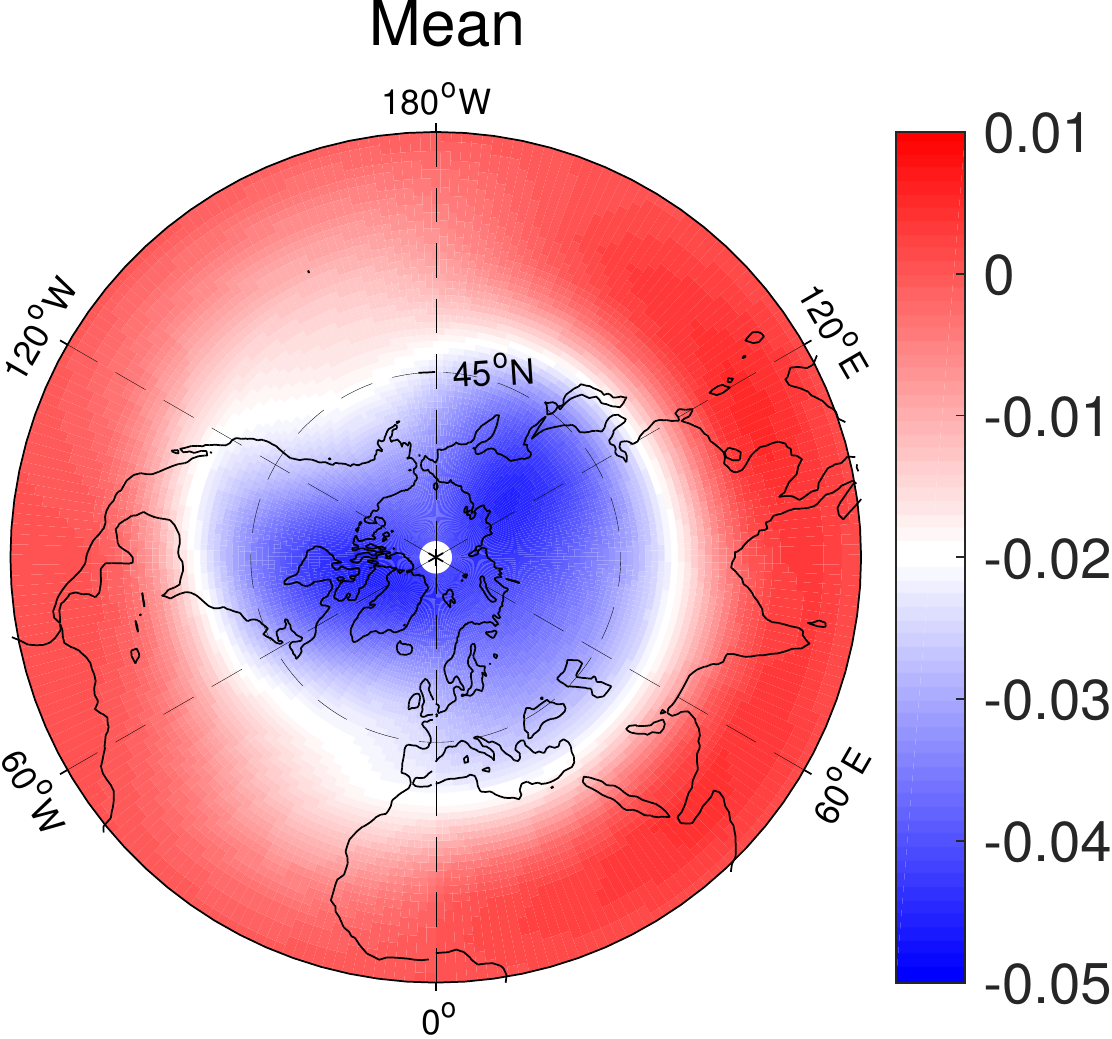}
\vspace{-0.5cm}
\caption{}
\label{T21:a}
\end{subfigure} 
\hspace{0.1cm}
\centering
\begin{subfigure}{0.3\textwidth}
\centering
\includegraphics[height=3.6cm]{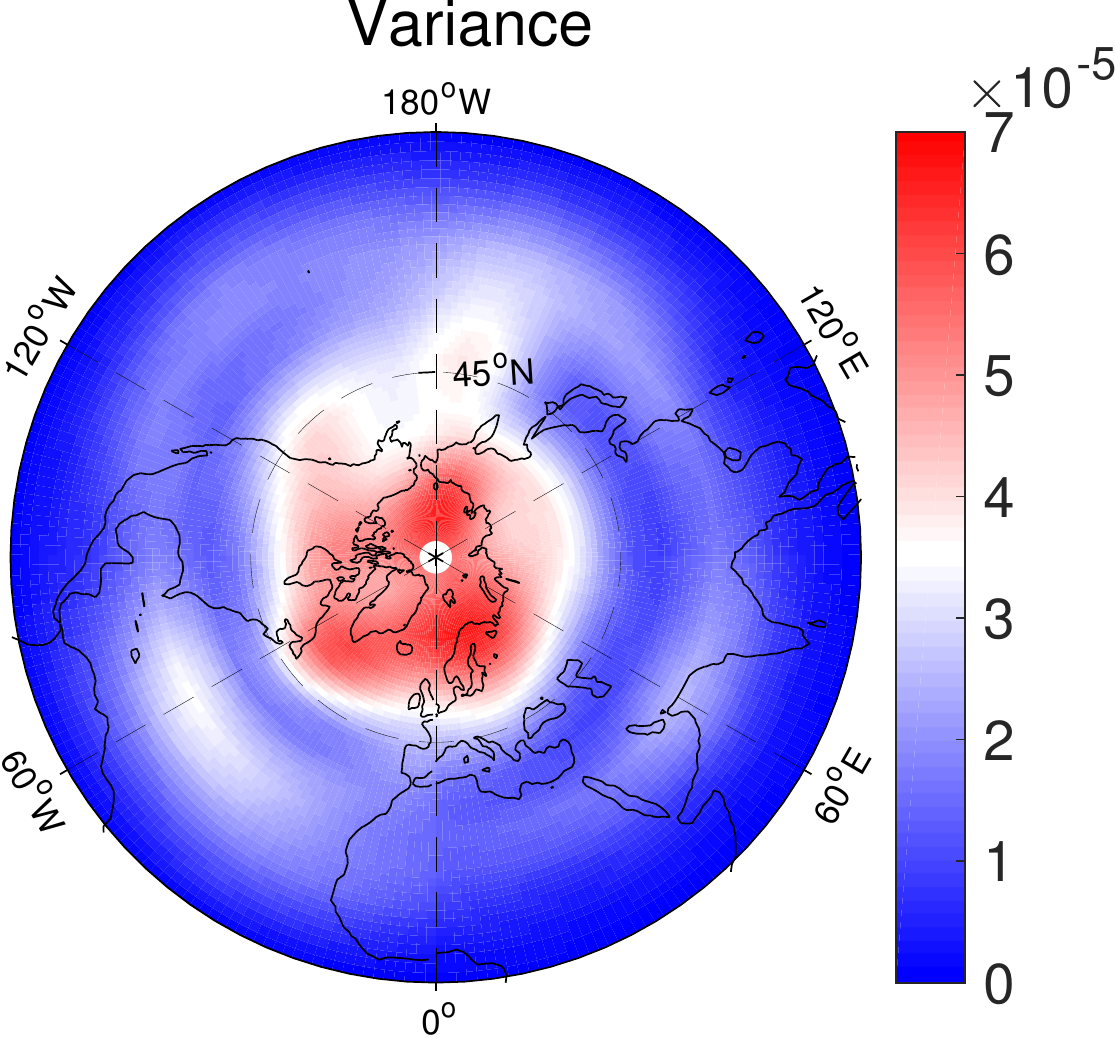}
\vspace{-0.5cm}
\caption{}
\label{T21:b}
\end{subfigure} 
\hspace{0.1cm}
\centering
\begin{subfigure}{.35\textwidth}
\centering
\includegraphics[height=3.6cm]{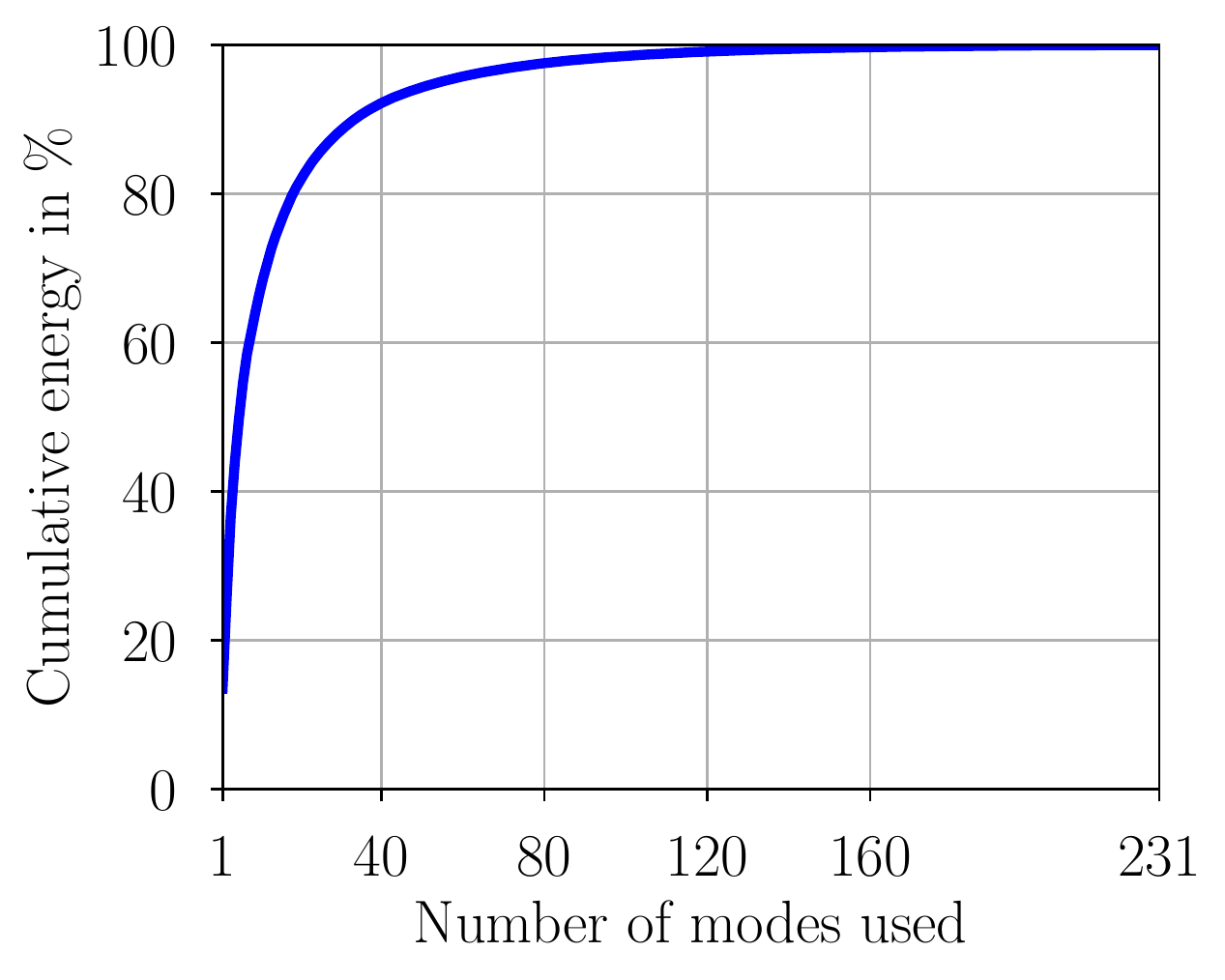}
\vspace{-0.5cm}
\caption{}
\label{T21:c}
\end{subfigure}
\caption{ \textbf{(a)} Mean of the Barotropic model at statistical steady state.
\textbf{(b)} Variance of the Barotropic model at statistical steady state. 
\textbf{(c)} Percentage of energy explained with respect to the modeled modes.
}
\label{fig:T_21}
\end{figure}

The training data are obtained by integrating the Eq. (\ref{eq:barotropic}) for $10^{5}$ days after an initial spin-up period of $1000$ days, using a fourth-order Adams-Bashforth integration scheme with a $45$-min time step in accordance with \cite{Zhong2017}, with $k_1=15$ days, while $k_2$ is selected such that wavenumber $21$ is damped at a time scale of $3$ days. In this way we end up with a time series $\zeta_t$ with $10^4$ samples. The spherical surface is discretized into a $D=64\times32$ mesh with equally spaces latitude and longitude. From the gathered data, $90\%$ is used for training and $10\%$ for validation. The mean and variance of the statistical steady state are shown in Figures \ref{T21:a} and \ref{T21:b}.

The dimensionality of the barotropic climate model truncated at wavenumber $21$ is $231$. In order to reduce it, we identify Empirical Orthogonal Functions (EOFs) $\phi_i$, $i \in \{1,\dots,231 \}$ that form an orthogonal basis of the reduced order space. The details of the method are described in the Appendix. EOF analysis has been used to identify individual realistic climatic modes such as the Arctic Oscillation (AO), the Pacific/North America (PNA) and the Tropical/Northern Hemisphere (TNH) pattern known as teleconnections \cite{Thompson2000, Mo1986}. Accurate prediction of these modes is of high practical importance as they feature realistic climate patterns. 
After projecting the state of the barotropic model to the EOFs, we take into account only the $r_{dim}$ coefficients corresponding to the most energetic EOFs that form the reduced order state $\mathbf{y}^{*}$. In our study, the dimensionality of the reduced space is $r_{dim}=30$, as $\mathbf{\phi}_{30}$ contains only $3.65\%$ of the energy of $\mathbf{\phi}_1$, while the $30$ most energetic modes contain approximately $82\%$ of the total energy, as depicted in Figure \ref{T21:c}.

\subsubsection{Training and Prediction}
\label{sec:trainingbarotropic}

The reduced order state that we want to predict using the LSTM are the $30$ components of $\mathbf{y}$. A \textit{stateless} LSTM with $h=140$ hidden units is considered, while the truncated back-propagation horizon is set to $d=10$. The prototypical system is less chaotic than the KS equation and the Lorenz 96, which enables us to use more hidden units. The reason is that as chaoticity is decreased trajectories sampled from the attractor as training and validation dataset become more interconnected and the task is inherently easier and less prone to overfitting. In the extreme case of a periodic system, the information would be identical. $500$ points are randomly and uniformly picked from the attractor as initial conditions for testing. A Gaussian ensemble with a small variance ($\sigma_{en}=0.001$) along each dimension is formed and marched using the reduced-order GPR, MSM, Mixed GPR-MSM and LSTM methods.

\subsubsection{Results}
\label{sec:barotropicresults}

\begin{figure}[!httb]
\centering
\begin{subfigure}{.47\textwidth}
\centering
\includegraphics[width=0.9\textwidth]{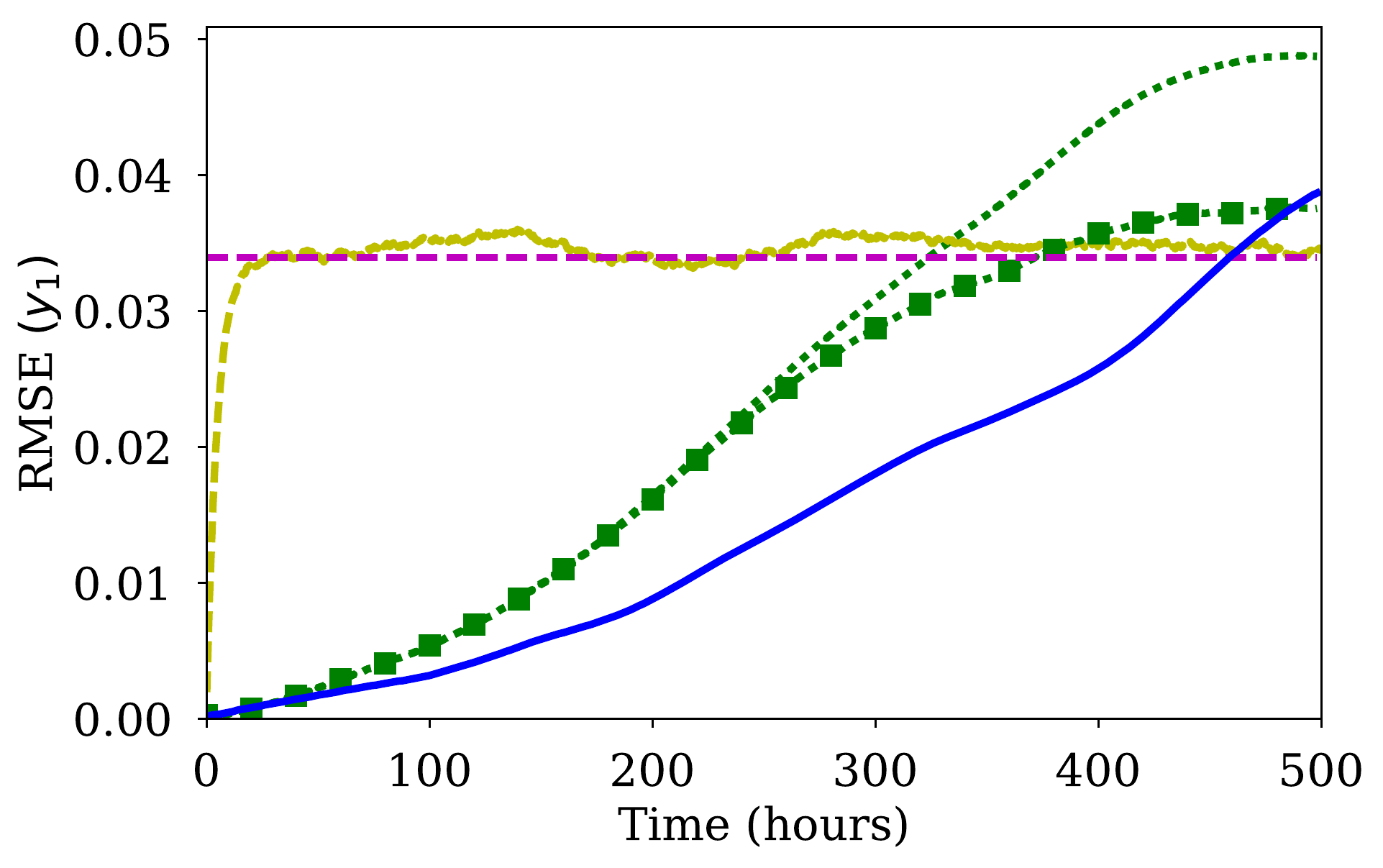}
\vspace{-0.75cm}
\caption{}
\label{barotropic_results:a}
\end{subfigure} 
\hspace{0.3cm}
\begin{subfigure}{.47\textwidth}
\centering
\includegraphics[width=0.9\textwidth]{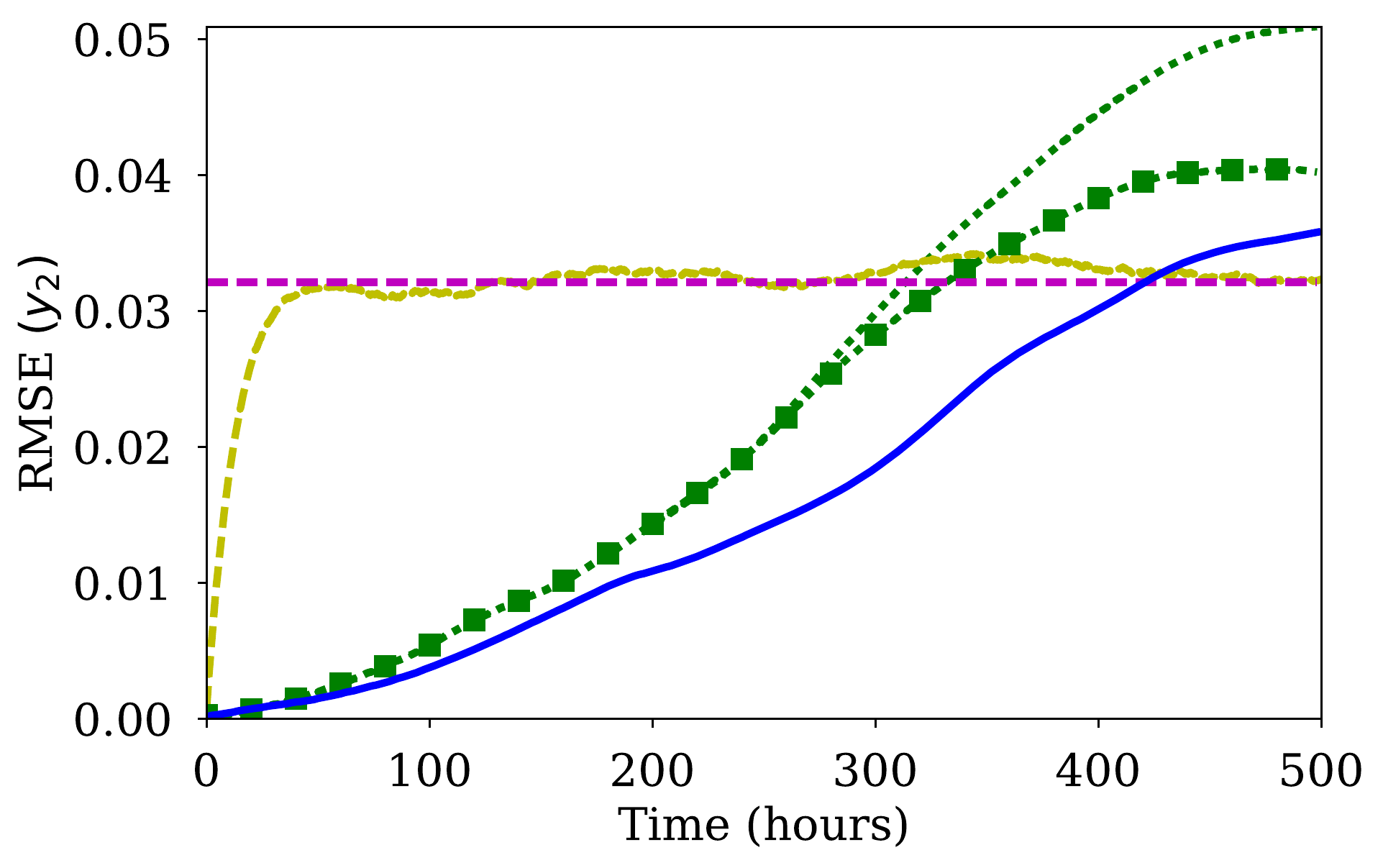}
\vspace{-0.75cm}
\caption{}
\label{barotropic_results:b}
\end{subfigure}
\centering
\begin{subfigure}{.47\textwidth}
\centering
\includegraphics[width=0.9\textwidth]{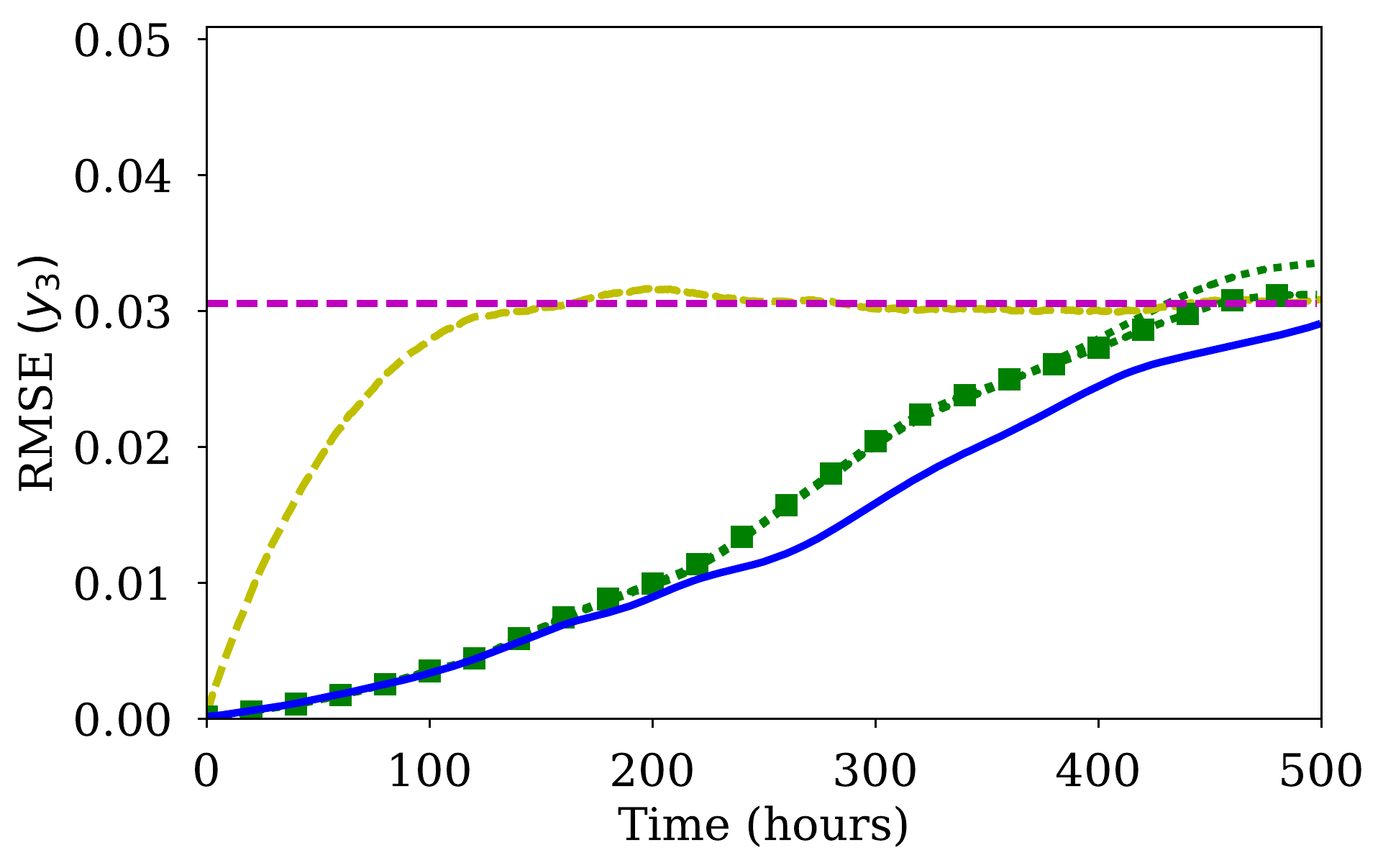}
\vspace{-0.75cm}
\caption{}
\label{barotropic_results:c}
\end{subfigure} 
\hspace{0.3cm}
\begin{subfigure}{.47\textwidth}
\centering
\includegraphics[width=0.9\textwidth]{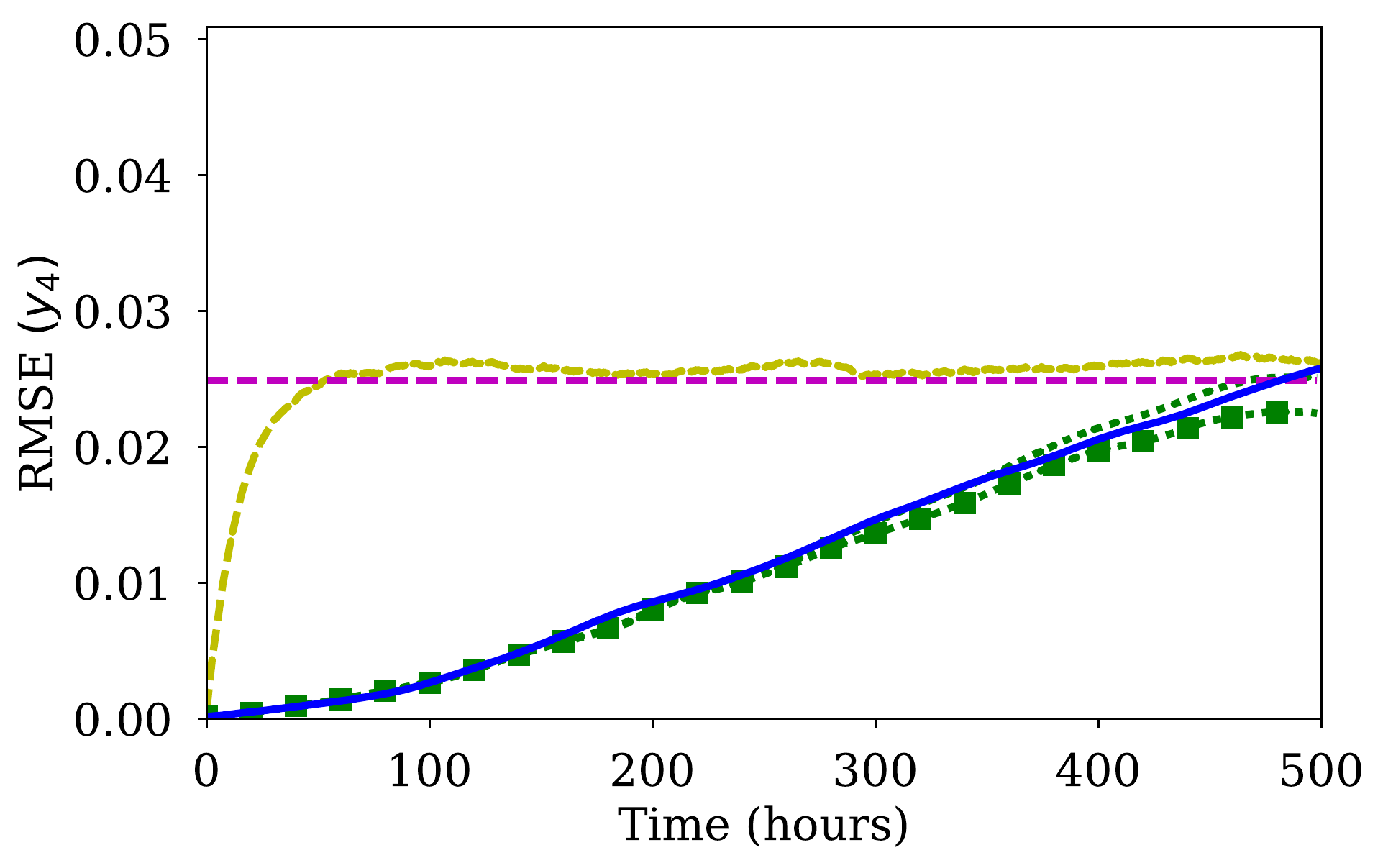}
\vspace{-0.75cm}
\caption{}
\label{barotropic_results:d}
\end{subfigure}
\caption{RMSE evolution of the four most energetic EOFs for the Barotropic climate model, average over $500$ initial conditions reported. 
\textbf{(a)} Most energetic EOF.
\textbf{(b)} Second most energetic EOF.
\textbf{(c)} Third most energetic EOF.
\textbf{(d)} Fourth most energetic EOF.
\\ 
$\sigma_{attractor}$\protect\magentalineDashed; MSM \protect\yellowlineDashed; GPR \protect\greenlineDotted; GPR-MSM \protect\greenlineDottedRectangle; LSTM \protect\blueline}
\label{fig:barotropic_results}
\end{figure}

The RMSE error of the four most energetic reduced order space variables $\mathbf{y}_i$ for $i \in \{1,\dots,4 \}$ is plotted in Figure \ref{fig:barotropic_results}. The LSTM takes $400-500$ $h$ to reach the attractor, while GPR based methods generally take $300-400$ $h$. In contrast, the MSM reaches the attractor already after $1$ hour. This implies that the LSTM can better capture the non-linear dynamics compared to GPR. Note that the barotropic model is much less chaotic than the Lorenz 96 system with $F=16$, where all methods show comparable prediction performance. Blended LSTM models with MSM are omitted here, as LSTM models only reach the attractor standard deviation towards the end of the simulated time and MSM-LSTM shows identical performance.

\section{A Comment on Computational Cost of Prediction}
\label{sec:complexity}

The computational cost of making a single prediction can be quantified by the number of operations (multiplications and additions) needed. In GPR based approaches the computational cost is of order $O(N^2)$, where $N$ is the number of samples used in training. For GPR methods illustrated in the previous section $N\approx 2500$. The GPR models the global dynamics by uniformly sampling the attractor and "carries" this training dataset at each time instant to identify the geometric relation between the input and the training dataset (modeled with a covariance matrix metric) and make (exact or approximate) probabilistic inference on the output.

In contrast, LSTM adjusts its parameters to reproduce the local dynamics. As a consequence, the inference computational complexity does not depend on the number of samples used for training. The inference complexity is roughly $O(d_i \cdot d \cdot h + d \cdot h^2)$, where $d_i$ is the dimension of each input, $d$ is the number of inputs and $h$ is the number of hidden units. This complexity is significantly smaller than GPR, which can be translated to faster prediction. However, it is logical that the LSTM is more prone to diverge from the attractor, as there is no guarantee that the infrequent training samples near the attractor limits where memorized. This remark explains the faster divergence of LSTM in the more turbulent regimes considered in Section \ref{sec:applications}.

\section{Conclusions}
\label{sec:conclusion}

We propose a data-driven method, based on long short-term memory networks, for modeling and prediction in the reduced space of chaotic dynamical systems. The LSTM uses the short-term history of the reduced order variable to predict the state derivative and uses it  for one-step prediction. The network is trained on time-series data and it requires no prior knowledge of the underlying governing equations. Using the trained network, long-term predictions are made by iteratively predicting one step forward. 

The features of the proposed technique are showcased through comparisons with GPR and MSM on benchmarked cases. Three applications are considered, the Lorenz 96 system, the Kuramoto-Sivashinsky equation and a barotropic climate model. The chaoticity of these systems ranges from weakly chaotic to fully turbulent, ensuring a complete simulation study. Comparison measures include the RMSE and ACC between the predicted trajectories and trajectories of the real dynamics. 

In all cases, the proposed approach performs better, in short-term predictions, as the LSTM is more efficient in capturing the local dynamics and complex interactions between the modes. However, the prediction error accumulates as we iteratively perform predictions and similar to GPR does not converge to the invariant measure. Furthermore in the cases of increased chaoticity the LSTM diverges faster than GPR. This may be attributed to the absence of certain attractor regions in the training data, insufficient training, and propagation of the exponentially increasing prediction error. To mitigate this effect, LSTM is also combined with MSM, following ideas presented in \cite{Zhong2017}, in order to guarantee convergence to the invariant measure. Blending LSTM or GPR with MSM leads to a deterioration in the short-term prediction performance but the steady-state statistical behavior is captured. The hybrid LSTM-MSM exhibits a slightly superior performance than GPR-MSM in all systems considered in this study.

In the Kuramoto-Sivashinsky equation LSTM can capture better the local dynamics compared to Lorenz 96 due to the lower intrinsic attractor dimensionality. LSTM is more accurate than GPR in the short-term, but especially in the chaotic regime $1/\nu=16$ forecasts of LSTM fly away from the attractor faster. LSTM-MSM counters this effect and long-term forecasts converge to the invariant measure at the expense of a compromise in the short-term forecasting accuracy. The higher short-term forecasting accuracy of LSTM can be attributed to the fact that it is a nonlinear approximator and can also capture correlations between modes in the reduced space. In contrast, GPR is a locally linear approximator modeling each mode independently in the output, assuming Gaussian correlations between modes in the input. LSTM and GPR show comparable forecasting accuracy in the barotropic model, as the intrinsic dimensionality is significantly smaller than Kuramoto-Sivashinsky and Lorenz 96 and both methods can effectively capture the dynamics.

Future directions include modeling the lower energy modes and interpolation errors using a stochastic component in the LSTM to improve the forecasting accuracy. Another possible research direction is to model the attractor in the reduced space using a mixture of LSTM models, one model for each region. The LSTM proposed in this work models the attractor globally. However, different attractor regions may exhibit very different dynamic behaviors, which cannot be simultaneously modeled using only one network. Moreover, these local models can be combined with a closure scheme compensating for truncation and modeling errors. This local modeling approach may further improve prediction performance.

\section{Data Accessibility}

The code and data used in this work are available at the link: \\
\centerline{https://polybox.ethz.ch/index.php/s/keH7PftvLmbkYU1}
\\
The password is {\tt rspa\_paper}. 
The {\tt TensorFlow} library and {\tt python3} were used for the implementation of LSTM architectures while {\tt Matlab} was used for Gaussian Processes. These packages need to be installed in order to run the codes.

\section{Authors' Contributions}

PRV conceived the idea of the blended LSTM-MSM scheme, implemented the neural network architectures and the simulations, interpreted the computational results, and wrote the manuscript. WB supervised the work and contributed to the implementation of the LSTM. ZYW implemented the GPR and made contributions to the manuscript. PK had the original idea of the LSTM scheme and contributed to the manuscript. PK and TPS contributed to the interpretation of the results and offered consultation. All authors gave final approval for publication.

\section{Competing Interests}

We have no competing interests.

\section{Funding} 

TPS and ZYW have been supported by an Air Force Office of Scientific Research grant FA9550-16-1-0231, an Office of Naval Research grant N00014-15-1-2381, and an Army Research Office grant 66710-EG-YIP. PK and PRV gratefully acknowledge support from the European Research Council (ERC) Advanced Investigator Award (No. 341117).

\section{Acknowledgments} 

We thank the two anonymous reviewers whose insightful comments helped us to enhance the manuscript.

\appendix

\section{Long short-term memory (LSTM)}

\subsection{Training and inference}

In this section, the LSTM training procedure is explained in detail. We assume that time series data stemming from a dynamical system is available in the form $D=\{ z_{t:N}, \dot{z}_{t:N} \}$, where $z_t \in \R^{d_i}$ is the state at time step $t$ and $\dot{z}_{t}$ is the derivative. The available time series data are divided into two separate sets, the training dataset and the validation dataset, i.e. $z_{t}^{train},  \,\dot{z}_{t}^{train}, \,t \in \{1, \cdots, N_{train} \},$ and $z_{t}^{val},  \, \dot{z}_{t}^{val}, \, t \in \{1, \cdots, N_{val} \}$. $N_{train}$ and $N_{val}$ are the number of training and validation samples respectively. We set the ratio to $N_{train}/N=0.8$.
This data is stacked as
\begin{equation}
\underbrace{
\i_t^{train} = \begin{pmatrix} 
  z_{t+d-1}^{train} \\ 
 z_{t+d-2}^{train} \\ 
  \vdots \\ 
  z_{t}^{train} \\ 
\end{pmatrix}
}_{\text{Input stack}}, \quad
\underbrace{
\o_t^{train} = \dot{z}_{t+d-1}^{train}
}_{\text{Output stack}},
\label{eq:stack}
\end{equation}
for $t \in \{1,2,\dots, N_{train} -d+1\}$, in order to form the training (and validation) input and output of the LSTM. These training samples are used to optimize the parameters of the LSTM (weights and biases) in order to learn the mapping $\i_t \to \o_t$. The loss function of each sample is
\begin{equation}
\mathcal{L}_{sample}(\i_t^{train}, \o_t^{train}, w) = ||\mathcal{F}^w(\underbrace{z^{train}_{t:t-d+1}}_{\i_t^{train}}) - \o_t^{train}||^2,
\end{equation}
while the total Loss is defined as
\begin{equation}
\mathcal{L}(D, w) = \frac{1}{S} \sum_{b=1}^{S} \mathcal{L}(\i_b^{train}, \o_b^{train}, w),
\end{equation}
where $S=N_{train}-d+1$ is the total number of samples. These samples can be further stacked together as batches of size $B$, with the loss of the batch defined as the mean loss of the samples belonging to the batch. Using only one sample for the loss gradient estimation may lead to noisy gradient estimates and slow convergence. Mini-batch optimization tackles this issue.

At the beginning of the training the weights are randomly initialized to $w^{0}$ using Xavier initialization. We also tried other initialization methods like drawing initial weights from random normal distributions, or initializing them to constant values, but they often led to saturation of the activation functions, especially for architectures with higher back-propagation horizon $d$. The training proceeds by optimizing the network weights iteratively for each batch. In order to perform this optimization step, a gradient descent optimizer can be used
\begin{equation}
w^{i+1} = w^{i} - \eta \nabla_w \mathcal{L}(\i_t^{train}, \o_t^{train}, w^{i}),
\end{equation}
where $\eta$ is the step-size parameter, $w^{i}$ are the weights before optimizing the batch $i$ and $w^{i+1}$ are the updated weights. Plain gradient descent optimization suffers from slow convergence in practice and convergence to local sub-optimal solutions. This approach is especially not well-suited for high dimensional problems in deep learning where the number of parameters (weights) to be optimized lie in a high-dimensional manifold with many local optima. Sparse gradients stemming from the mini-batch-optimization lead also to slow convergence as previously computed gradients are ignored. Recent advances in stochastic optimization led to the invention of adaptive schemes that efficiently cope with the aforementioned problems.

In our work, we used the Adam stochastic optimization method. Adam exploits previously computed gradients using moments. The weights are initialized to $w^{0}$ and the moment vectors to $m^{0}_1$ and $m^{0}_2$. At each step the updates in the Adam optimizer are
\begin{equation}
\begin{aligned}
&g = \nabla_w \mathcal{L}(\i_t^{train}, \o_t^{train}, w^{i})\\
&m^{i+1}_1 = \beta_1 m^{i}_1 + (1-\beta_1) \, g  \\
&m^{i+1}_2 = \beta_2 m^{i}_2 + (1-\beta_2) \, g^2  \\
&\hat{m}_1 =m^{i+1}_1/(1-\beta_1^{i})  \\
&\hat{m}_2 =m^{i+1}_2/(1-\beta_2^{i})  \\
&w_{i+1} =w_{i} - \eta \,  \hat{m}_1 / (\sqrt{\hat{m}_2}+\epsilon),
\end{aligned}
\label{eq:adam}
\end{equation}
where $\beta_1, \beta_2, \epsilon,$ and $\eta$ are hyper-parameters, $g^2$ is the point-wise square of the gradient and $\beta_1^{i}$ is the parameter $\beta_1$ in the $i$\textsuperscript{th} power, where $i$ is the iteration number. After updating the weights using the Adam procedure (\ref{eq:adam}) for every batch, a \textit{training epoch} is completed. Many such epochs are performed until the total training loss reaches a plateau. After each epoch the loss is evaluated also in the validation data set, in order to avoid overfitting. The validation loss is used as a proxy of the generalization error. The training is stopped when the validation error is not decreasing for $30$ consecutive epochs or the maximum of $1000$ epochs is reached. In our work we used $\beta_1=0.9$, $\beta_2=0.999$, $\epsilon=1e-8$. We found that our results were robust towards the selection of these hyper-parameters. To speed up convergence speed, a higher initial learning rate $\eta=0.001$ was used and the models were then refined with $\eta=0.0001$.
 
\subsection{Weighting the loss function}

In the training procedure described above the loss function for each sample is given by
\begin{equation}
\mathcal{L}_{sample}(\i_t, \o_t, w) = ||\mathcal{F}^w(\underbrace{z_{t:t-d+1}}_{\i_t}) - \o_t||^2.
\end{equation}
However, in the applications considered in this paper the neural network output $\mathcal{F}^w$ is a multidimensional vector and represents a prediction of the derivative of the reduced order state of a dynamical system. In a dynamical system, specific reduced order states are more important than others as they may explain a bigger portion of the total energy. This importance can be introduced in the loss function by assigning different weights in different outputs of the neural network. The loss of each sample takes then the following form
 \begin{equation}
\mathcal{L}_{sample}(\i_t^{j}, \o_t^{j}, w) = \frac{1}{d_o} \sum_{j=1}^{d_o} w_j \Big(\mathcal{F}^w(\underbrace{z^{j}_{t:t-d+1}}_{\i_{t}^j}) - \o_{t}^{ j} \Big)^2,
\end{equation}
where $d_o$ is the output dimension and weights $w_j$ are selected according to the significance of each output component, e.g. energy of each component in the physical system.

\subsection{LSTM architecture}

\begin{figure}[!httb]
\centering
\includegraphics[width=0.9\textwidth]{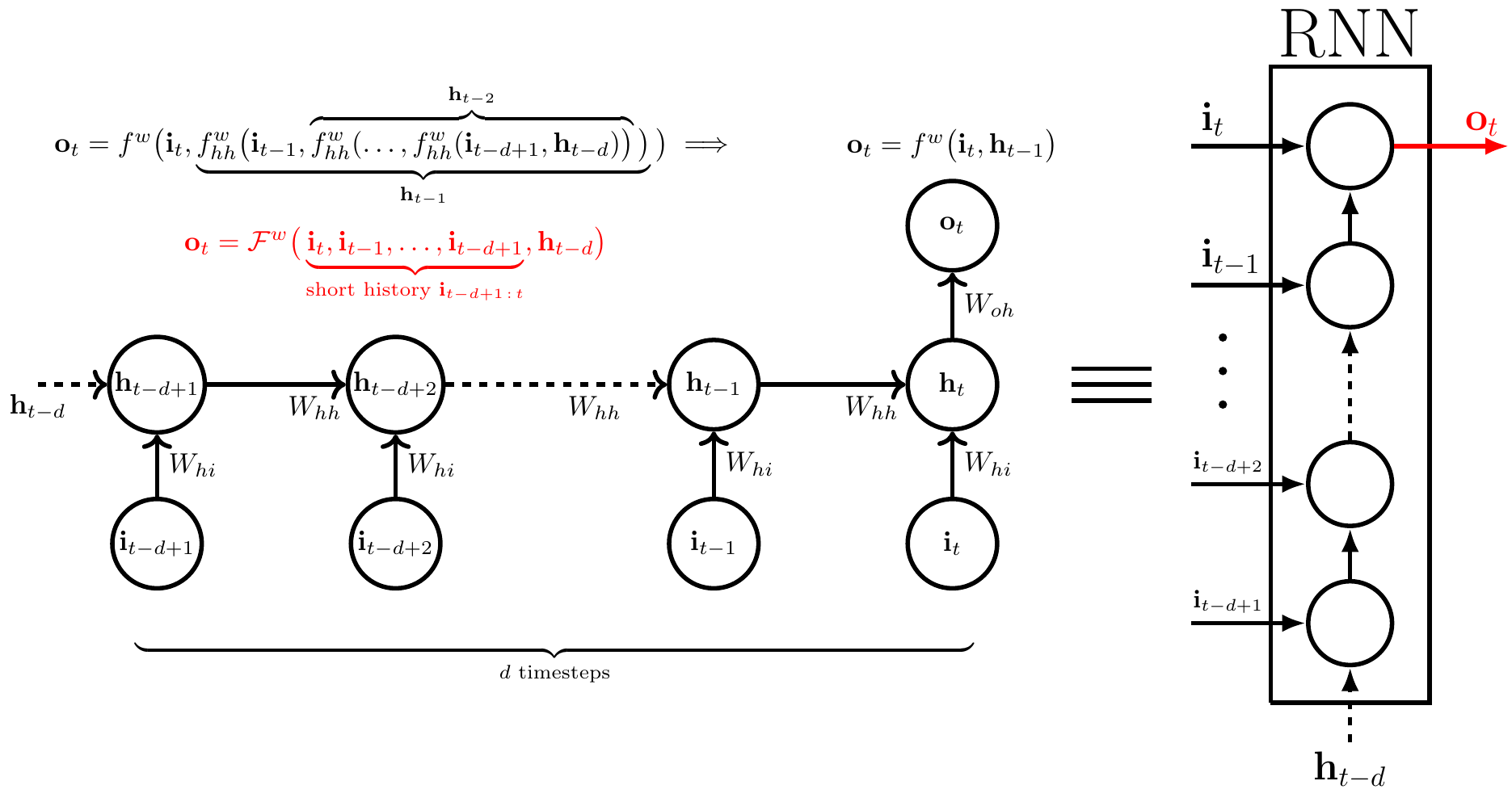}
\caption{An RNN unfolded $d$ timesteps. In mathematical terms, unfolding is equivalent with iteratively applying $f^w_{hh}$ to $\mathbf{h}_{t-d}$ and finally feeding the result to the output function $f^w$. The output of the RNN is thus a function of the $d$ previous inputs $\i_{t:t-d+1}$ and the initialization of the hidden states $\mathbf{h}_{t-d}$. This function is denoted with $\mathcal{F}^w$. For the RNN the hidden state mapping has the simple form $f_{hh}^w(\mathbf{i}_{t}, \mathbf{h}_{t-1})=\sigma_h(W_{hi} \mathbf{i}_{t} + W_{hh} \mathbf{h}_{t-1})$, while the output mapping is $f^w(\mathbf{i}_t,\mathbf{h}_{t-1})=\sigma_o(W_{oh} \mathbf{h}_t)=
\sigma_o \big(W_{oh} f_{hh}^w(\mathbf{i}_t,\mathbf{h}_{t-1}) \big)
$. The same argumentation holds for LSTM, though the form of $f_{hh}^w$, $f^w$ and $\mathcal{F}^w$ are more complicated.}
\label{fig:RNN_unfolded_function_tt}
\end{figure}

An RNN unfolded $d$ temporal time steps in the past is illustrated in Figure \ref{fig:RNN_unfolded_function_tt}. The following discussion on \textit{Stateless} and \textit{Stateful} RNNs generalizes to LSTMs, with the only difference that the hidden state consists of $\h_t, C_t$ instead of solely $\h_t$ and the functions coupling the hidden states with the input as well as the output with the hidden states are more complicated.

In \textit{Stateless} RNNs the hidden states at the truncation layer $d$, $\h_{t-d}$ are initialized always with $0$. As a consequence, $\o= \mathcal{F}^w( \i_{t:t-d+1})$ and only the short-term history is used to perform a prediction. The only difference when using LSTM cells is that the function $\mathcal{F}^w$ has a more complex structure and additionally $\h_{t-d}, C_{t-d}=0$. 

In contrast, in \textit{Stateful} RNNs the states $\h_{t-d} \neq 0$. In this case, these states can be initialized by "teacher forcing" the RNN using data from a longer history in the past. For example, assuming $\i_{t-d:t-2d+1}$ is known, we can set $\h_{t-2d}=0$, and compute $\h_{t-d}$ using the given history $\i_{t-d:t-2d+1}$ ignoring the outputs. This value can then be used to predict $\o_t=\mathcal{F}^{w} (\i_{t:t-d+1}, \h_{t-d})$ as in Figure \ref{fig:RNN_unfolded_function_tt}. This approach has two disadvantages.

\begin{itemize}
\item In order to be able to forecast starting from various initial conditions, even with "teacher forcing" some initialization of the hidden states is imperative. This initialization introduces additional error, which is not the case for the \textit{Stateless} RNN.
\item In the \textit{Stateful} RNN a longer history needs to be known in order to initialize the hidden states with "teacher forcing". Even though more data needs to be available, we did not observe any prediction accuracy improvement in the cases considered. This follows from the chaotic nature of the systems, as information longer than some time-steps in the past are irrelevant for the prediction.
\end{itemize}

\section{Lorenz 96}
\label{sec:appendixlorenz}

The most energetic Fourier modes in the Lorenz 96 system for different forcing regimes $F \in \{4,6,8,16 \}$ are given in Table \ref{tab:mostenergeticmodes}. These modes are used in order to construct the reduced order phase space.

\renewcommand{\arraystretch}{1.2}
\begin{table}[h!]
\centering
\begin{tabular}{|c|c||c|c|}  \hline
Forcing & Wavenumbers $k$ & Forcing & Wavenumbers $k$  \\  \hline 
$F=4$ & 7,10,14,9,17,16 & $F=8$ & 8,9,7,10,11,6 \\ \hline
$F=6$ & 8,7,9,10,11,6 & $F=16$ & 8,9,10,7,11,6 \\ \hline
\end{tabular}
\caption{Most energetic Fourier modes used in the reduced order phase space}
\label{tab:mostenergeticmodes}
\end{table}
\renewcommand{\arraystretch}{1}

\section{Kuramoto-Sivashinsky equation}

\subsection{Dimensionality reduction}
\label{sec:appendixdimreckuramoto}

\begin{figure}[!httb]
\centering
\begin{subfigure}{.46\textwidth}
\centering
\includegraphics[width=1\textwidth]{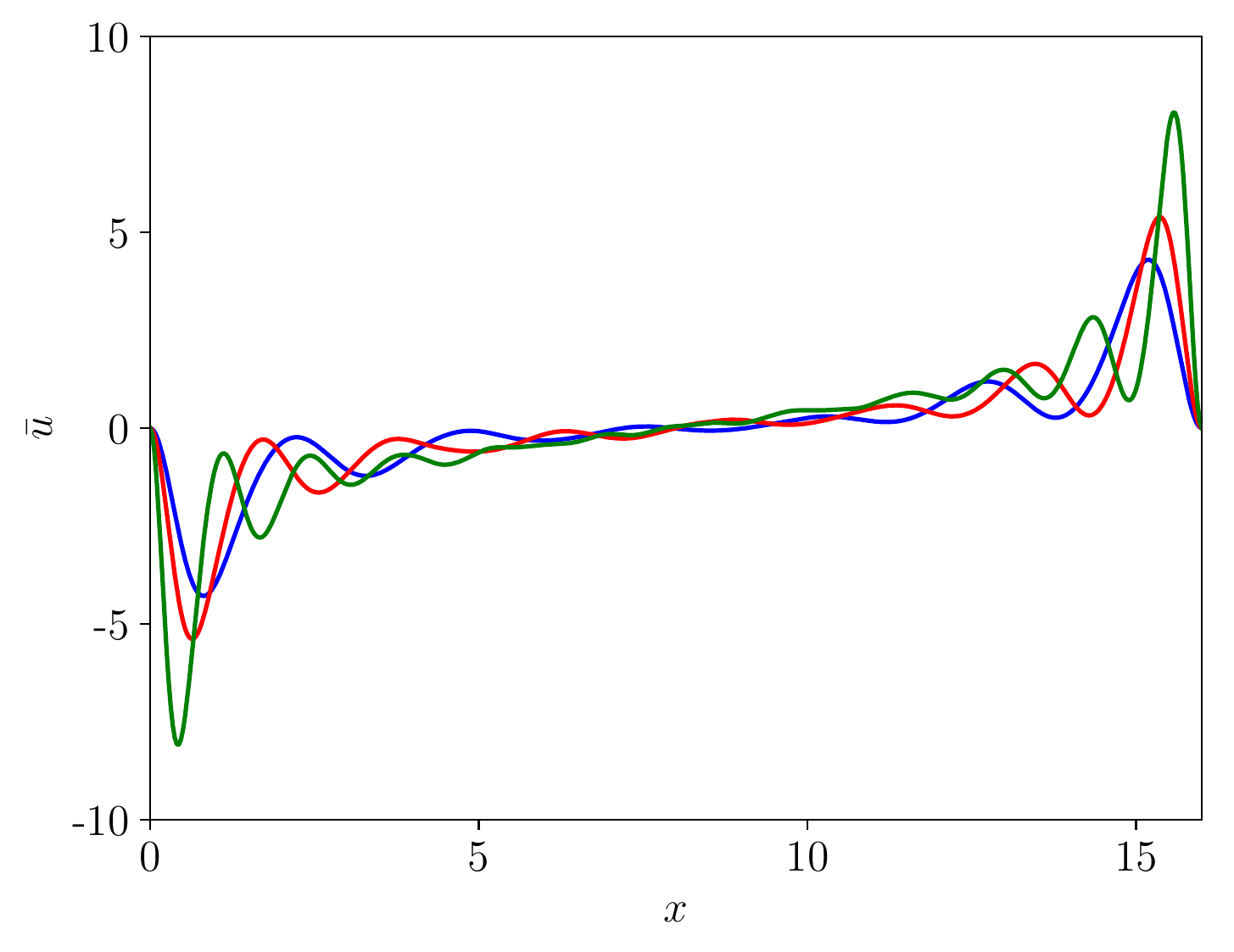}
\end{subfigure} 
\hspace{0.6cm}
\begin{subfigure}{.46\linewidth}
\includegraphics[width=1\textwidth]{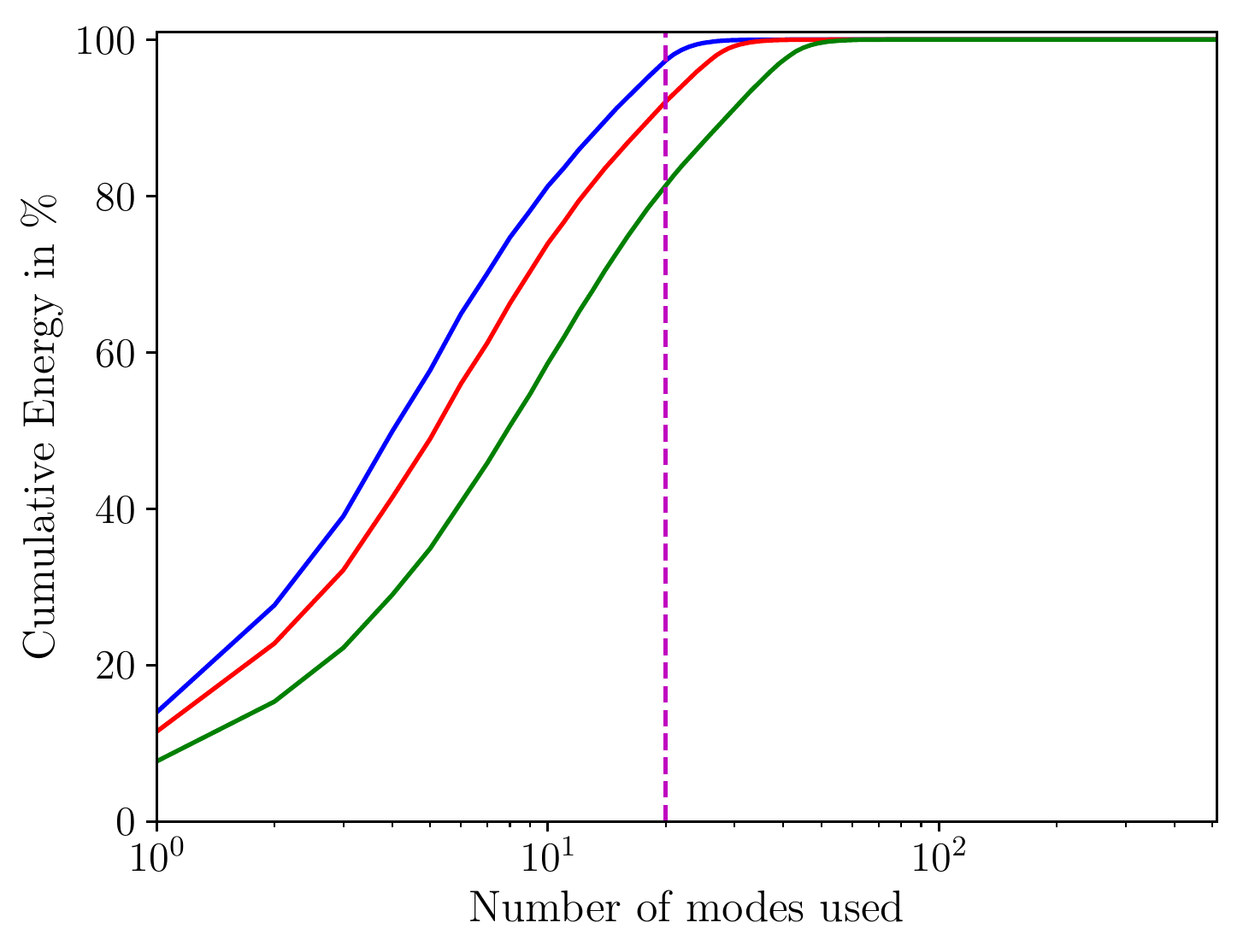}
\end{subfigure}
\caption{Temporal average $\overline{\mathbf{u}}$ and cumulative mode (PCA) energy for different values of $\nu$ in the Kuramoto-Sivashinsky system. \\ $1/\nu=10$ \protect\blueline ; $1/\nu=16$ \protect\redline ; $1/\nu=36$ \protect\greenline ; 20 modes \protect\magentalineDashed}
\label{fig:Plot_U_E_Kuramoto}
\end{figure}

The temporal average of the state of the Kuramoto-Sivashinsky equation and the cumulative energy are plotted in Figure \ref{fig:Plot_U_E_Kuramoto}. As $\nu$ declines, chaoticity in the system rises and higher oscillations of the mean towards the Dirichlet boundary conditions are observed in Figure \ref{fig:Plot_U_E_Kuramoto}, while the number of modes needed to capture most of the energy is higher.
 
\subsection{Results}

\begin{figure}[!httb]
\centering
\begin{subfigure}{.45\textwidth}
\centering
\includegraphics[width=0.95\textwidth]{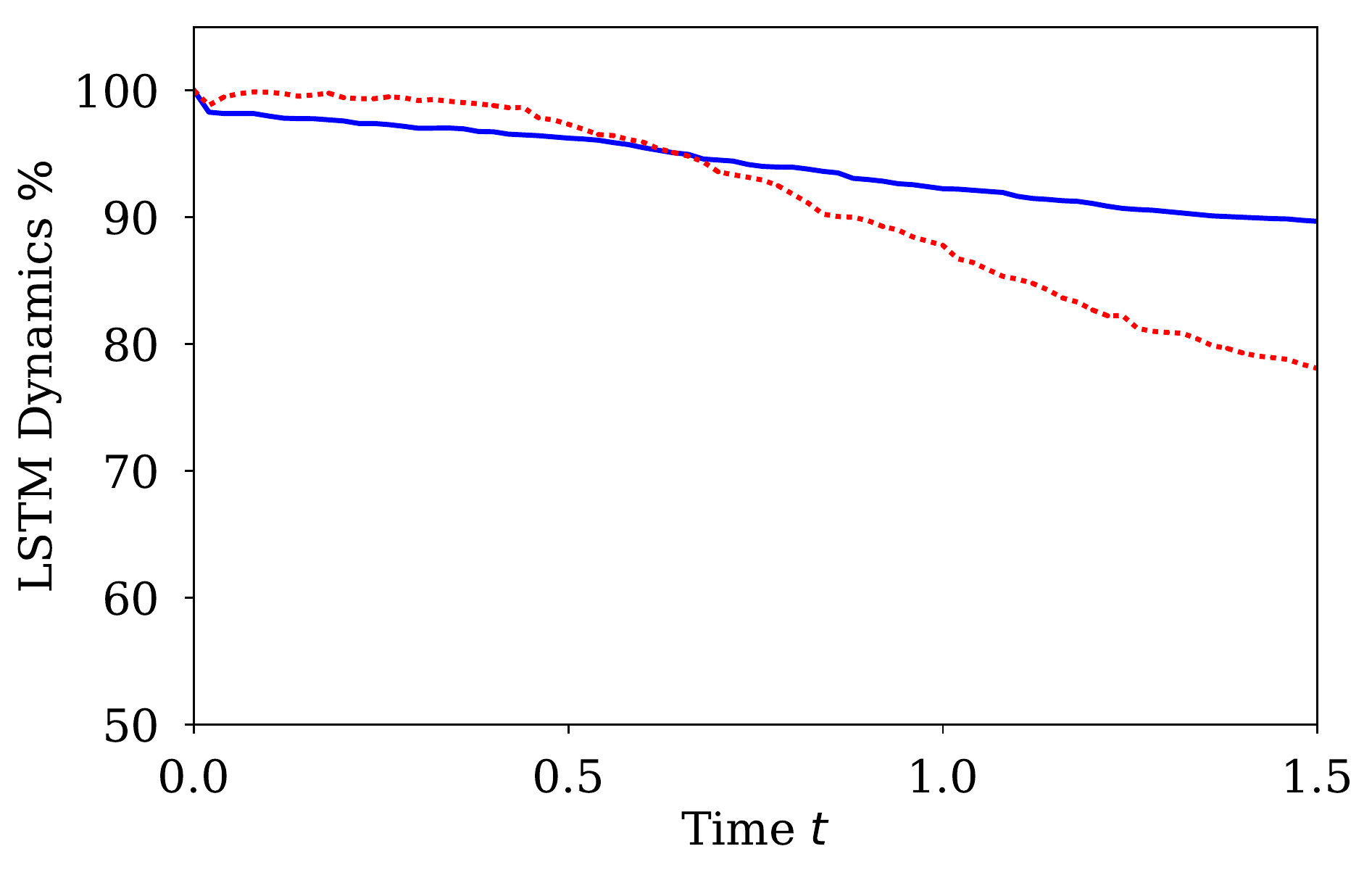}
\vspace{-0.5cm}
\caption{}
\label{KS_MSM_Quotient:a}
\end{subfigure}
\begin{subfigure}{.45\textwidth}
\centering
\includegraphics[width=0.95\textwidth]{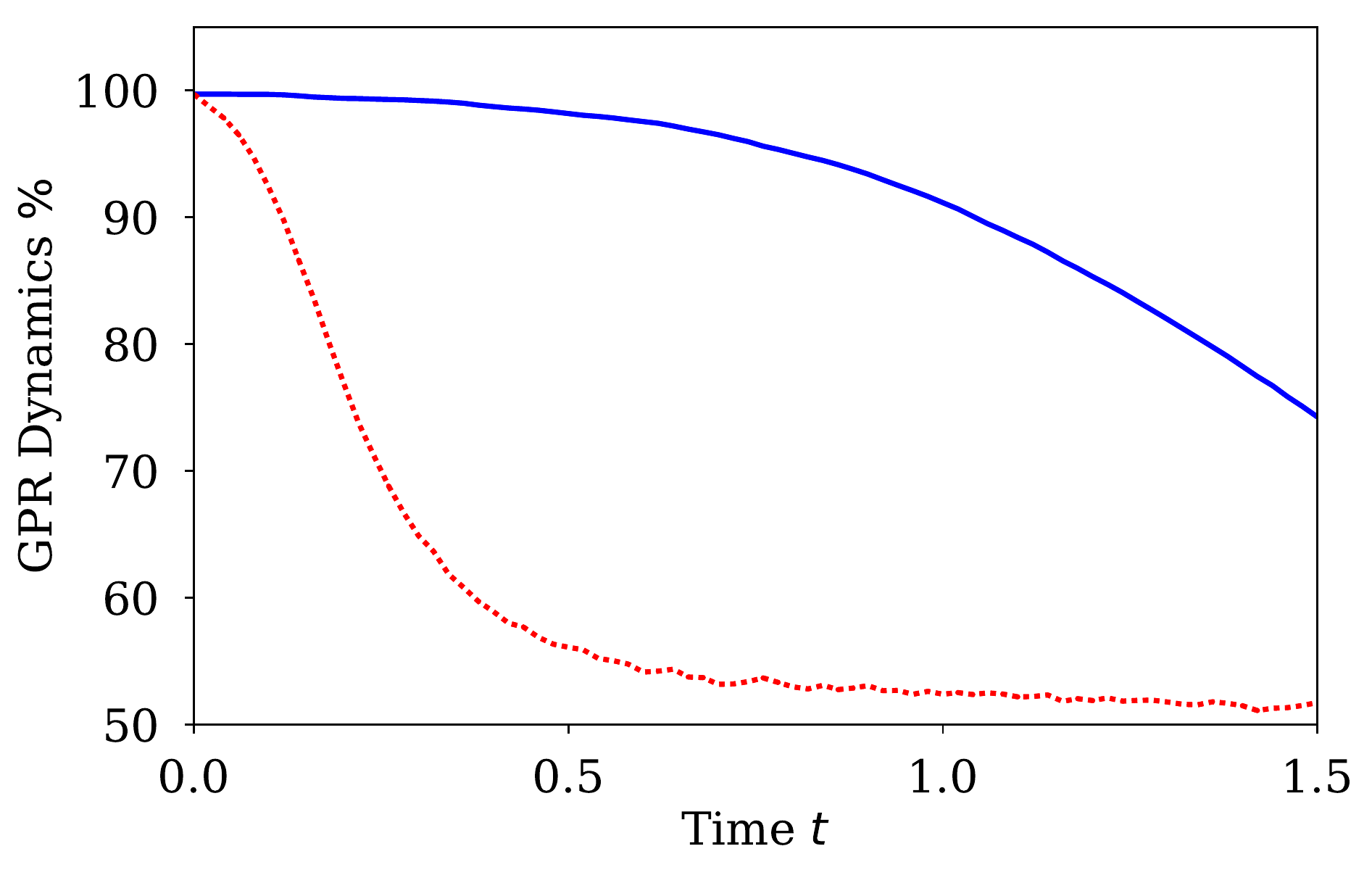}
\vspace{-0.5cm}
\caption{}
\label{KS_MSM_Quotient:b}
\end{subfigure}
\caption{\textbf{(a)} Ratio of LSTM-MSM ensemble members modeled by the LSTM dynamics for the Kuramoto-Sivashinsky ($T=1.5$). \textbf{(b)} The same for GPR in the hybrid GPR-MSM. (Mean over 1000 initial conditions) \\ 
$1/\nu=10$ \protect\blueline ; $1/\nu=16$ \protect\redlineDotted}
\label{fig:KS_MSM_Quotient}
\end{figure}

For the hybrid LSTM-MSM, the ratio of the ensemble members that are modeled with LSTM is plotted with respect to time in Figure \ref{KS_MSM_Quotient:a}. The quotient drops slower for $1/\nu=10$ in the long run as the intrinsic dimensionality of the attractor is smaller and trajectories diverge slower. However, in the beginning the LSTM ratio is higher for $1/\nu=16$ as the MSM drives initial conditions close to the boundary faster towards the attractor due to the higher damping coefficients compared to the case $1/\nu=10$. This explains the initial knick in the graph for $1/\nu=16$. The slow damping coefficients for $1/\nu=10$ do not allow the MSM to drive the trajectories back to the attractor in a faster pace than the diffusion caused by the LSTM forecasting. Compared with GPR plotted in Figure \ref{KS_MSM_Quotient:b}, the ratio drops slower.

\section{Barotropic model}
\label{sec:appendixdimrecbarotropic}

In this section we describe the method used to reduce the dimensionality of the Barotropic climate model. First, the original problem dimension of $231$ is reduced using a generalized version of the classical multidimensional scaling method. Then, we construct Empirical Orthogonal Functions (EOFs) that form an orthogonal basis of the reduced order space and project the dynamics to them.

The classical multidimensional scaling procedure tries to identify an embedding with a lower dimensionality such that the pairwise inner products of the dataset are preserved. Assuming that the dataset consists of points $\mathbf{\zeta}_i$, $i\in \{1,\dots,N \}$, whose reduced order representation is denoted with $\mathbf{y}_i$, the procedure is equivalent with the solution of the following optimization problem
\begin{equation}
\underset{\mathbf{y}_1,\dots,\mathbf{y}_N}{\text{minimize}} \sum_{i<j} \big( \langle \mathbf{\zeta}_i, \mathbf{\zeta}_j  \rangle_{\mathbf{\zeta}} - 
\langle \mathbf{y}_i,\mathbf{y}_j \rangle_{\mathbf{y}} \big)^2,
\label{eq:dms}
\end{equation}
where $\langle \cdot, \cdot  \rangle_{\mathbf{\zeta}}$, and $\langle \cdot, \cdot  \rangle_{\mathbf{y}}$ denote some well defined inner product of the original space $\mathbf{\zeta}$ and the embedding space $\mathbf{y}$ respectively. Problem (\ref{eq:dms}) minimizes the total squared error between pairwise products. In case both products are the scalar products, the solution of (\ref{eq:dms}) is equivalent with PCA. Assuming only $\langle \cdot, \cdot  \rangle_{\mathbf{y}}$ is the scalar product, problem (\ref{eq:dms}) also accepts an analytic solution. Let $W_{ij}=\langle \mathbf{\zeta}_i, \mathbf{\zeta}_j  \rangle_{\mathbf{\zeta}}$ be the coefficients of the Gram matrix, $ |k_1| \geq |k_2| \geq \dots \geq |k_N|$ its eigenvalues sorted in descending absolute value and $\mathbf{u}_1, \mathbf{u}_2, \dots, \mathbf{u}_N$ the respective eigenvectors. The optimal $d$-dimensional embedding for a point $\mathbf{\zeta}_n$ is given by
\begin{equation}
\mathbf{y}_n = 
\begin{pmatrix}
k_1^{1/2} \mathbf{u}_{1}^n \\
k_2^{1/2} \mathbf{u}_{2}^n \\
\vdots \\
k_d^{1/2} \mathbf{u}_{d}^n \\
\end{pmatrix}	,
\label{eq:yn}
\end{equation}
where $\mathbf{u}_{m}^n$ denotes the $n^{th}$ component of the $m^{th}$ eigenvector. The optimality of (\ref{eq:yn}) can be proven by the Eckart-Young-Mirsky theorem, as problem (\ref{eq:dms}) is equivalent with finding the best $d$ rank approximation in the Frobenius norm. In our problem, the standard kinetic energy product is used to preserve the nonlinear symmetries of the system dynamics:
\begin{equation}
\langle \mathbf{\zeta}_i, \mathbf{\zeta}_j  \rangle_{\mathbf{\zeta}} = \int_{\mathcal{S}} \nabla \psi_i \cdot \nabla \psi_j d \, \mathcal{S}
= - \int_{\mathcal{S}} \zeta_i  \psi_j d \, \mathcal{S} = - \int_{\mathcal{S}} \zeta_j  \psi_i d \, \mathcal{S},
\label{eq:kineticenergyproduct}
\end{equation}
where the last identities are derived using partial integration and the fact that $\mathbf{\zeta} = \Delta \mathbf{y}$.

\begin{figure}[!ht]
\centering
\includegraphics[width=1\textwidth]{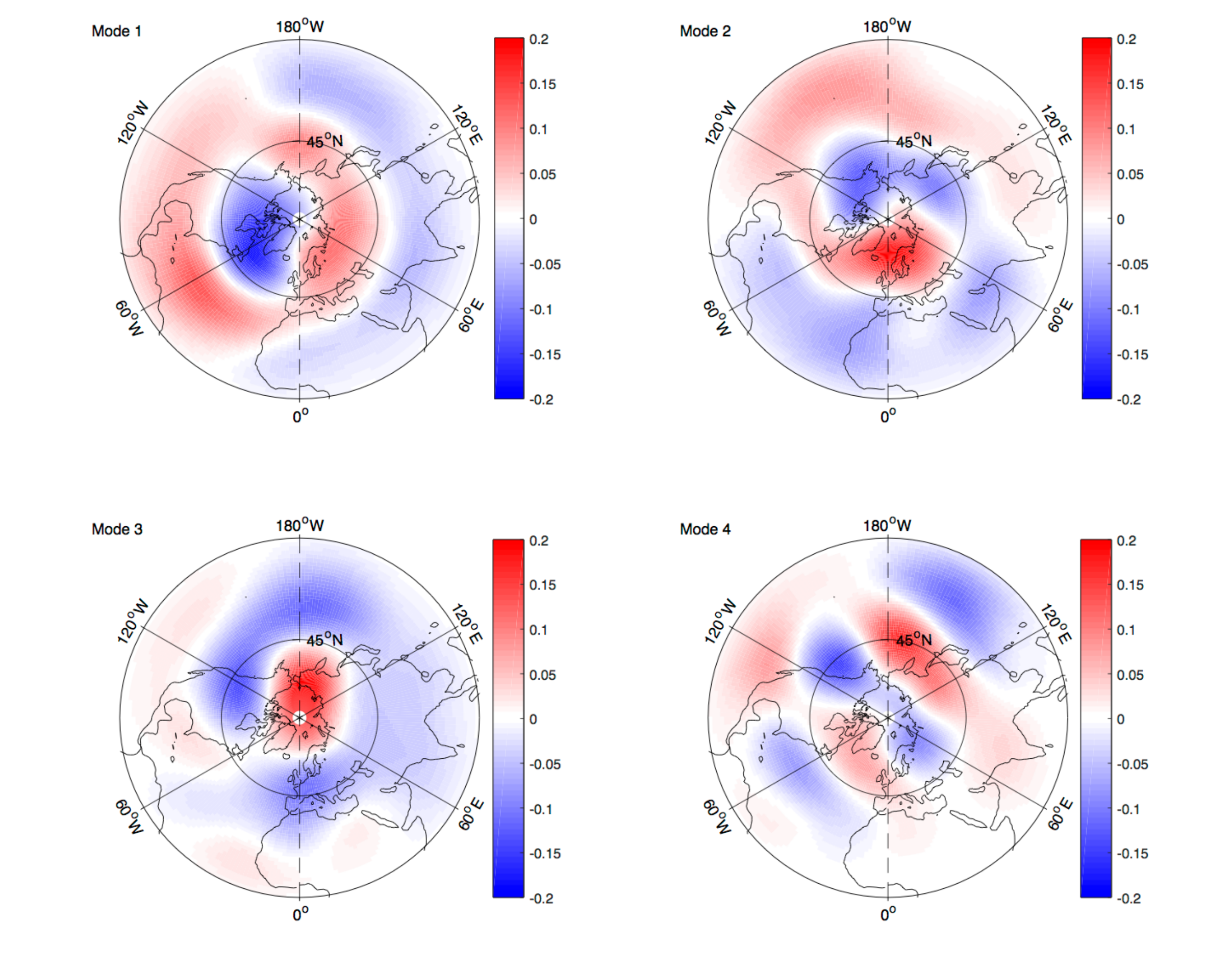}
\caption{The four most energetic empirical orthogonal functions of the barotropic model}
\label{fig:T21_EOFs}
\end{figure}

Solution (\ref{eq:yn}) is only optimal w.r.t. the $N$ training data points used to construct the Gram matrix. In order to calculate the embedding for a new point, it is convenient to compute the EOFs which form an orthogonal basis of the reduced order space $\mathbf{y}$. The EOFs are given by
\begin{equation}
\mathbf{\phi}_m = \sum_{n=1}^{N} k_m^{-1/2} \mathbf{u}_m^{n} \mathbf{\zeta}_n,
\label{eq:eofs}
\end{equation}
where $m$ runs from $1$ to $d$. The EOFs are sorted in descending order according to their energy level. The first four EOFs are plotted in Figure \ref{fig:T21_EOFs}. EOF analysis has been used to identify individual realistic climatic modes such as the Arctic Oscillation (AO) known as {\tt teleconnections}. The first EOF is characterized by a center of action over the Arctic that is surrounded by a zonal symmetric structure in mid-latitudes. This pattern resembles the Arctic Oscillation/Northern Hemisphere Annular Mode (AO/NAM) and explains approximately $13.5\%$ of the total energy. The second, third and fourth EOFs are quantitatively very similar to the East Atlantic/West Russia, the Pacific/North America (PNA) and the Tropical/Northern Hemisphere (TNH) patterns end account for $11.4\%$, $10.4\%$ and $7.1\%$ of the total energy respectively. Since these EOFs feature realistic climate teleconnections, performing accurate predictions of them is of high practical importance.

As a consequence of the orthogonality of the EOFs w.r.t. the kinetic energy product, the reduced representation $\mathbf{y}^{*}$ of a new state $\mathbf{\zeta}^{*}$ can be recovered from
\begin{equation}
\mathbf{y}^{*} = 
\begin{pmatrix}
\langle \mathbf{\zeta}^{*}, \mathbf{\phi}_1  \rangle_{\mathbf{\zeta}} \\
\langle \mathbf{\zeta}^{*}, \mathbf{\phi}_2  \rangle_{\mathbf{\zeta}} \\
\vdots \\
\langle \mathbf{\zeta}^{*}, \mathbf{\phi}_d  \rangle_{\mathbf{\zeta}} \\
\end{pmatrix}.
\label{eq:psistar}
\end{equation}
Note that only the $d$ coefficients corresponding to the most energetic EOFs form the reduced order state $\mathbf{y}^{*}$. In essence, the EOFs act as an orthogonal basis of the reduced order space and the state obtained from classical multidimensional scaling $\mathbf{\zeta}^{*}$ is projected to this basis.

\section{Sensitivity to noise in the data}

In this section, we evaluate the robustness of the proposed approach to noise. For this purpose, the training data are perturbed with different noise levels. We add Gaussian noise sampled from $N(0, \sigma_{noise})$ where the noise standard deviation is proportional to the attractor standard deviation $\sigma_{attractor}$ of each system, i.e. $\sigma_{noise}=k \, \sigma_{attractor}$. We note that $\sigma_{attractor}$ is computed from the training data, as the standard deviation of the samples of the reduced order state of the system. Different noise levels $k$ are considered.

\subsection{Lorenz 96}

\begin{figure}[!httb]
\centering
\begin{subfigure}{.31\textwidth}
\centering
\includegraphics[width=1\textwidth]{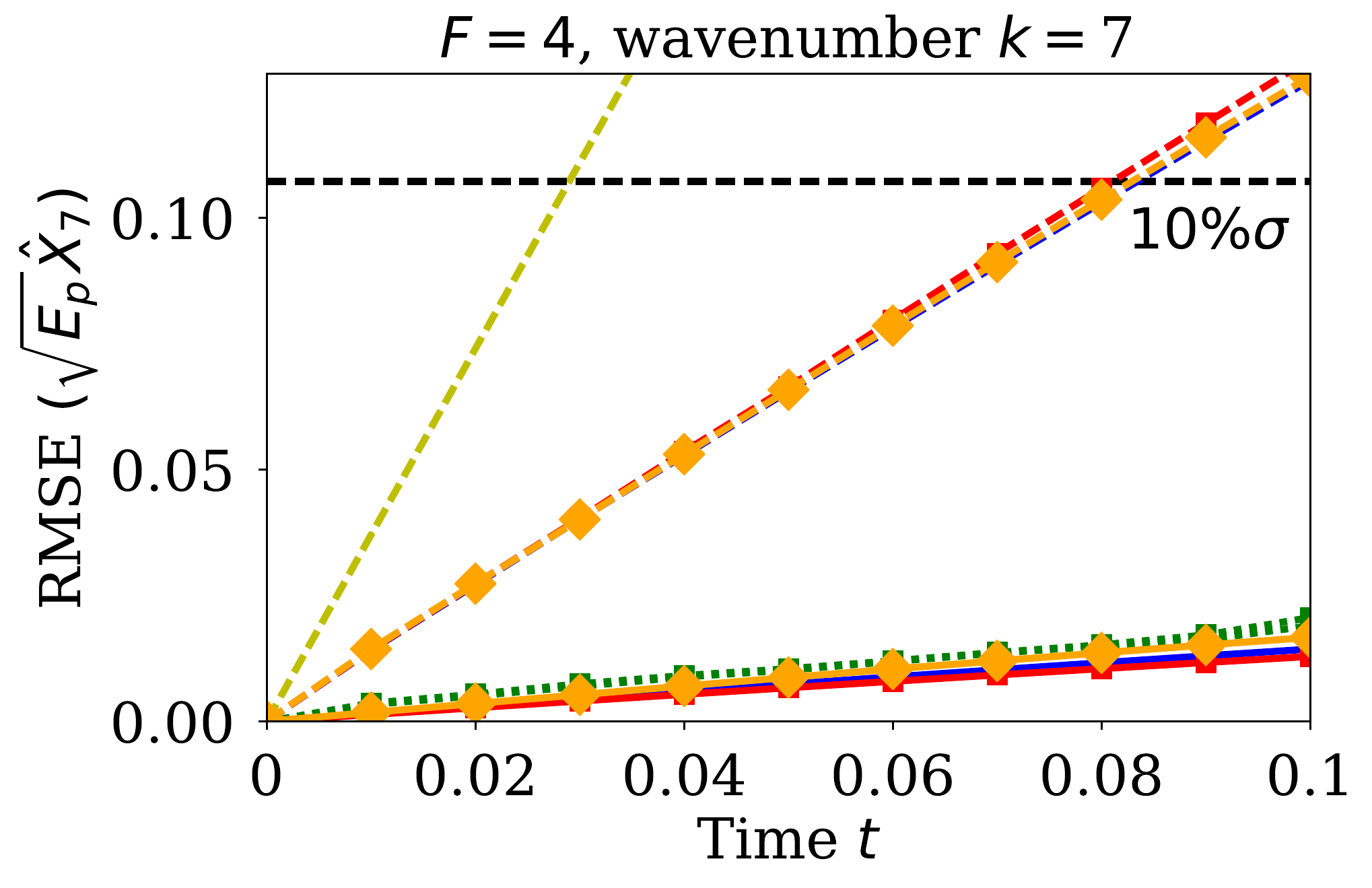}
\vspace{-0.75cm}
\caption{}
\label{results_lorenz_noise:a}
\end{subfigure} 
\hspace{0.3cm}
\begin{subfigure}{.31\textwidth}
\centering
\includegraphics[width=1\textwidth]{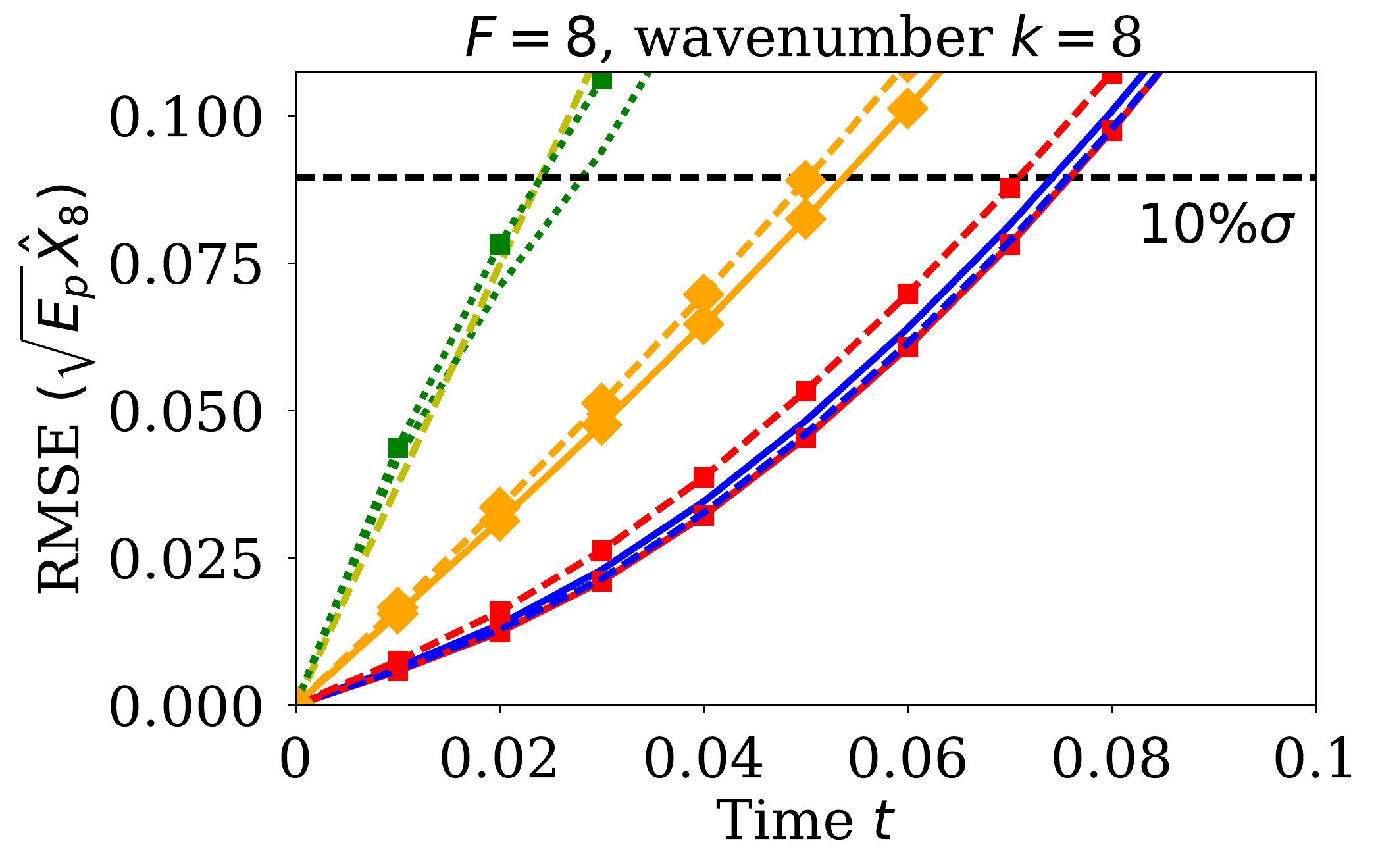}
\vspace{-0.75cm}
\caption{}
\label{results_lorenz_noise:b}
\end{subfigure}
\centering
\begin{subfigure}{.31\textwidth}
\centering
\includegraphics[width=1\textwidth]{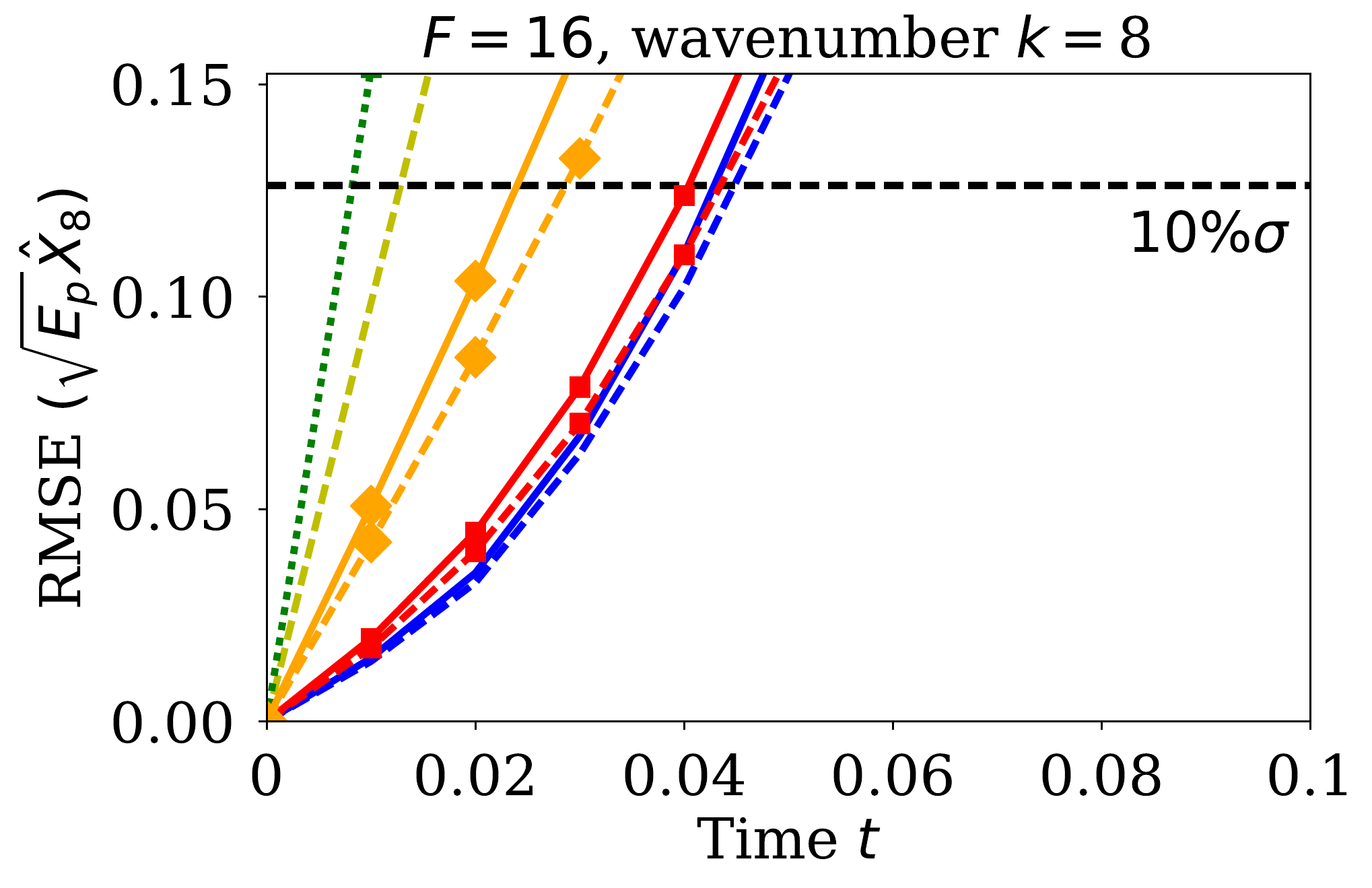}
\vspace{-0.75cm}
\caption{}
\label{results_lorenz_noise:c}
\end{subfigure}
\begin{subfigure}{.31\textwidth}
\centering
\includegraphics[width=1\textwidth]{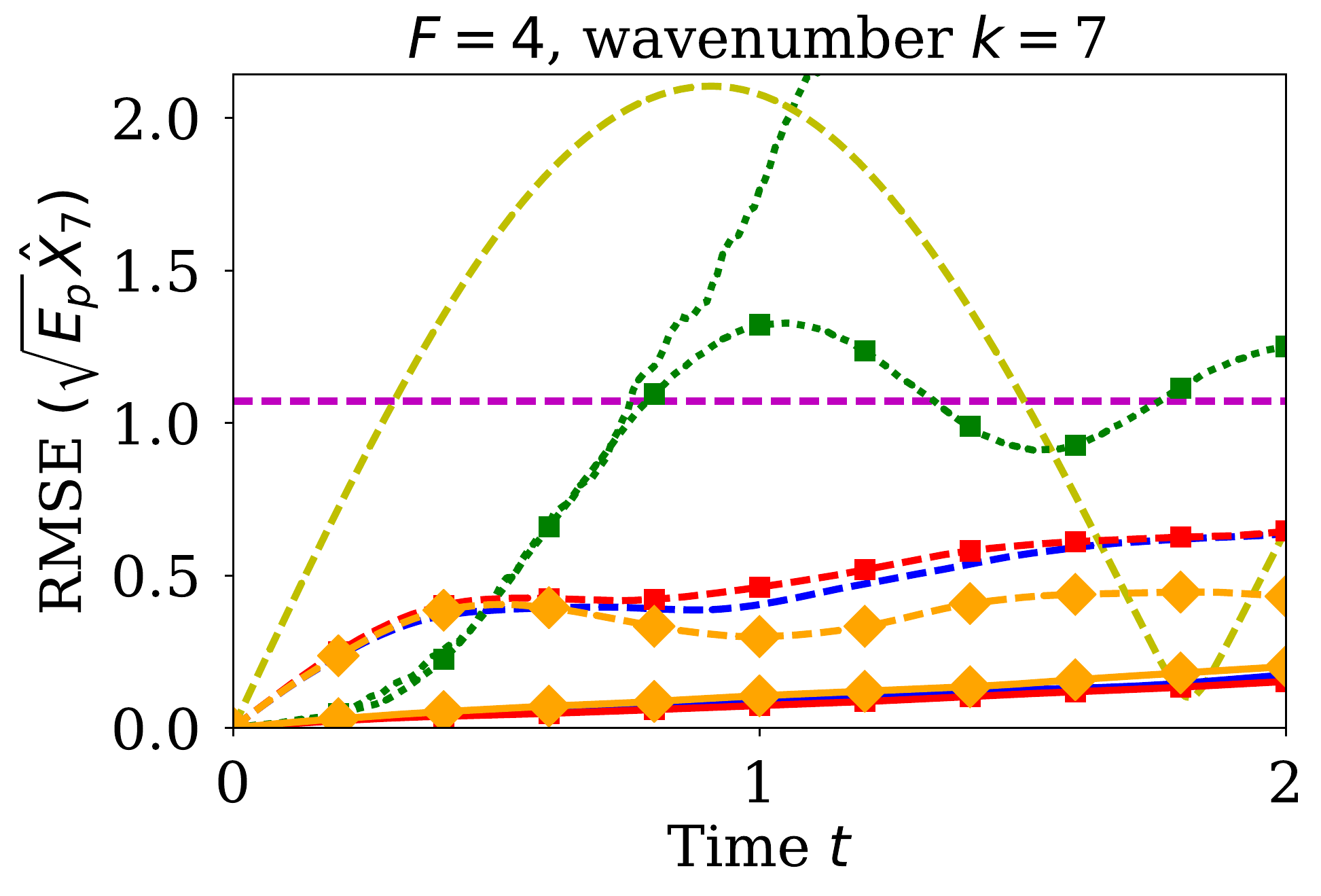}
\vspace{-0.75cm}
\caption{}
\label{results_lorenz_noise:d}
\end{subfigure} 
\hspace{0.3cm}
\begin{subfigure}{.31\textwidth}
\centering
\includegraphics[width=1\textwidth]{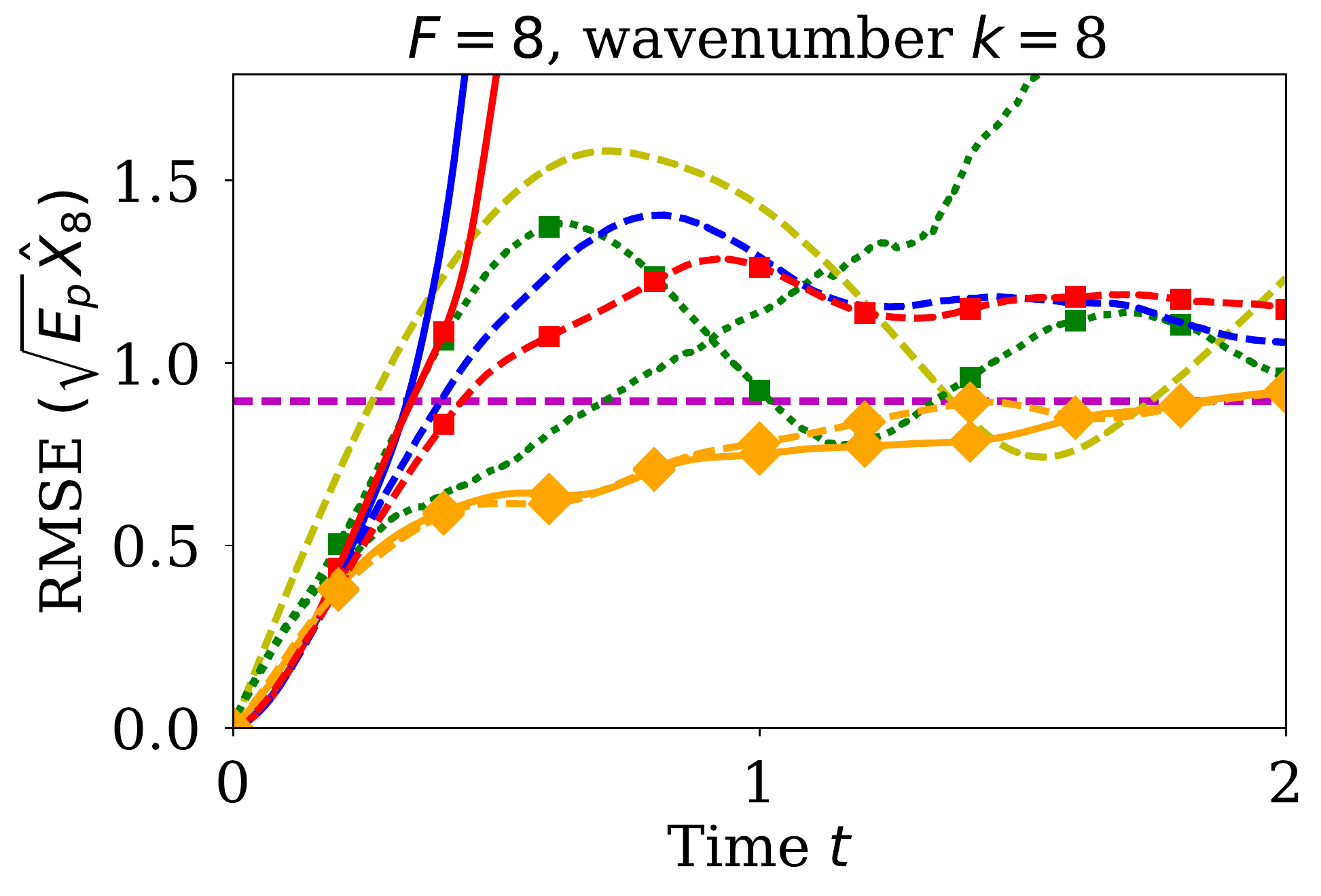}
\vspace{-0.75cm}
\caption{}
\label{results_lorenz_noise:e}
\end{subfigure}
\centering
\begin{subfigure}{.31\textwidth}
\centering
\includegraphics[width=1\textwidth]{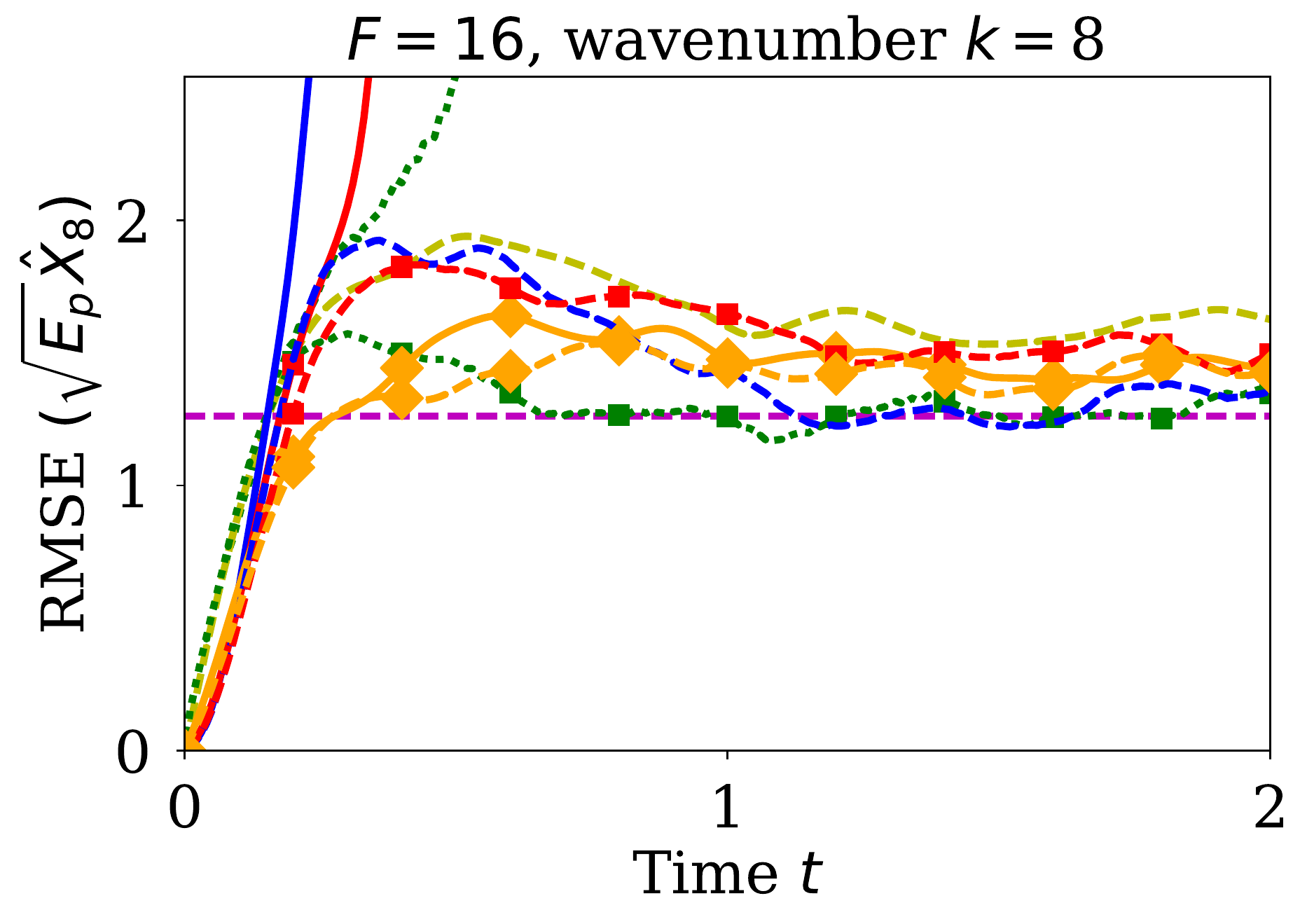}
\vspace{-0.75cm}
\caption{}
\label{results_lorenz_noise:f}
\end{subfigure}
\begin{subfigure}{.31\textwidth}
\centering
\includegraphics[width=1\textwidth]{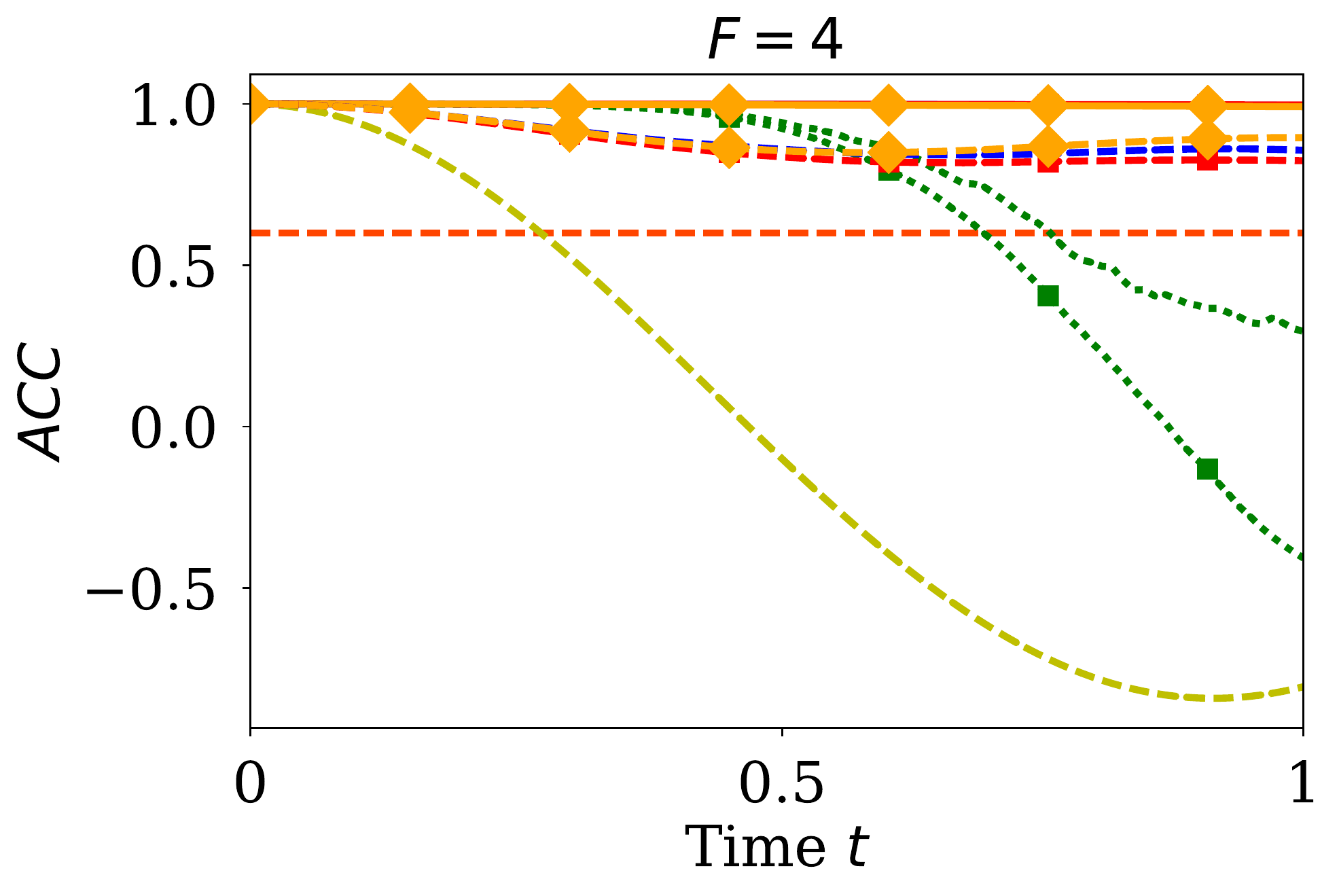}
\vspace{-0.75cm}
\caption{}
\label{results_lorenz_noise:g}
\end{subfigure} 
\hspace{0.3cm}
\begin{subfigure}{.31\textwidth}
\centering
\includegraphics[width=1\textwidth]{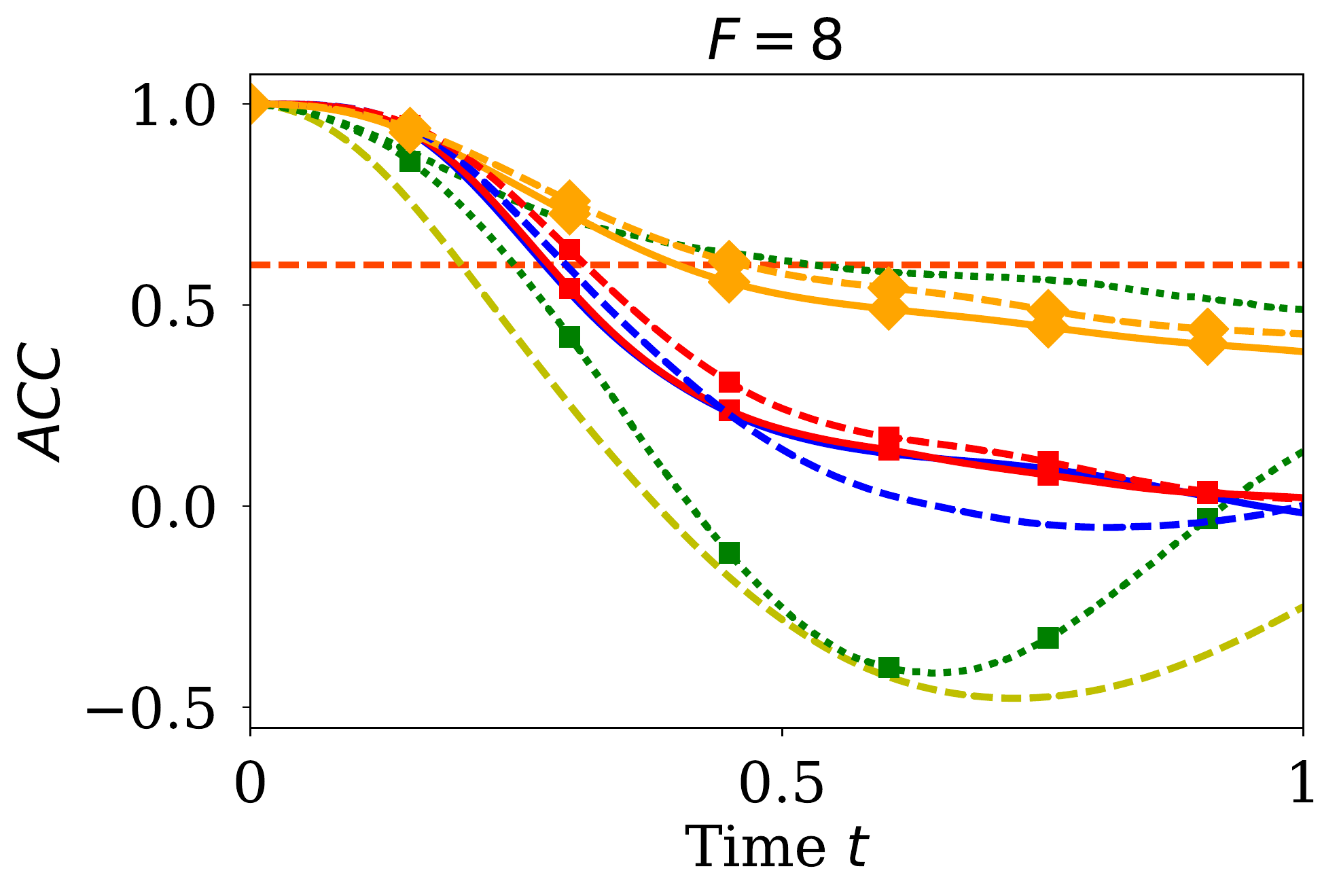}
\vspace{-0.75cm}
\caption{}
\label{results_lorenz_noise:h}
\end{subfigure}
\centering
\begin{subfigure}{.31\textwidth}
\centering
\includegraphics[width=1\textwidth]{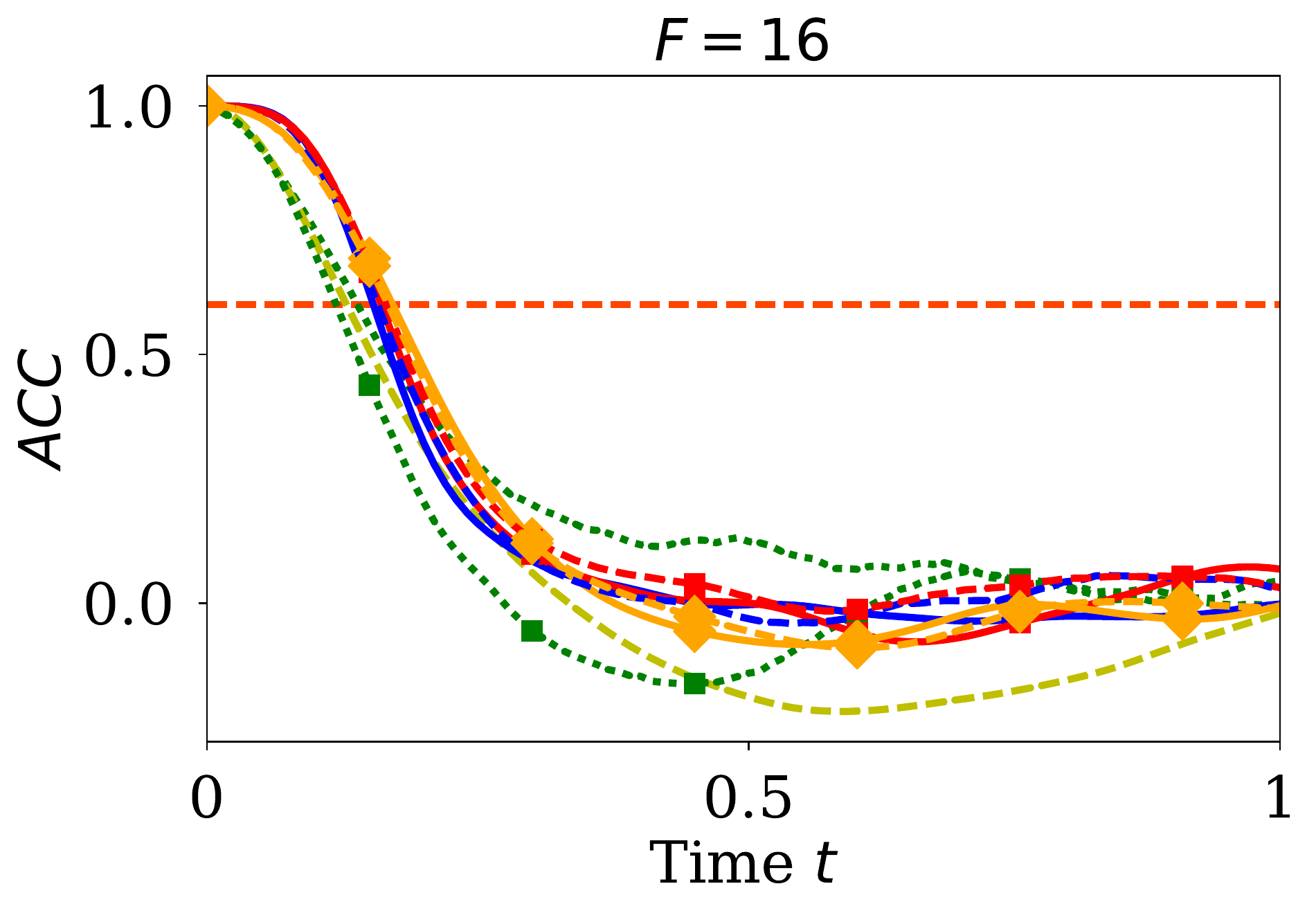}
\vspace{-0.75cm}
\caption{}
\label{results_lorenz_noise:i}
\end{subfigure}
\caption{\textbf{(a), (b), (c)} Short term RMSE evolution of the most energetic mode for forcing regimes $F=4,8,16$ respectively of the Lorenz 96 system. \textbf{(d), (e), (f)} Long term RMSE evolution. \textbf{(g), (h), (i)} Evolution of the ACC coefficient.  (In all plots average over $1000$ initial conditions is reported). \\
$10\% \, \sigma_{attractor}$  \protect \blacklineDashed; 
$\sigma_{attractor}$ \protect\magentalineDashed; 
$ACC=0.6$ threshold \protect\redlineDashed; 
MSM\protect\yellowlineDashed; 
GPR\protect \greenlineDotted;   \\
GPR-MSM\protect\greenlineDottedRectangle; 
LSTM $k=0\permil$\protect\blueline; 
LSTM-MSM $k=0\permil$\protect\bluelineDashed;  \\
LSTM $k=10\permil$\protect\redlineRectangle; 
LSTM-MSM $k=10\permil$\protect\redlineDashedRectangle;
LSTM $k=200\permil$\protect\orangelineDiamond;  \\
LSTM-MSM $k=200\permil$\protect\orangelineDashedDiamond
}
\label{fig:results_lorenz_noise}
\end{figure}

\begin{figure}[!httb]
\centering
\begin{subfigure}{.4\textwidth}
\centering
\includegraphics[width=1\textwidth]{./Lorenz96/RMSE_ST_8_mode_0_Noise.pdf}
\vspace{-0.75cm}
\caption{}
\label{results_lorenz_F8_Noise:a}
\end{subfigure} 
\hspace{0.3cm}
\begin{subfigure}{.4\textwidth}
\centering
\includegraphics[width=1\textwidth]{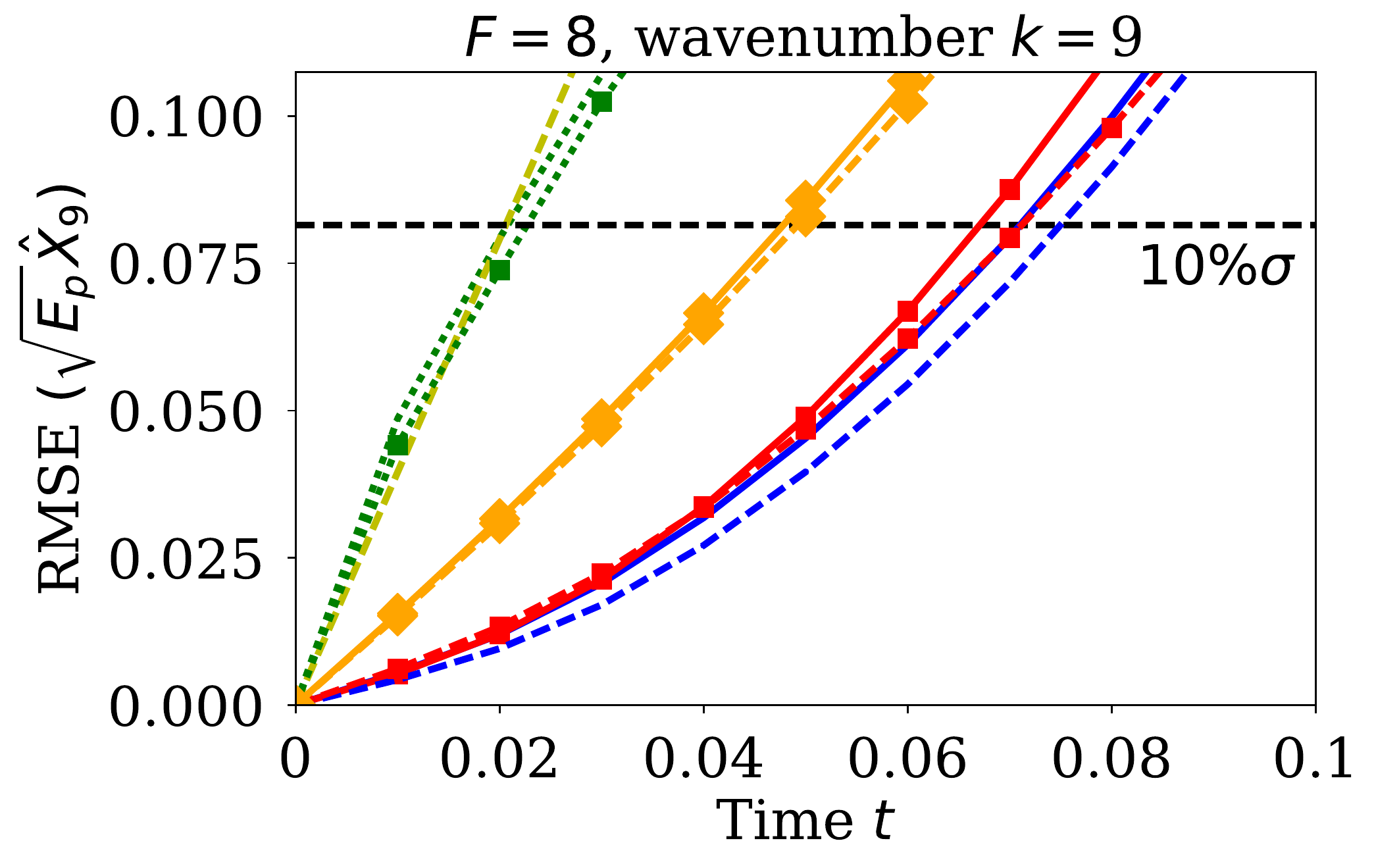}
\vspace{-0.75cm}
\caption{}
\label{results_lorenz_F8_Noise:b}
\end{subfigure}
\centering
\begin{subfigure}{.4\textwidth}
\centering
\includegraphics[width=1\textwidth]{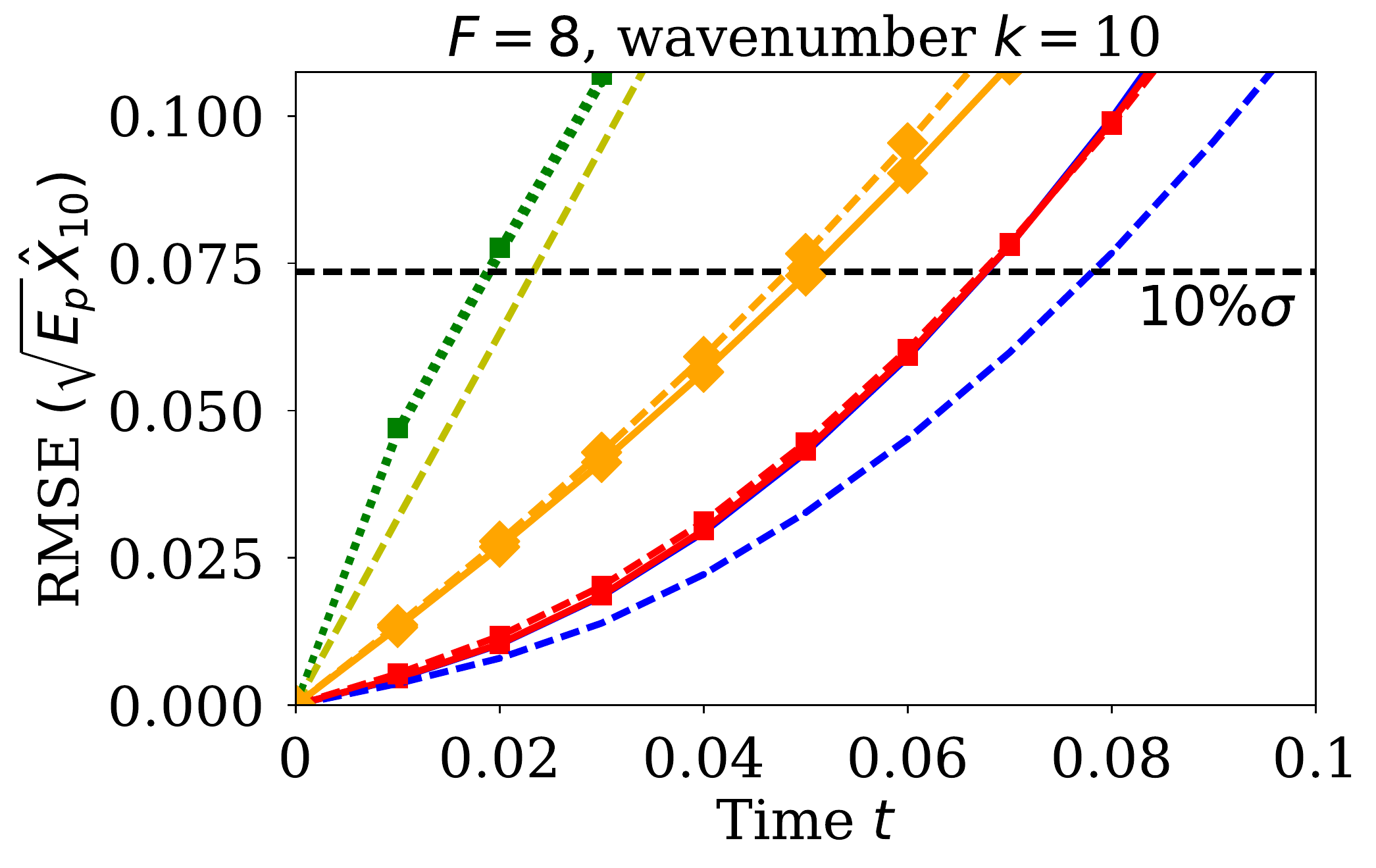}
\vspace{-0.75cm}
\caption{}
\label{results_lorenz_F8_Noise:c}
\end{subfigure} 
\hspace{0.3cm}
\begin{subfigure}{.4\textwidth}
\centering
\includegraphics[width=1\textwidth]{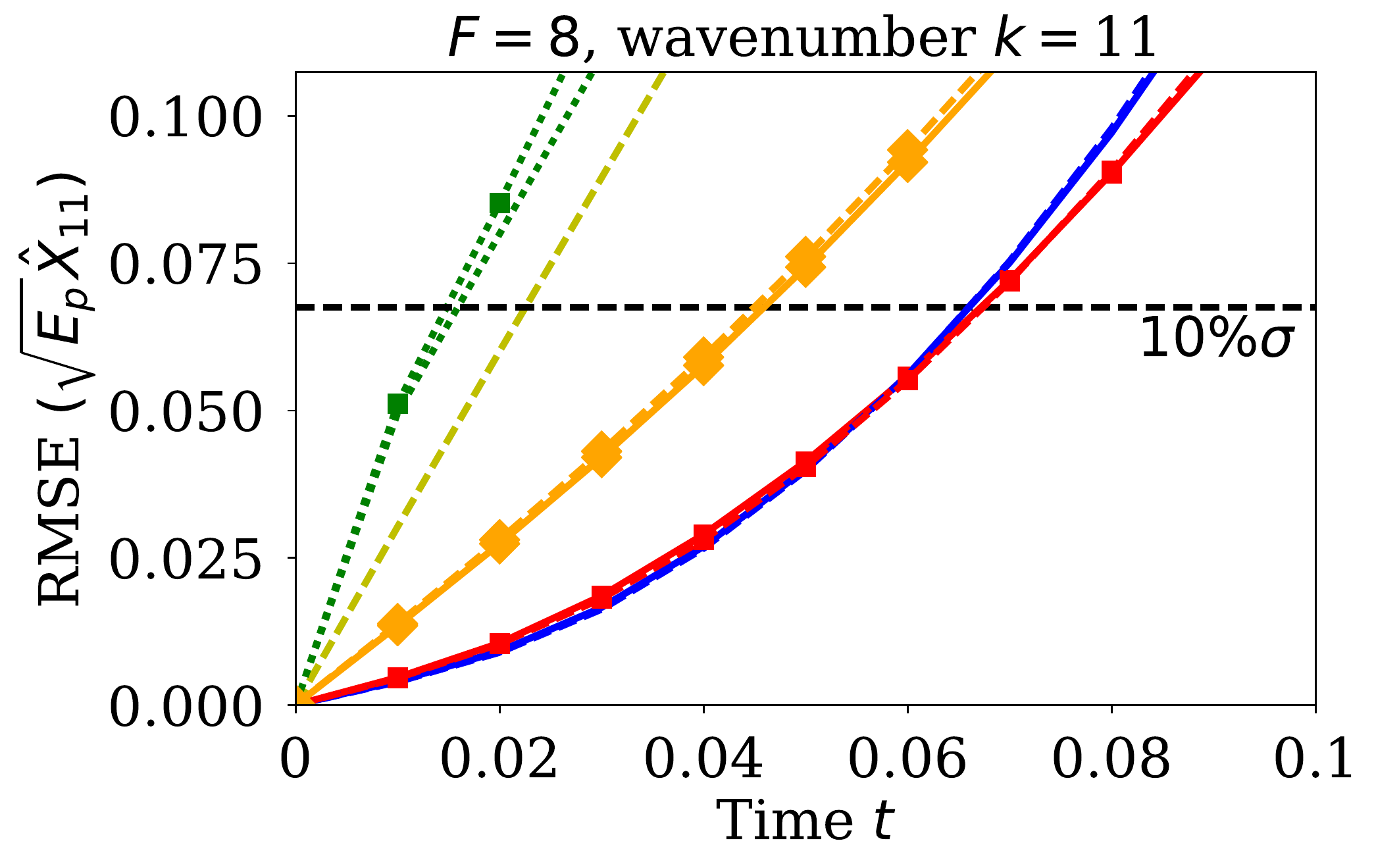}
\vspace{-0.75cm}
\caption{}
\label{results_lorenz_F8_Noise:d}
\end{subfigure}
\caption{RMSE prediction error evolution of four energetic modes for the Lorenz 96 system with forcing $F=8$. \textbf{(a)} Most energetic mode $k=8$. \textbf{(b)} Low energy mode $k=9$. \textbf{(c)} Low energy mode $k=10$. \textbf{(d)} Low energy mode $k=11$. (In all plots average over $1000$ initial conditions reported)\\
 $10\% \, \sigma_{attractor}$  \protect \blacklineDashed; 
 MSM\protect\yellowlineDashed; 
GPR\protect \greenlineDotted; 
GPR-MSM\protect\greenlineDottedRectangle; 
LSTM $k=0\permil$\protect\blueline; 
LSTM-MSM $k=0\permil$\protect\bluelineDashed;
LSTM $k=10\permil$\protect\redlineRectangle; 
LSTM-MSM $k=10\permil$\protect\redlineDashedRectangle;
LSTM $k=200\permil$\protect\orangelineDiamond; 
LSTM-MSM $k=200\permil$\protect\orangelineDashedDiamond
 }
\label{fig:results_lorenz_F8_Noise}
\end{figure}

In the following, we analyze the influence of noise in the training data for the Lorenz 96 system. In parallel with the main body of the paper, we plot the RMSE error evolution of the most energetic mode (first row of Figure \ref{fig:results_lorenz_noise}) for short term till $T=0.1$, the same for time $T=2$ (second row of Figure \ref{fig:results_lorenz_noise}) and the ACC (third row of Figure \ref{fig:results_lorenz_noise}). The columns of Figure \ref{fig:results_lorenz_noise} correspond to different chaotic regimes of the Lorenz 96 system. For the forcing $F=4$ and noise levels $k \in \{0.01, 0.2\}$, noise does not affect the prediction performance of the LSTM. This can be attributed to the fact that the attractor dimensionality is really low in this case and the amount of data is enough to capture the dynamics despite the noisy training data. However, for $F=8$ and $F=16$ adding noise leads to slight deterioration of the short term prediction accuracy for the noise level $k=0.01$, as illustrated by the last two Figures in the first row of Figure \ref{fig:results_lorenz_noise}. As a consequence, the method can be considered robust against noise. Increasing the noise level to $k=0.2$ corresponding to a noise standard deviation equal to $20 \%$ of the attractor standard deviation leads to an important deterioration in short term prediction performance. The deterioration in the short term prediction performance can be seen in the short term RMSE error evolution of the fourth most energetic modes for the forcing regime $F=8$ plotted in Figure \ref{fig:results_lorenz_F8_Noise}.

\subsection{Kuramoto-Sivashinsky equation}

\begin{figure}[!httb]
\centering
\begin{subfigure}{.45\textwidth}
\centering
\includegraphics[width=1\textwidth]{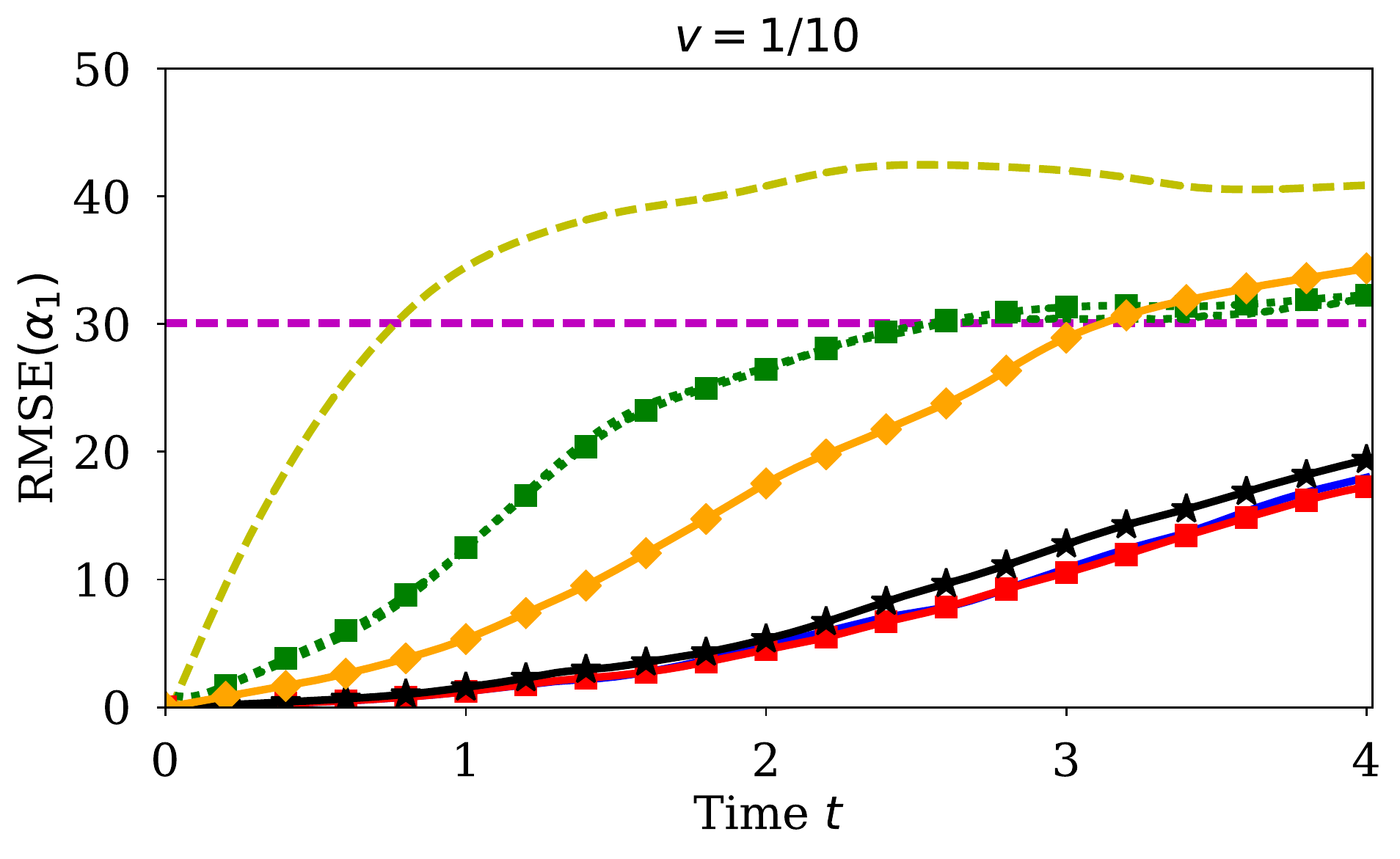}
\vspace{-0.75cm}
\caption{}
\label{results_kuramoto_noise:a}
\end{subfigure} 
\hspace{0.3cm}
\begin{subfigure}{.45\textwidth}
\centering
\includegraphics[width=1\textwidth]{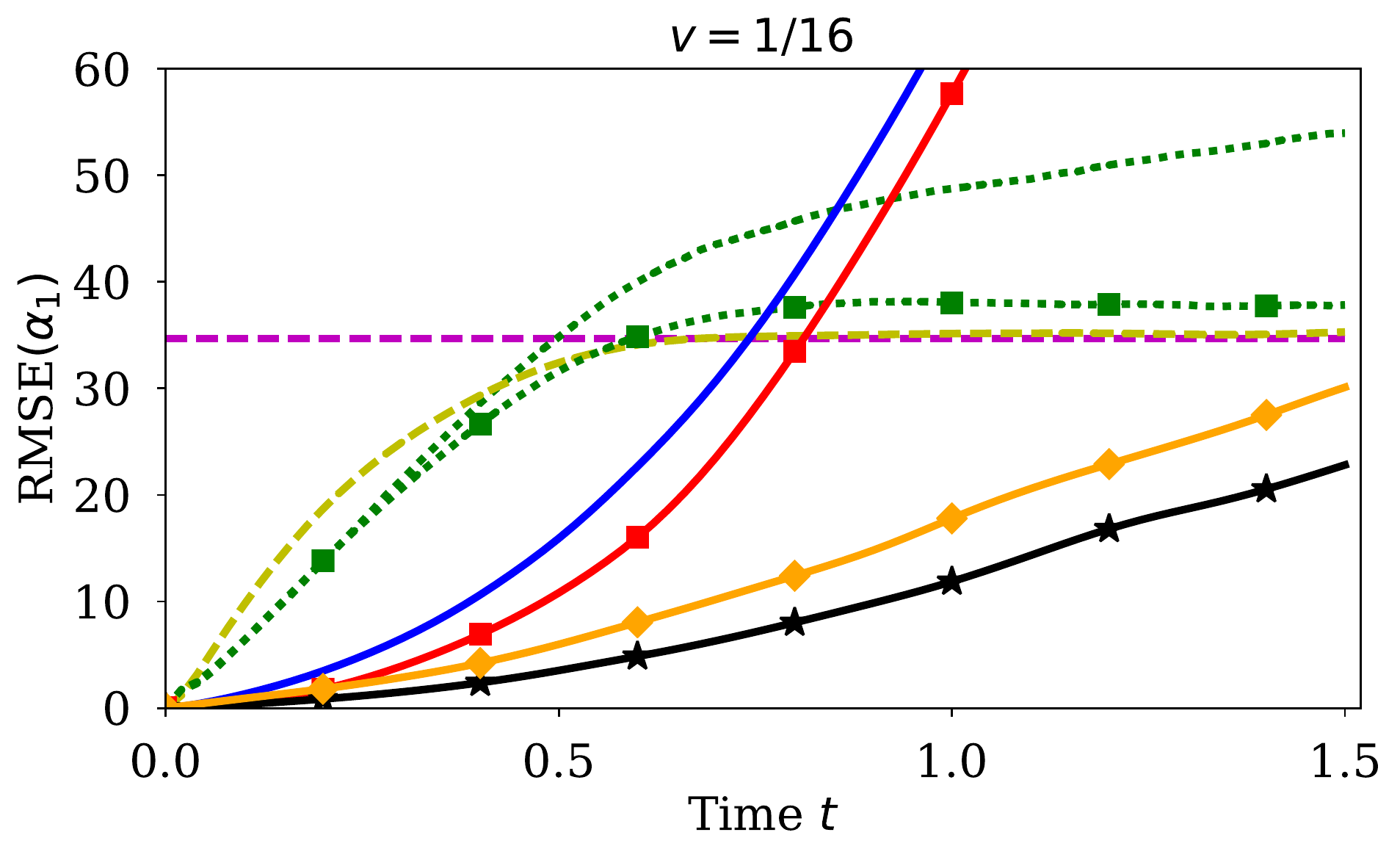}
\vspace{-0.75cm}
\caption{}
\label{results_kuramoto_noise:b}
\end{subfigure}
\centering
\begin{subfigure}{.45\textwidth}
\centering
\includegraphics[width=1\textwidth]{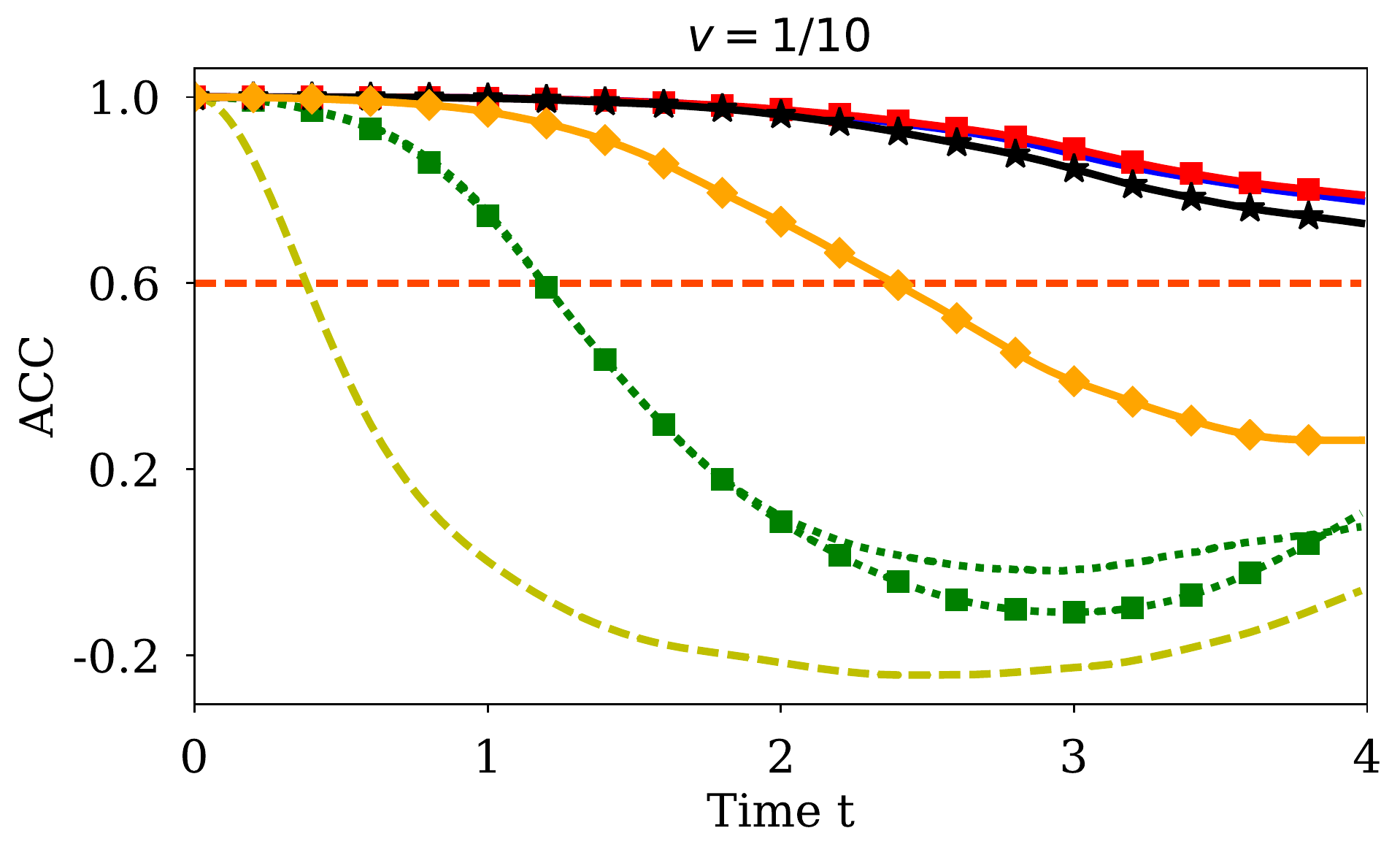}
\vspace{-0.75cm}
\caption{}
\label{results_kuramoto_noise:c}
\end{subfigure} 
\hspace{0.3cm}
\begin{subfigure}{.45\textwidth}
\centering
\includegraphics[width=1\textwidth]{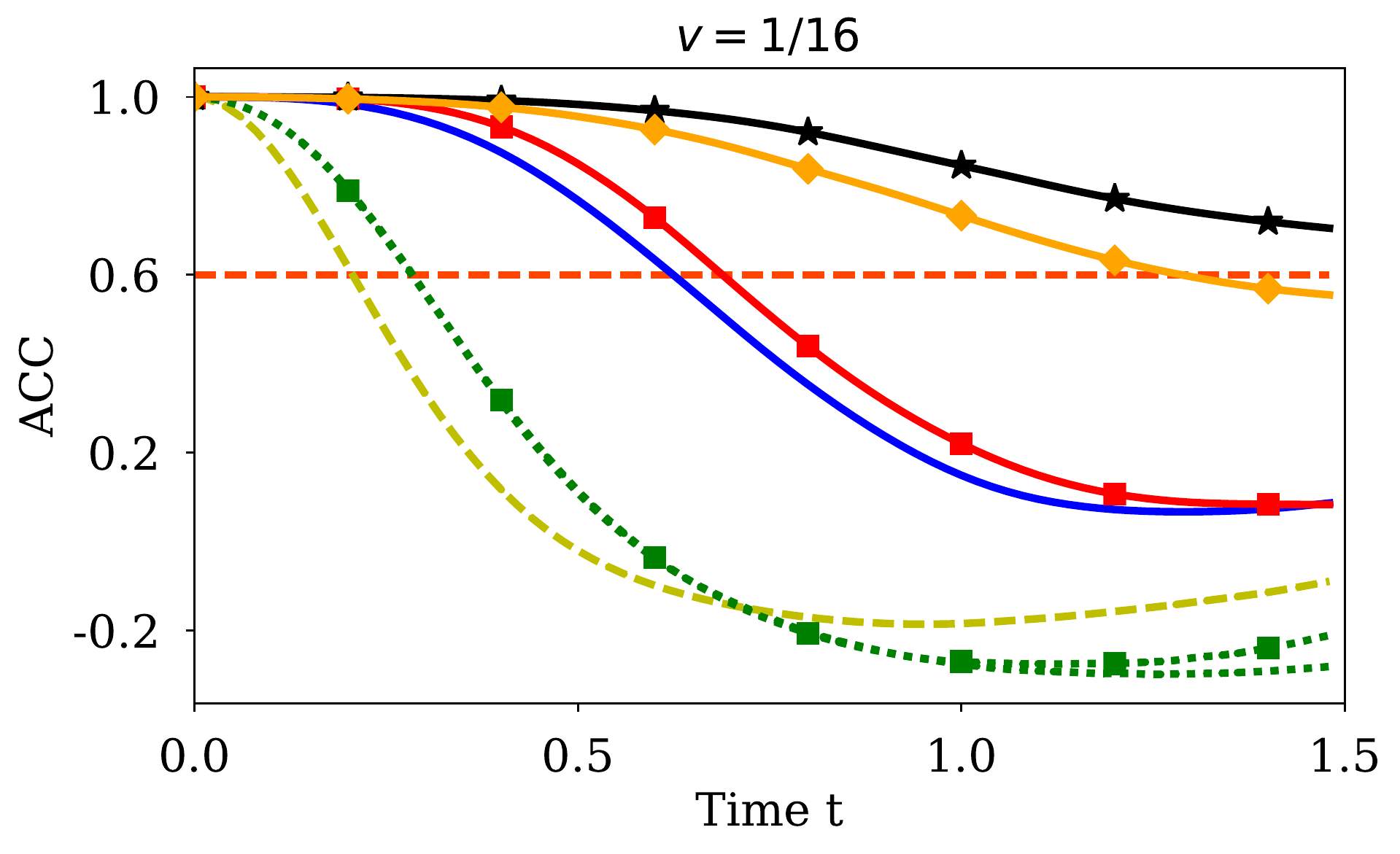}
\vspace{-0.75cm}
\caption{}
\label{results_kuramoto_noise:d}
\end{subfigure}
\caption{Training data of LSTM are perturbed with standard deviation $\sigma_{noise}=k \, \sigma_{attractor}$. Three different noise levels $k \in \{0.001, 0.01, 0.2\}$ are considered.
\textbf{(a)}, \textbf{(b)}  RMSE evolution of the most energetic mode of the K-S equation with $1/\nu=10$ and $1/\nu=16$.
\textbf{(c)}, \textbf{(d)}  ACC evolution of the most energetic mode of the K-S equation with $1/\nu=10$ and $1/\nu=16$.
(In all plots, average value over $1000$ initial conditions is reported) \\
$\sigma_{attractor}$\protect\magentalineDashed; $ACC=0.6$ threshold\protect\redlineDashed; MSM\protect\yellowlineDashed; GPR\protect\greenlineDotted; GPR-MSM\protect\greenlineDottedRectangle; 
LSTM $k=0 \permil$\protect\blueline; 
LSTM $k=1 \permil$ \protect \redlineRectangle;
LSTM $k=10 \permil$ \protect \blacklineStar;
LSTM $k=200 \permil$ \protect \orangelineDiamond
}
\label{fig:results_kuramoto_noise}
\end{figure}

In Figure \ref{fig:results_kuramoto_noise} we plot the RMSE error evolution for the most energetic mode and the ACC of the Kuramoto-Sivashinsky equation for two different chaotic regimes $1/\nu \in \{10,16 \}$. Three different noise levels $k \in \{0.001, 0.01, 0.2\}$ are considered. For the low chaotic regime $1/\nu=10$, predictability performance is robust against noise, as the error evolution changes slightly with $k \in \{0.001, 0.01 \}$. Only when the training data are polluted with noise with a standard deviation bigger than $20 \%$ of the attractor standard deviation is the predictability performance greatly deteriorated. On the contrary, adding noise to the training data in the input improves the predictability performance of LSTM for the chaotic regime $\nu = 1/16$. This can be attributed to the fact that in this chaotic regime correlation patterns are much less prominent, and the LSTM is more prone to overfit. As a consequence, adding noise to the input forces the neural network to learn only robust patterns in the data that can be generalized. Short term prediction performance is deteriorated slightly, but in the long term, the LSTM is more robust against accumulation of errors. This behavior has to be further investigated in future work.

\subsection{Barotropic model}

\begin{figure}[!httb]
\centering
\begin{subfigure}{.46\textwidth}
\centering
\includegraphics[width=1\textwidth]{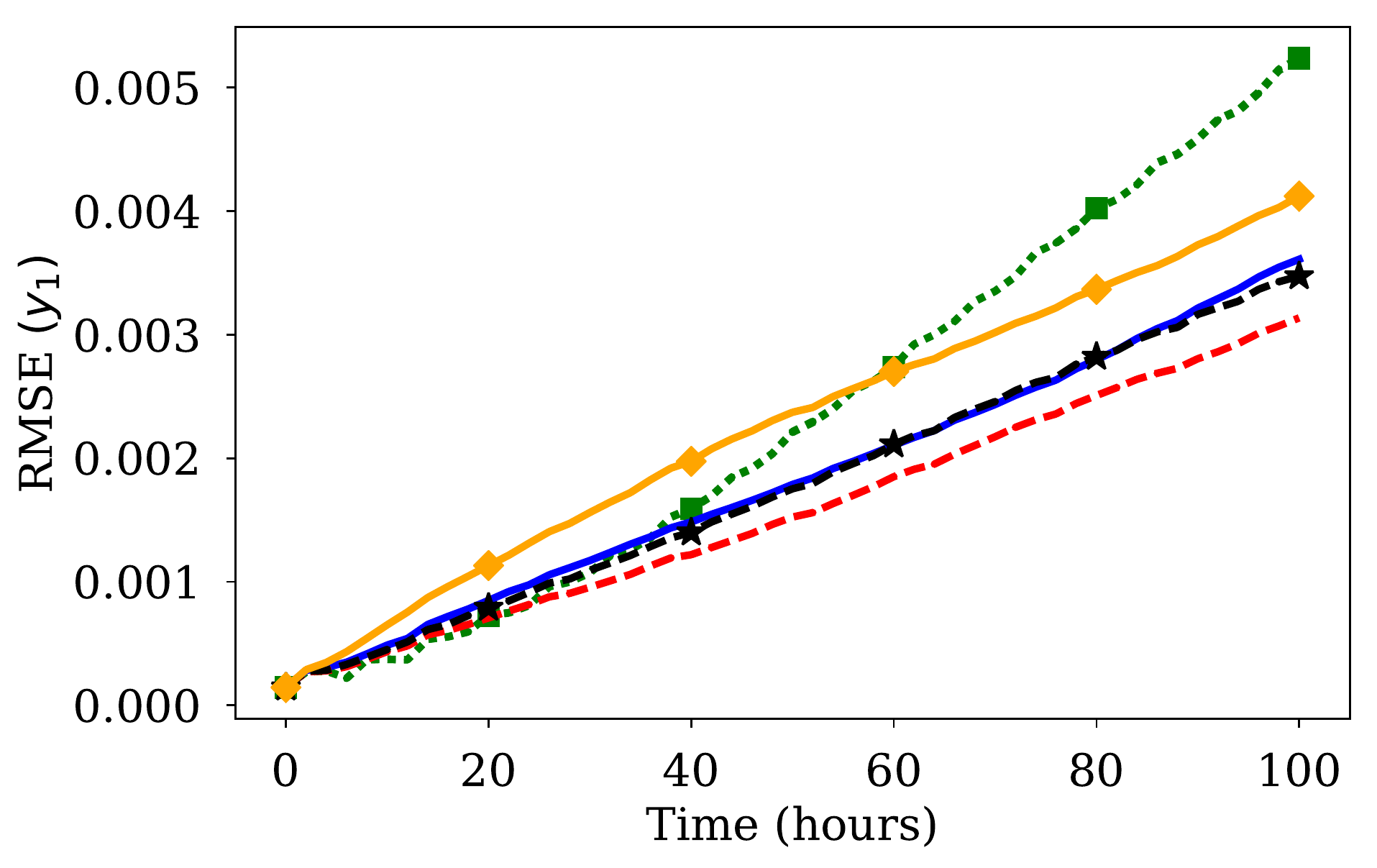}
\vspace{-0.75cm}
\caption{}
\label{barotropic_results_noise:a}
\end{subfigure} 
\hspace{0.2cm}
\begin{subfigure}{.46\textwidth}
\centering
\includegraphics[width=1\textwidth]{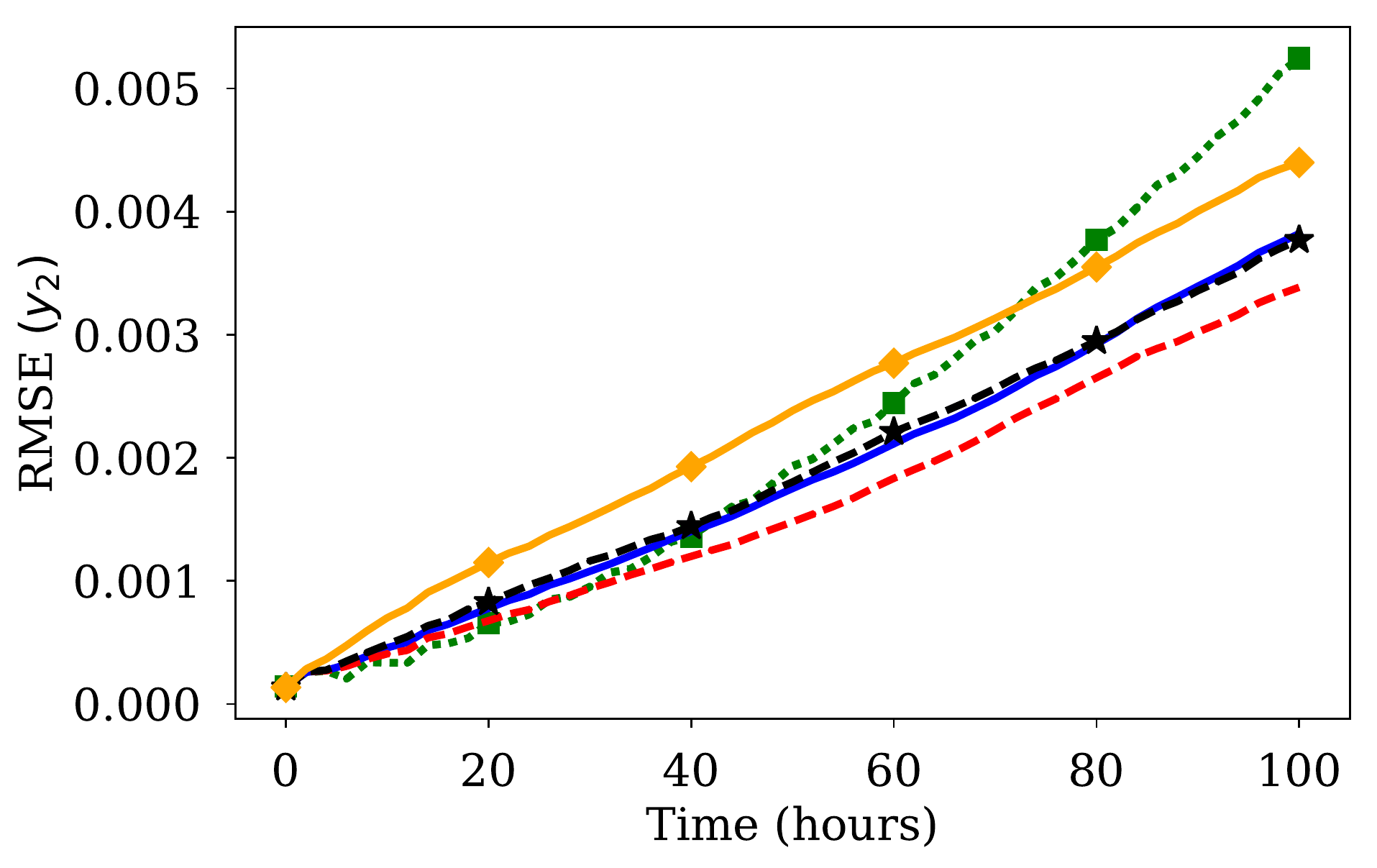}
\vspace{-0.75cm}
\caption{}
\label{barotropic_results_noise:b}
\end{subfigure}
\centering
\begin{subfigure}{.46\textwidth}
\centering
\includegraphics[width=1\textwidth]{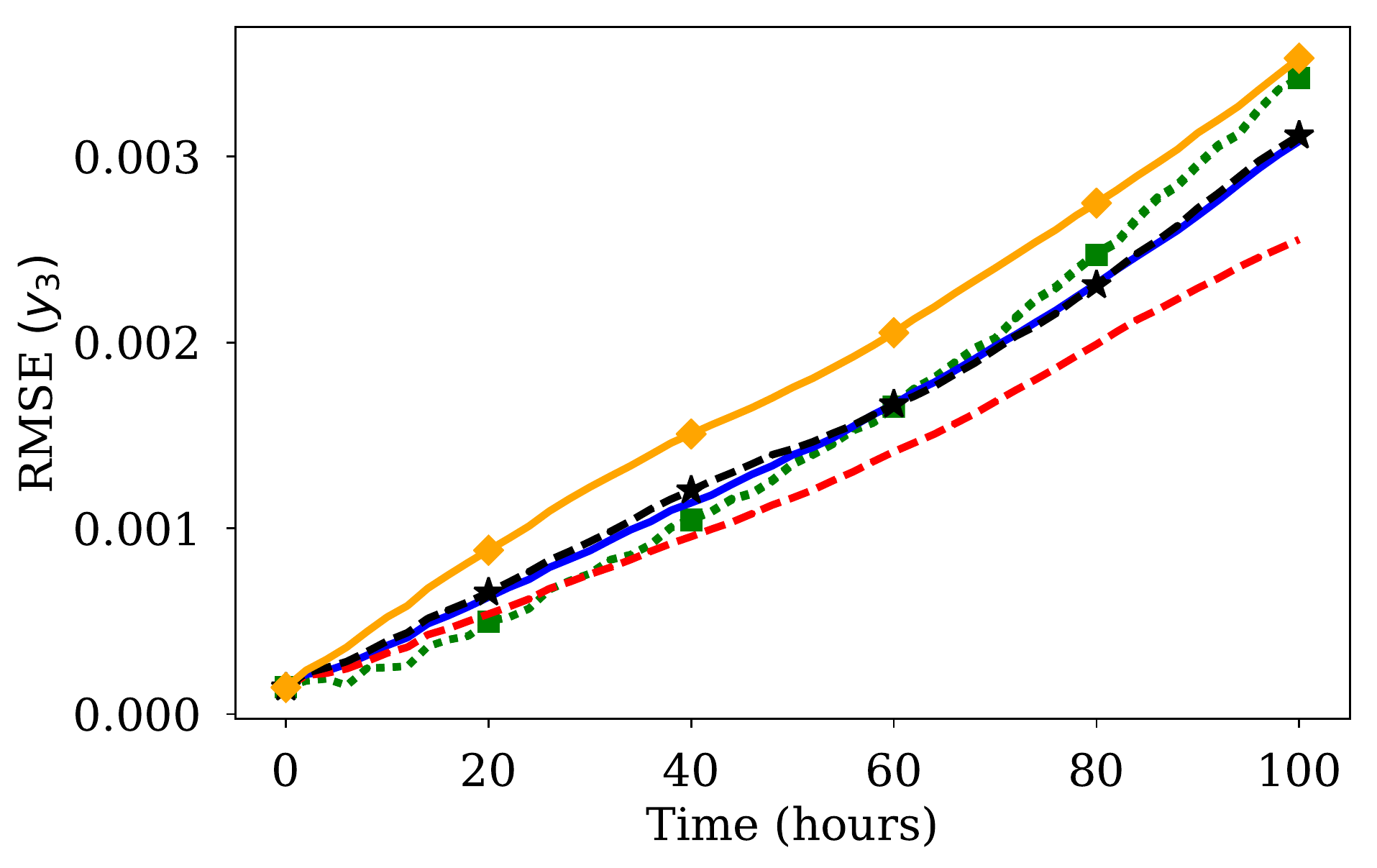}
\vspace{-0.75cm}
\caption{}
\label{barotropic_results_noise:c}
\end{subfigure} 
\hspace{0.2cm}
\begin{subfigure}{.46\textwidth}
\centering
\includegraphics[width=1\textwidth]{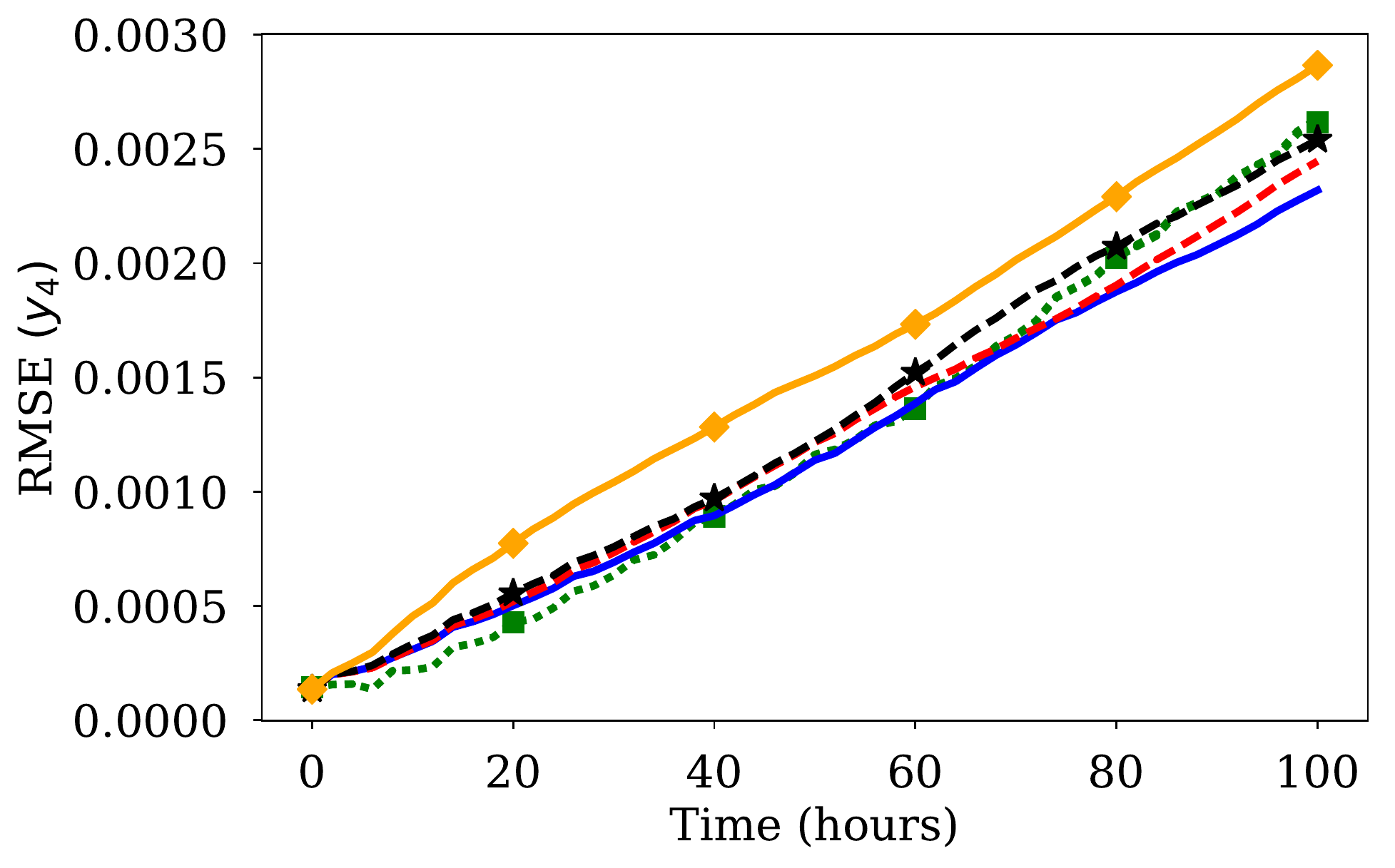}
\vspace{-0.75cm}
\caption{}
\label{barotropic_results_noise:d}
\end{subfigure}
\caption{RMSE evolution of the four most energetic EOFs for the Barotropic climate model, average over $500$ initial conditions reported. Training data are perturbed with Gaussian noise with standard deviation $\sigma_{noise}=k \, \sigma_{attractor}$. LSTM results for different noise levels $k$ are plotted.
\textbf{(a)} Most energetic EOF.
\textbf{(b)} Second most energetic EOF.
\textbf{(c)} Third most energetic EOF.
\textbf{(d)} Fourth most energetic EOF.
\\ 
GPR \protect\greenlineDotted; GPR-MSM \protect\greenlineDottedRectangle; LSTM $k=0 \permil$ \protect\blueline;
LSTM $k=1 \permil$ \protect \redlineDashed;
LSTM $k=10 \permil$ \protect \blacklineDashedStar;
LSTM $k=200 \permil$ \protect \orangelineDiamond;
}
\label{fig:barotropic_results_noise}
\end{figure}

In Figure \ref{fig:barotropic_results_noise} we plot the RMSE error evolution for the four most energetic EOFs of the Barotropic model. Three different noise levels $k \in \{0.001, 0.01, 0.2\}$ are considered. Only for the highest noise level is the prediction performance deteriorated. For low noise levels, the prediction performance can be increased ($k=0.001$), as the noise may regularize the Back-propagation procedure during training with stochastic methods. Adding noise to the input of neural networks can be used as a practical heuristic to increase their accuracy and can also be seen as a form of dropout in the input layer of the LSTM. The results indicate that the prediction performance of the LSTM is robust for the noise levels $k \in \{0.001, 0.01\}$. 

\clearpage

\section{Trajectory examples}

In this section we present examples of predicted trajectories of the reduced order state and compare them with the ground-truth trajectory. Moreover, we also compare the equivalent trajectories in the original space by replacing the unmodeled PCA modes with zero and projecting back to the original space. As a reference system, we pick the Kuramoto-Sivashinsky (KS) equation.

\begin{figure}[!httb]
\centering
\begin{subfigure}{.3\textwidth}
\centering
\includegraphics[width=1\textwidth]{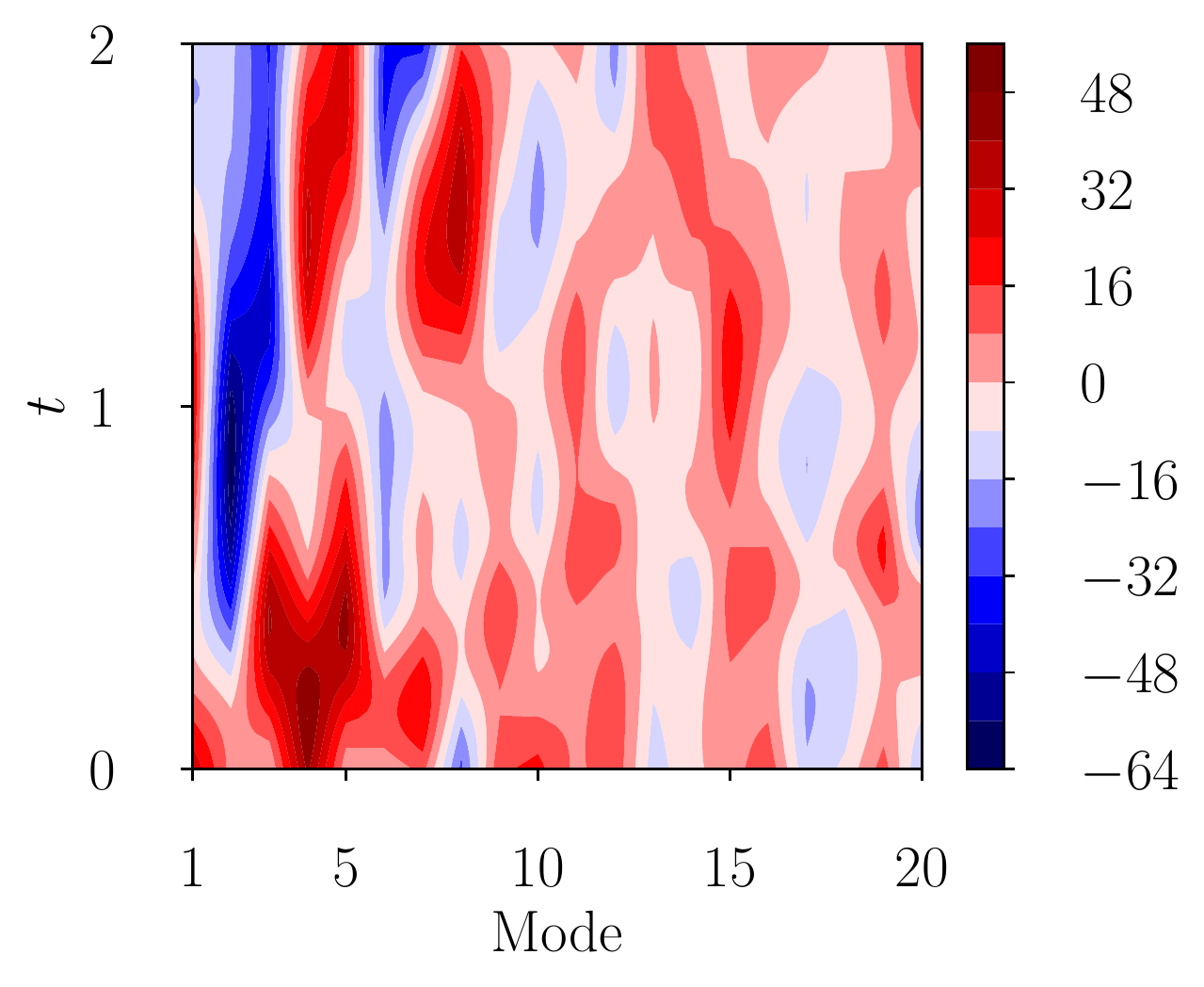}
\vspace{-0.75cm}
\caption{}
\label{ks_example_traj:a}
\end{subfigure} 
\begin{subfigure}{.3\textwidth}
\centering
\includegraphics[width=1\textwidth]{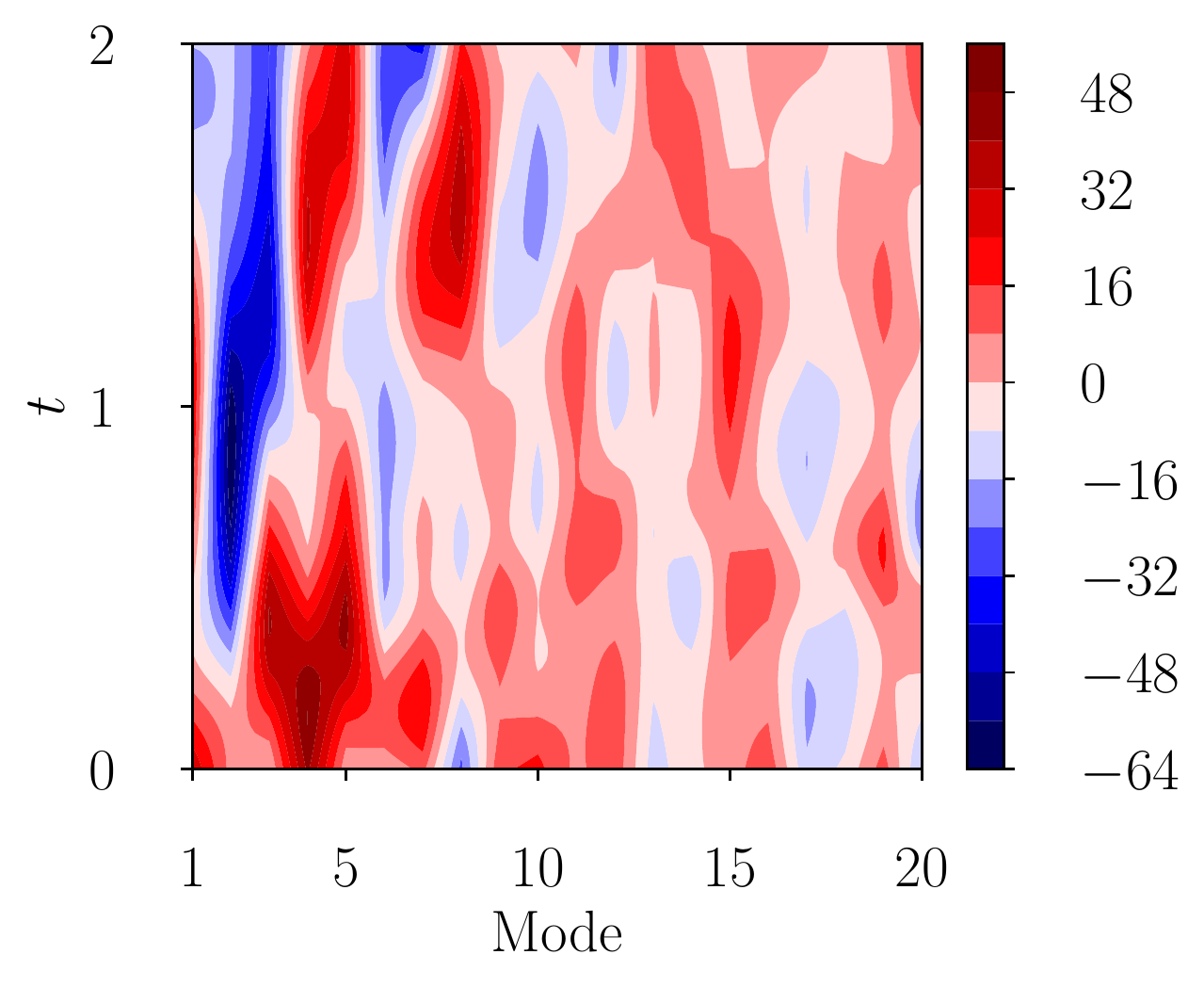}
\vspace{-0.75cm}
\caption{}
\label{ks_example_traj:b}
\end{subfigure}
\centering
\begin{subfigure}{.3\textwidth}
\centering
\includegraphics[width=1\textwidth]{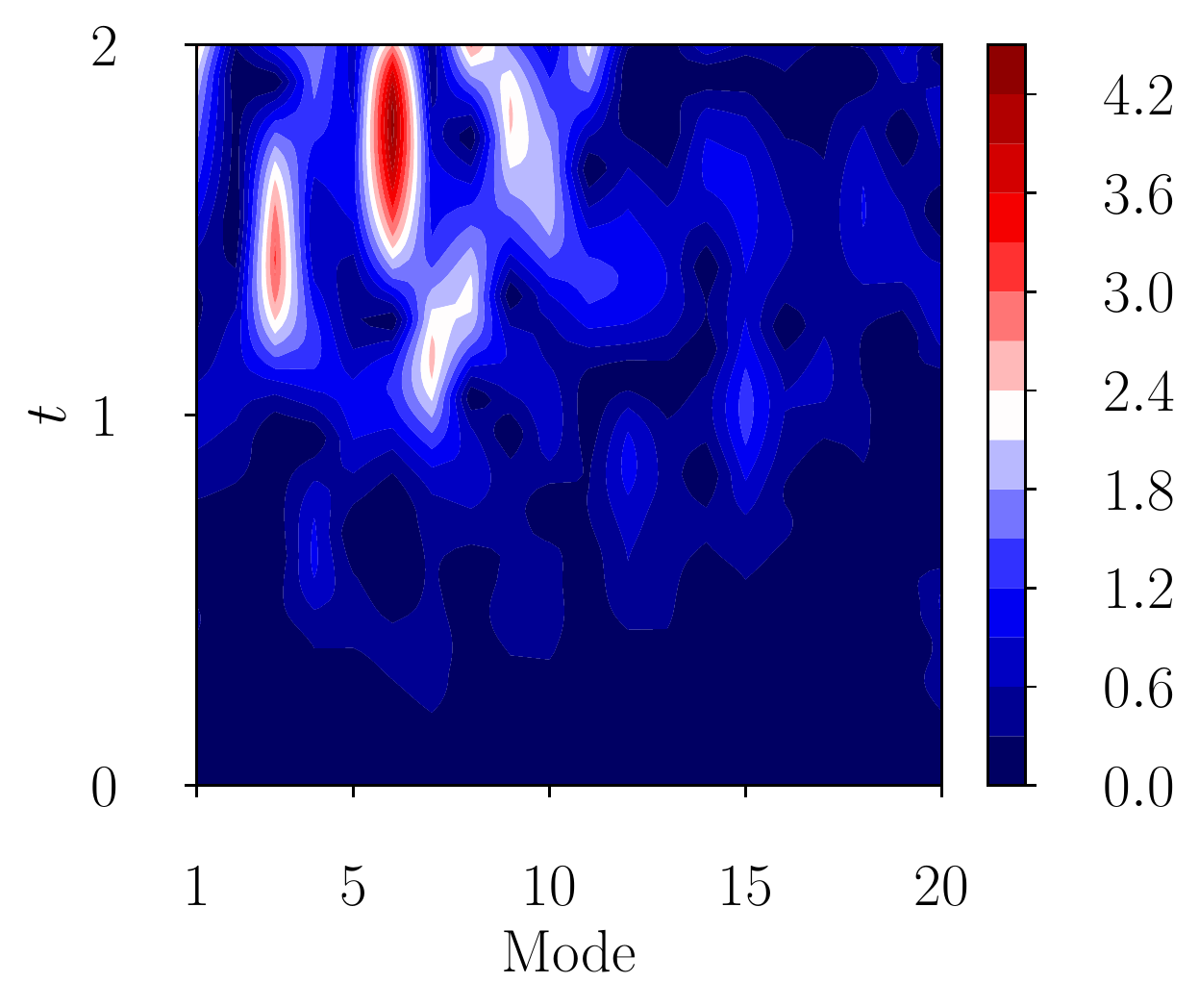}
\vspace{-0.75cm}
\caption{}
\label{ks_example_traj:c}
\end{subfigure} 
\begin{subfigure}{.3\textwidth}
\centering
\includegraphics[width=1\textwidth]{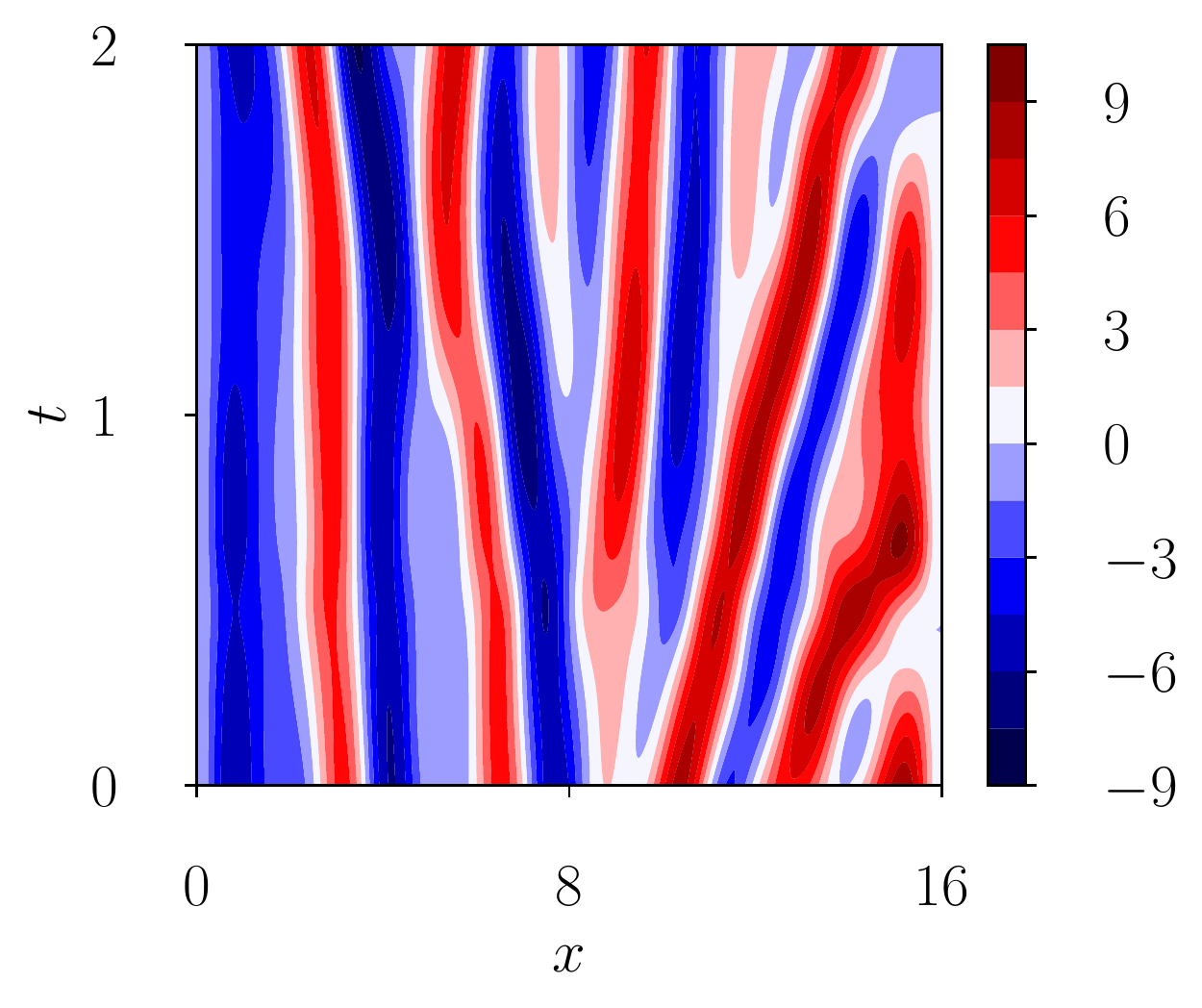}
\vspace{-0.75cm}
\caption{}
\label{ks_example_traj:d}
\end{subfigure} 
\begin{subfigure}{.3\textwidth}
\centering
\includegraphics[width=1\textwidth]{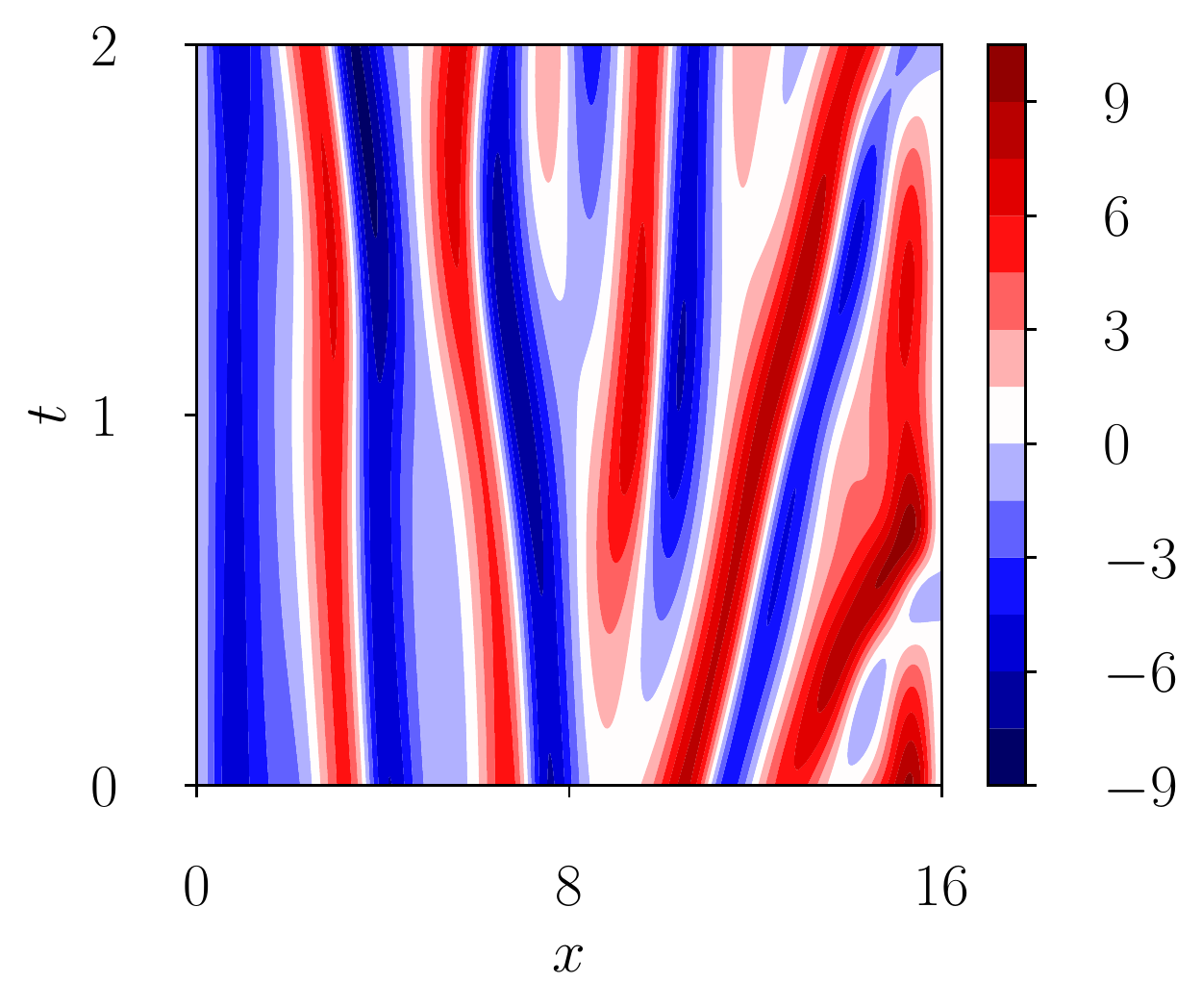}
\vspace{-0.75cm}
\caption{}
\label{ks_example_traj:e}
\end{subfigure}
\centering
\begin{subfigure}{.3\textwidth}
\centering
\includegraphics[width=1\textwidth]{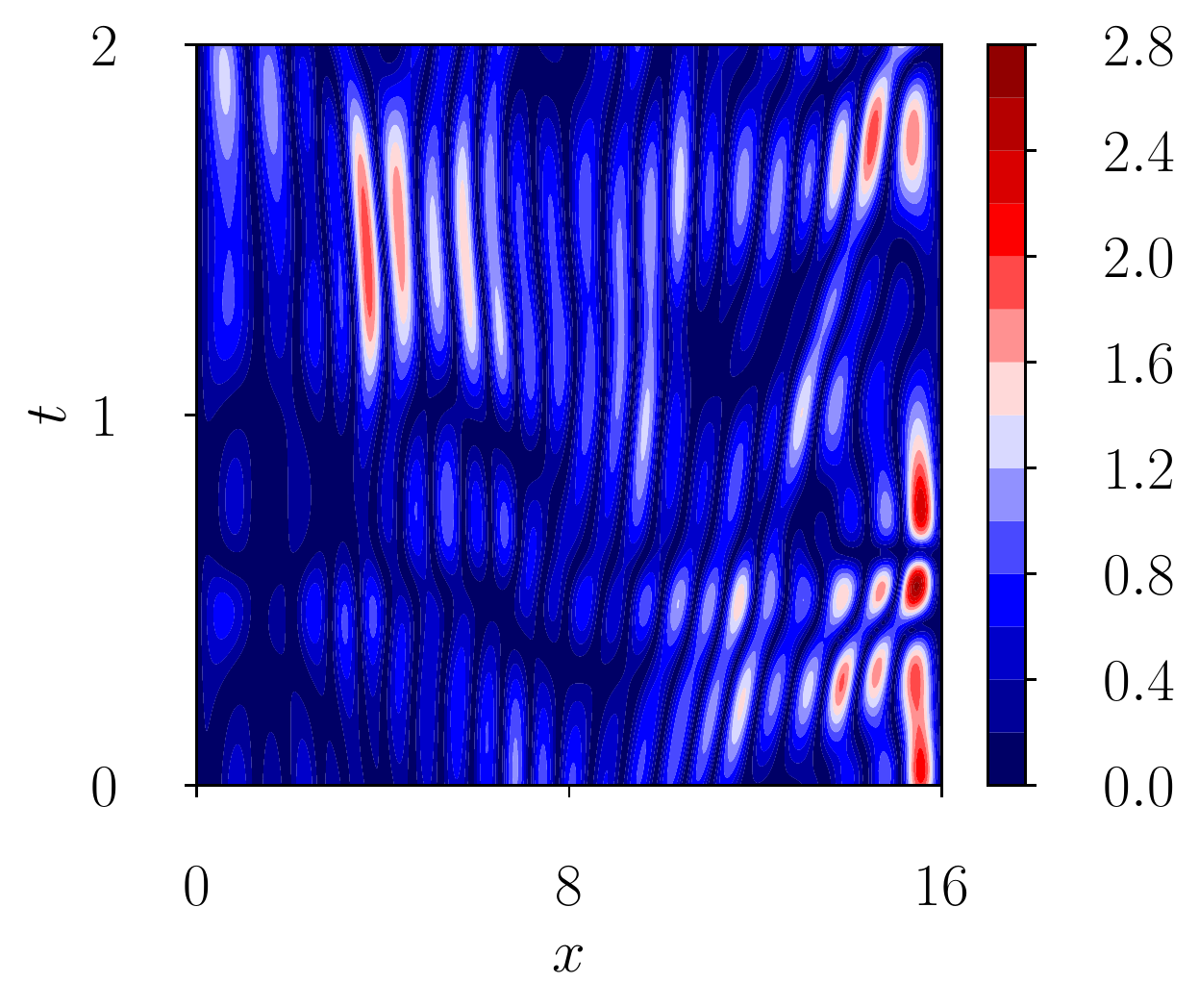}
\vspace{-0.75cm}
\caption{}
\label{ks_example_traj:f}
\end{subfigure} 
\caption{\textbf{(a)} Predicted evolution of the reduced order state for the KS equation with $\nu=1/10$. \textbf{(b)} True evolution of the reduced order. \textbf{(c)} Evolution of the root mean squared error. \textbf{(d)}-\textbf{(e)} The same for the original state dimension computed by projecting to the original space replacing the unmodeled modes with zeros.
}
\label{fig:ks_example_traj}
\end{figure}

\begin{figure}[!httb]
\centering
\begin{subfigure}{.3\textwidth}
\centering
\includegraphics[width=1\textwidth]{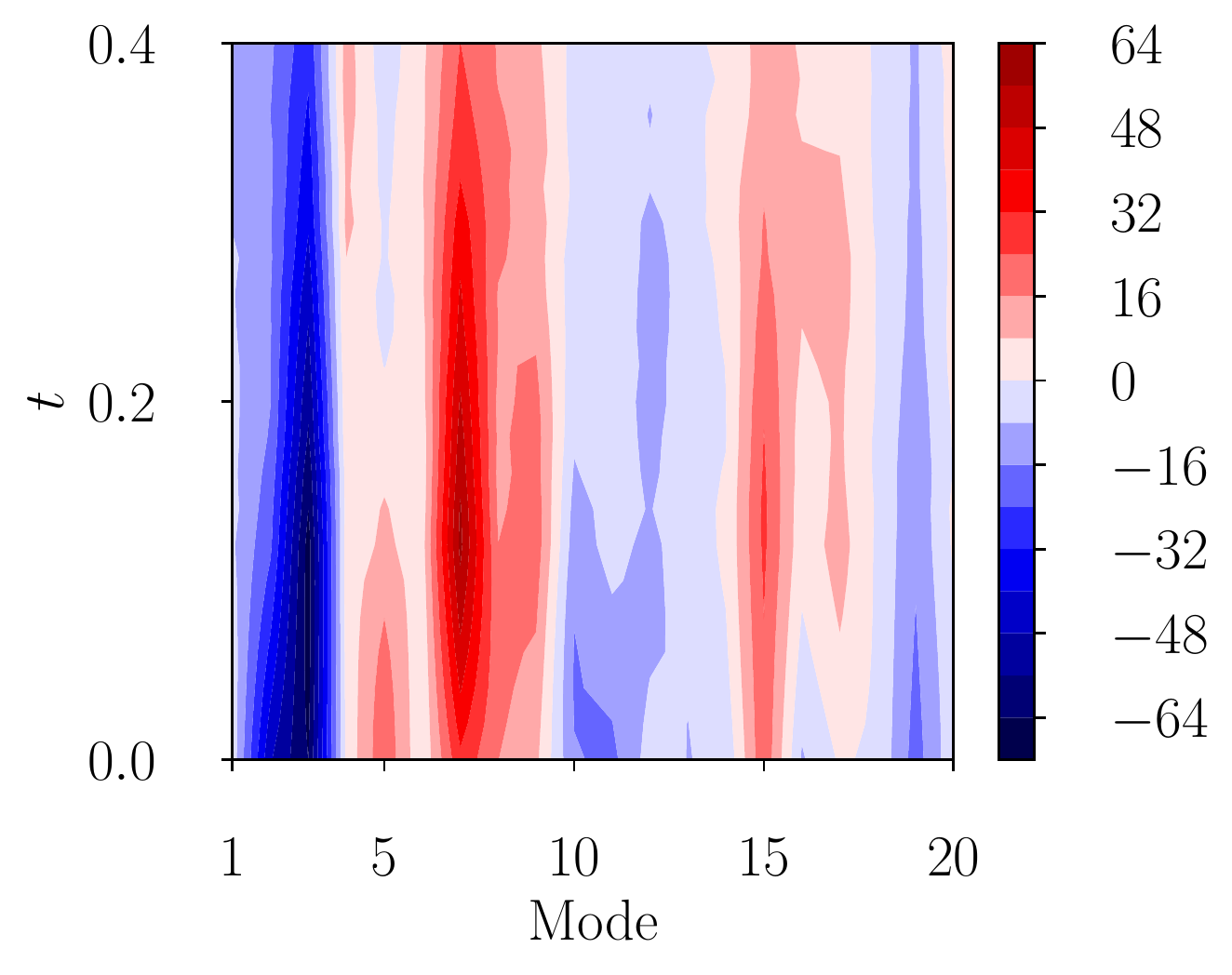}
\vspace{-0.75cm}
\caption{}
\label{ks_example_traj16:a}
\end{subfigure} 
\begin{subfigure}{.3\textwidth}
\centering
\includegraphics[width=1\textwidth]{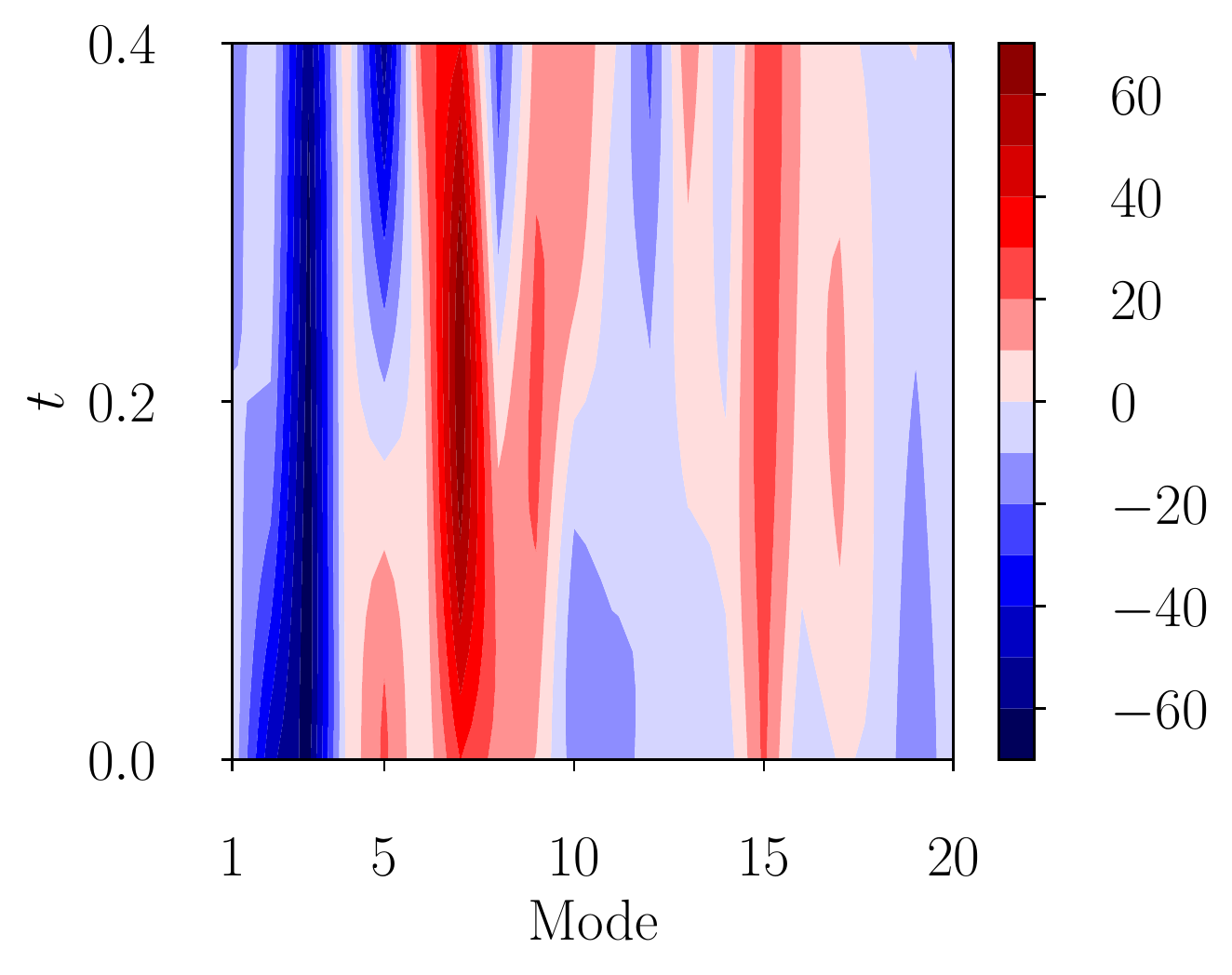}
\vspace{-0.75cm}
\caption{}
\label{ks_example_traj16:b}
\end{subfigure}
\centering
\begin{subfigure}{.3\textwidth}
\centering
\includegraphics[width=1\textwidth]{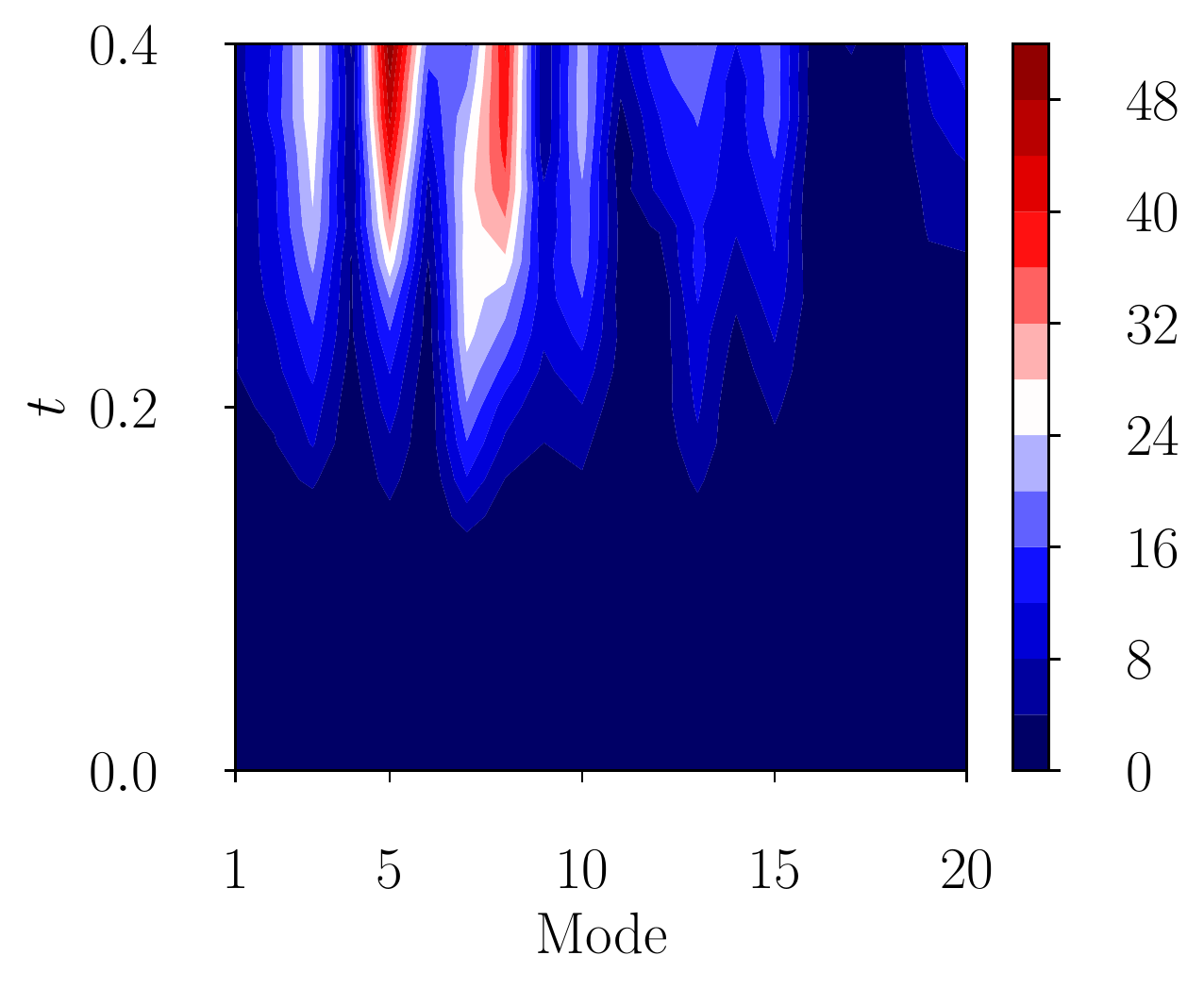}
\vspace{-0.75cm}
\caption{}
\label{ks_example_traj16:c}
\end{subfigure} 
\begin{subfigure}{.3\textwidth}
\centering
\includegraphics[width=1\textwidth]{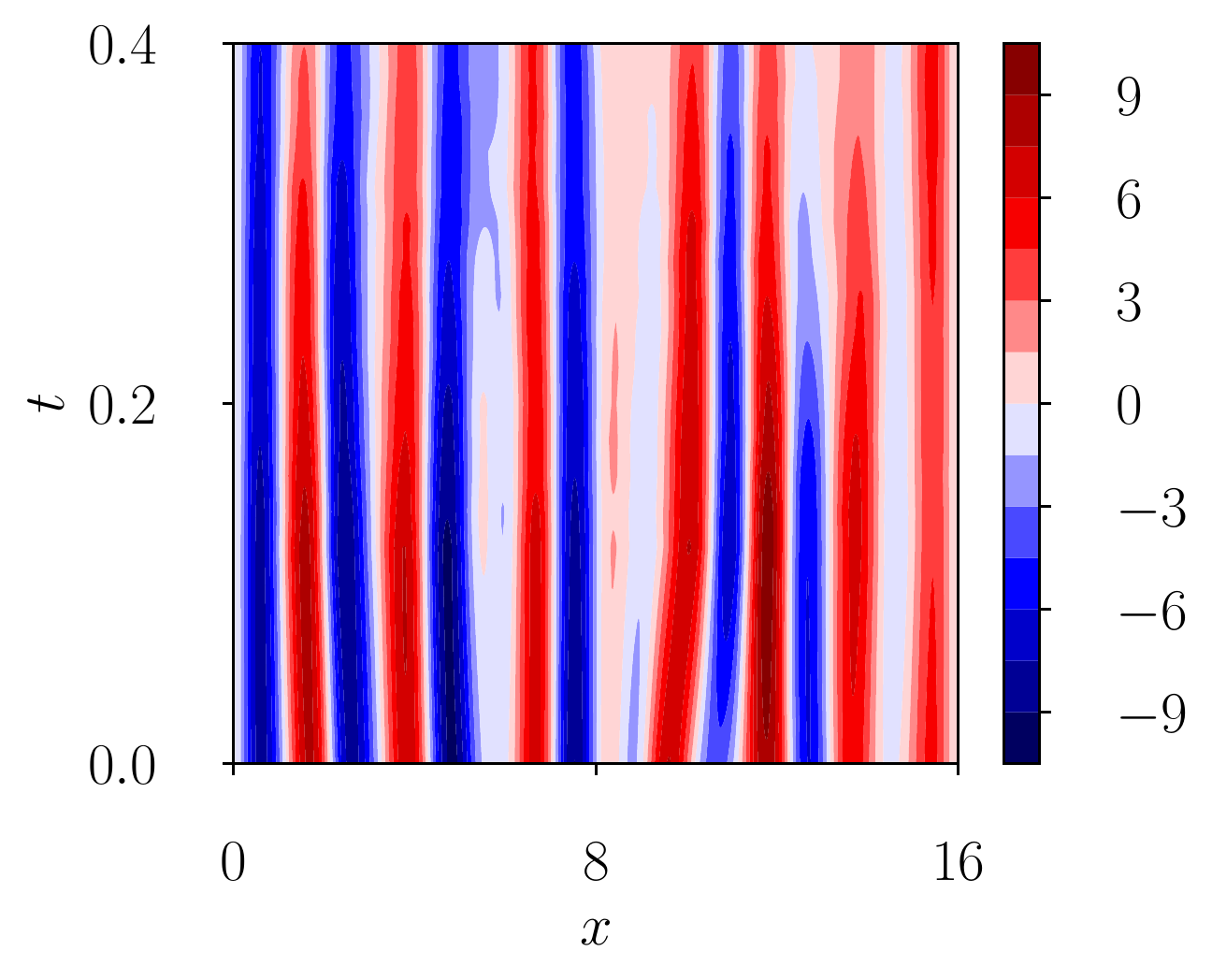}
\vspace{-0.75cm}
\caption{}
\label{ks_example_traj16:d}
\end{subfigure} 
\begin{subfigure}{.3\textwidth}
\centering
\includegraphics[width=1\textwidth]{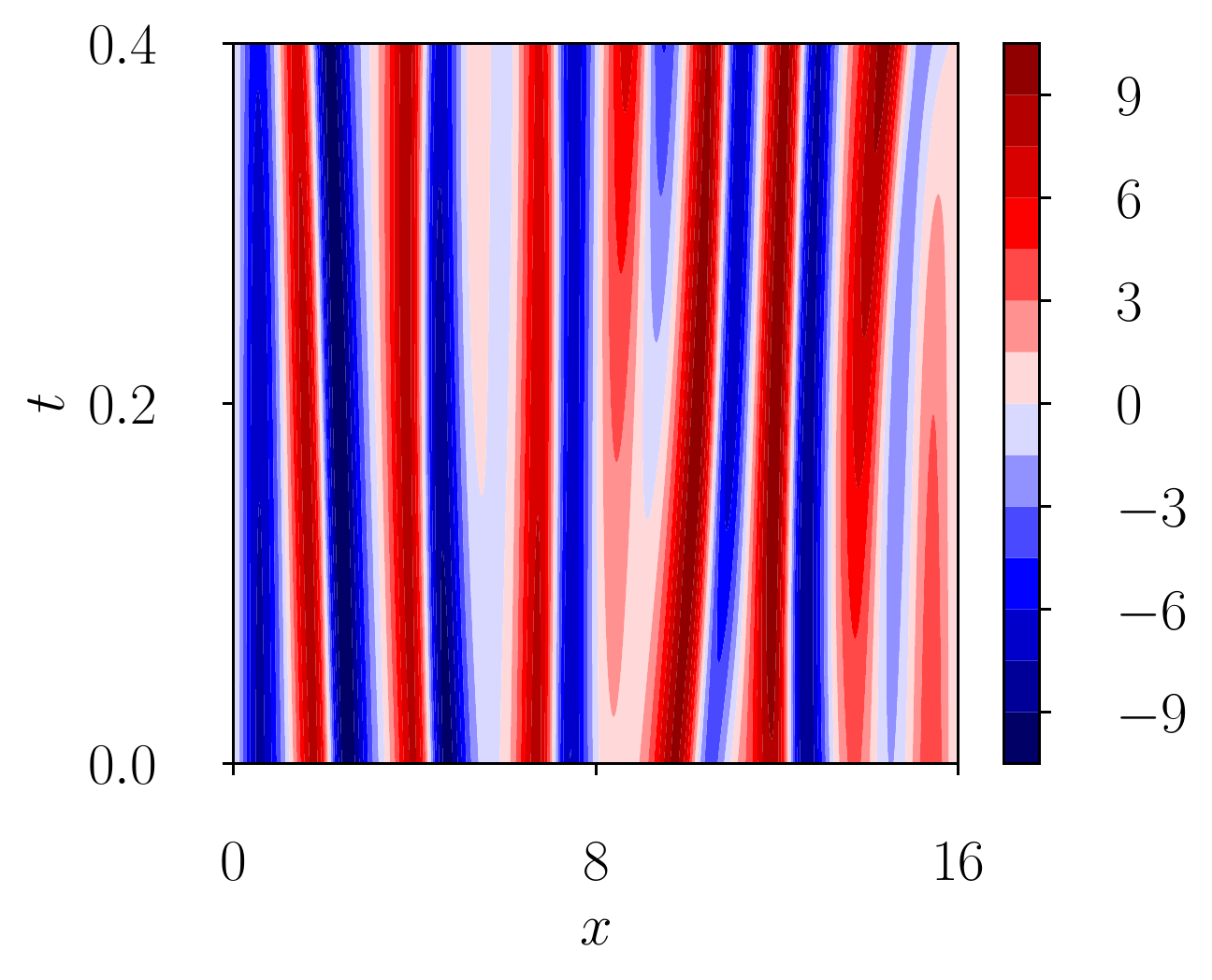}
\vspace{-0.75cm}
\caption{}
\label{ks_example_traj16:e}
\end{subfigure}
\centering
\begin{subfigure}{.3\textwidth}
\centering
\includegraphics[width=1\textwidth]{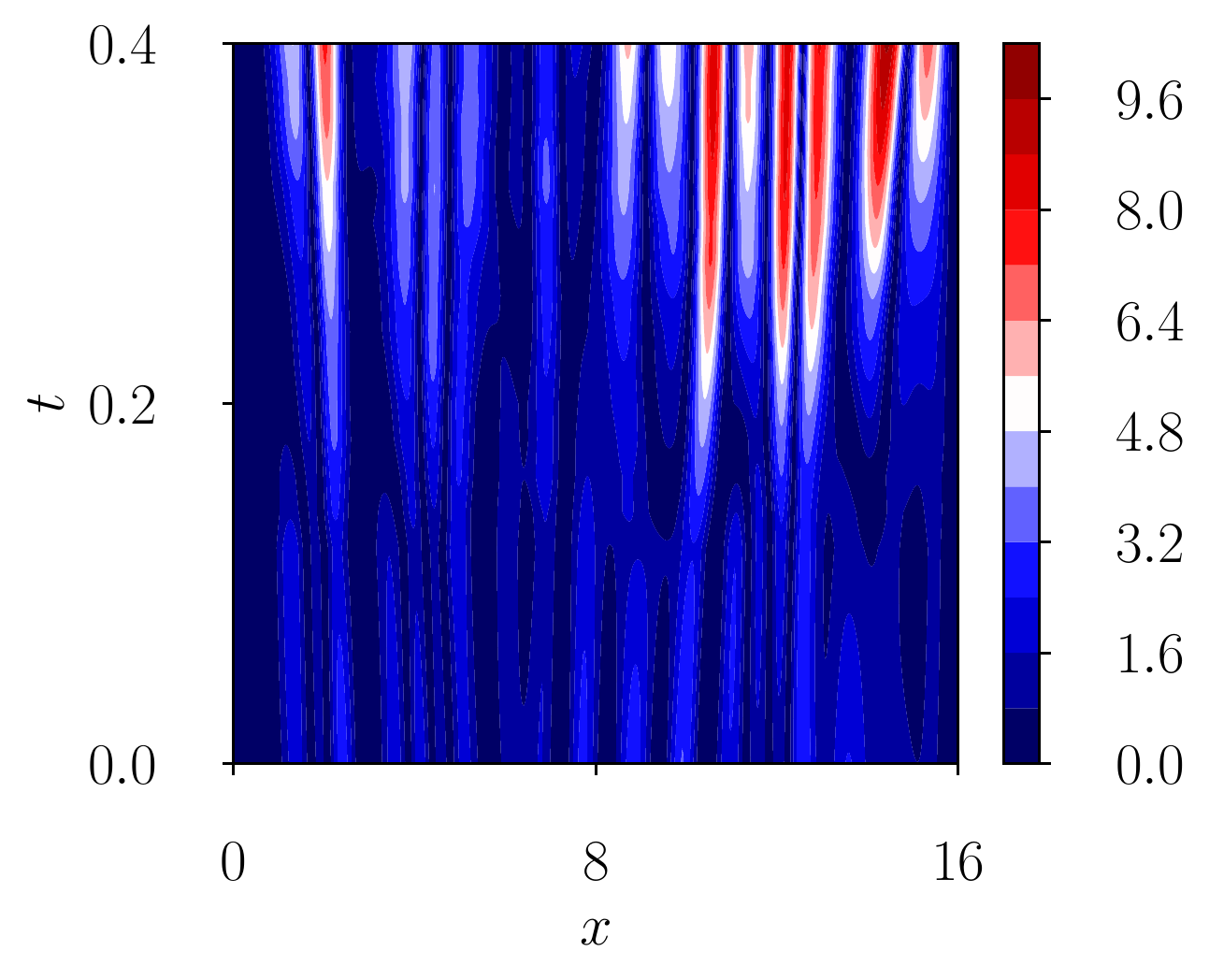}
\vspace{-0.75cm}
\caption{}
\label{ks_example_traj16:f}
\end{subfigure} 
\caption{\textbf{(a)} Predicted evolution of the reduced order state for the KS equation with $\nu=1/16$. \textbf{(b)} True evolution of the reduced order. \textbf{(c)} Evolution of the root mean squared error. \textbf{(d)}-\textbf{(e)} The same for the original state dimension computed by projecting to the original space replacing the unmodeled modes with zeros.
}
\label{fig:ks_example_traj16}
\end{figure}

The LSTM models for both $\nu=1/10$ and $\nu=1/16$ have $h=100$ hidden units and the back-propagation horizon was set to $d=50$. An example of a trajectory obtained starting from a known short-term history of the reduced order state is plotted in Figure \ref{ks_example_traj:a}. Moreover, the true trajectory obtained from simulating the original system, along with the evolution of the RMSE error for $\nu=1/10$ are plotted in Figures \ref{ks_example_traj:b} and \ref{ks_example_traj:c}. After projecting to the original space we get the error evolution given in \ref{ks_example_traj:f}. This error stems not only from the prediction error associated with the LSTM model used to perform forecasts but also with the error associated with the unmodeled dynamics as we only model $r_{dim}=20$ modes. The energy included in the unmodeled modes is higher for $\nu=1/16$ and the system exhibits higher Lyapunov exponents, as a consequence performing forecasts is a more challenging task. This can be observed in Figure \ref{fig:ks_example_traj16}   where the same plots are given for $\nu=1/16$. Note that the plotted time horizon is $T=0.4$ for $\nu=1/16$ compared to $T=2$ for  $\nu=1/10$. 

\begin{figure}[!httb]
\centering
\begin{subfigure}{.9\textwidth}
\centering
\includegraphics[width=1\textwidth]{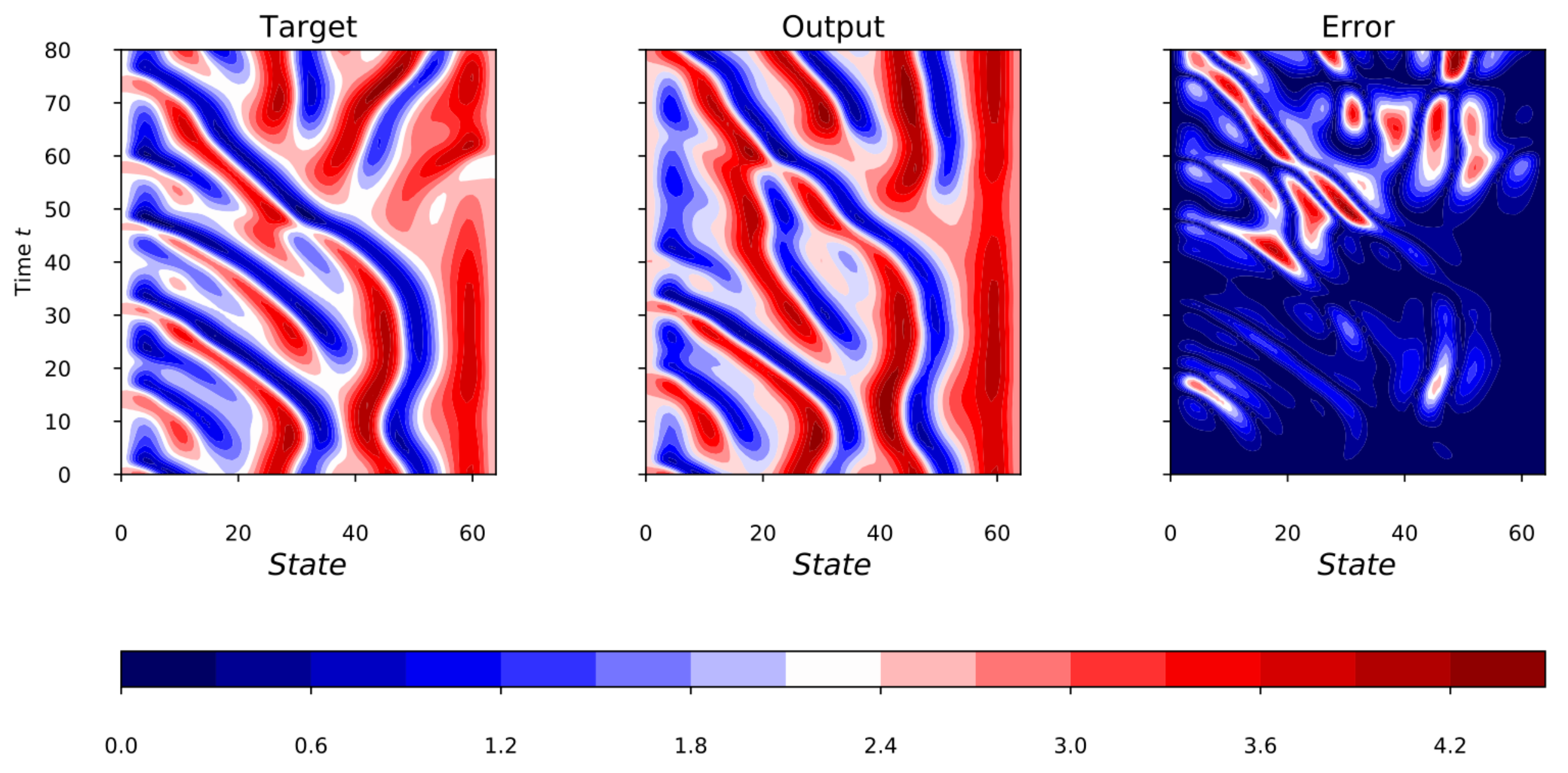}
\vspace{-0.75cm}
\caption{}
\label{ks_example_traj1:a}
\end{subfigure} 
\begin{subfigure}{.9\textwidth}
\centering
\includegraphics[width=1\textwidth]{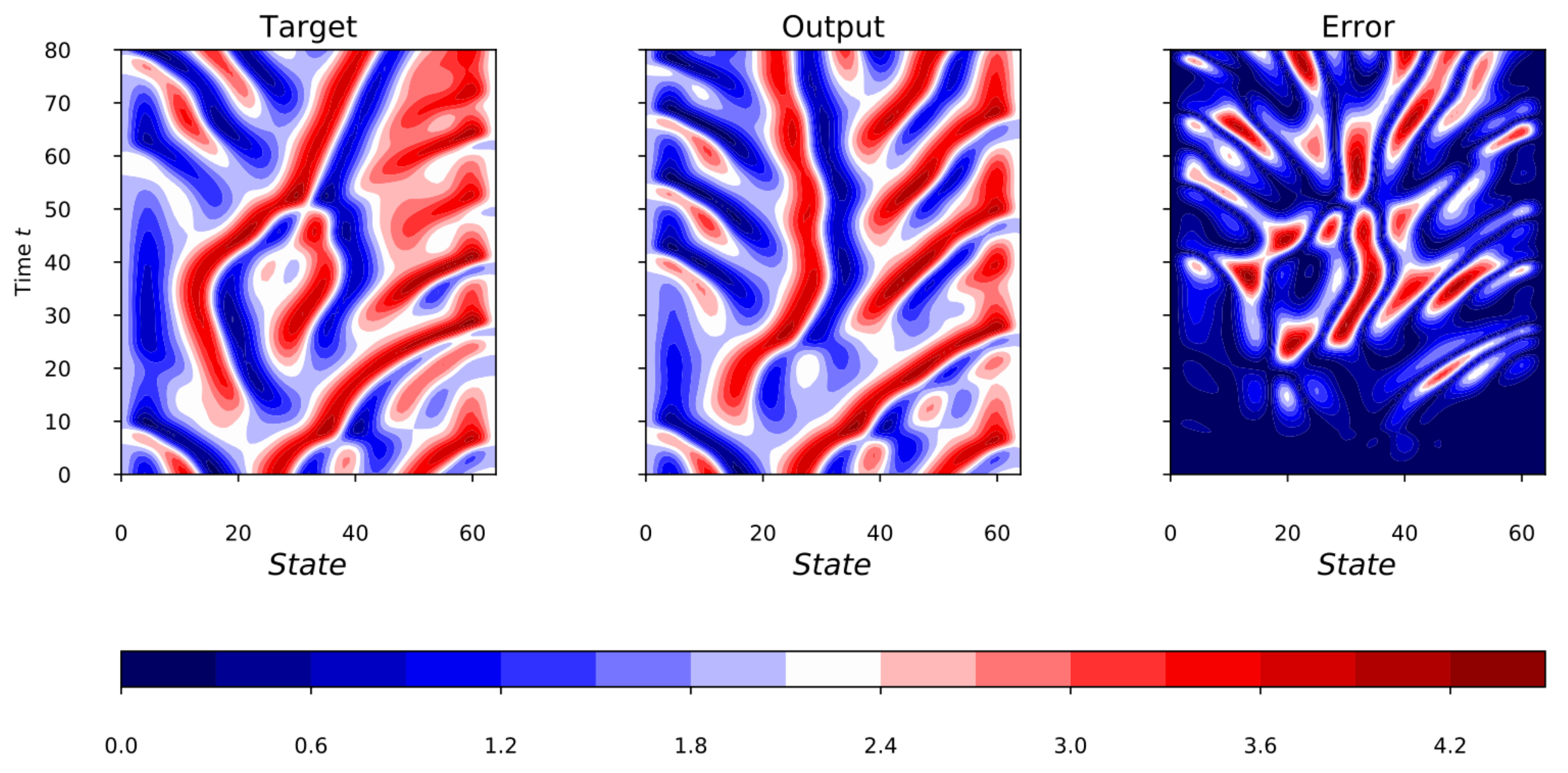}
\vspace{-0.75cm}
\caption{}
\label{ks_example_traj1:b}
\end{subfigure}
\centering
\begin{subfigure}{.9\textwidth}
\centering
\includegraphics[width=1\textwidth]{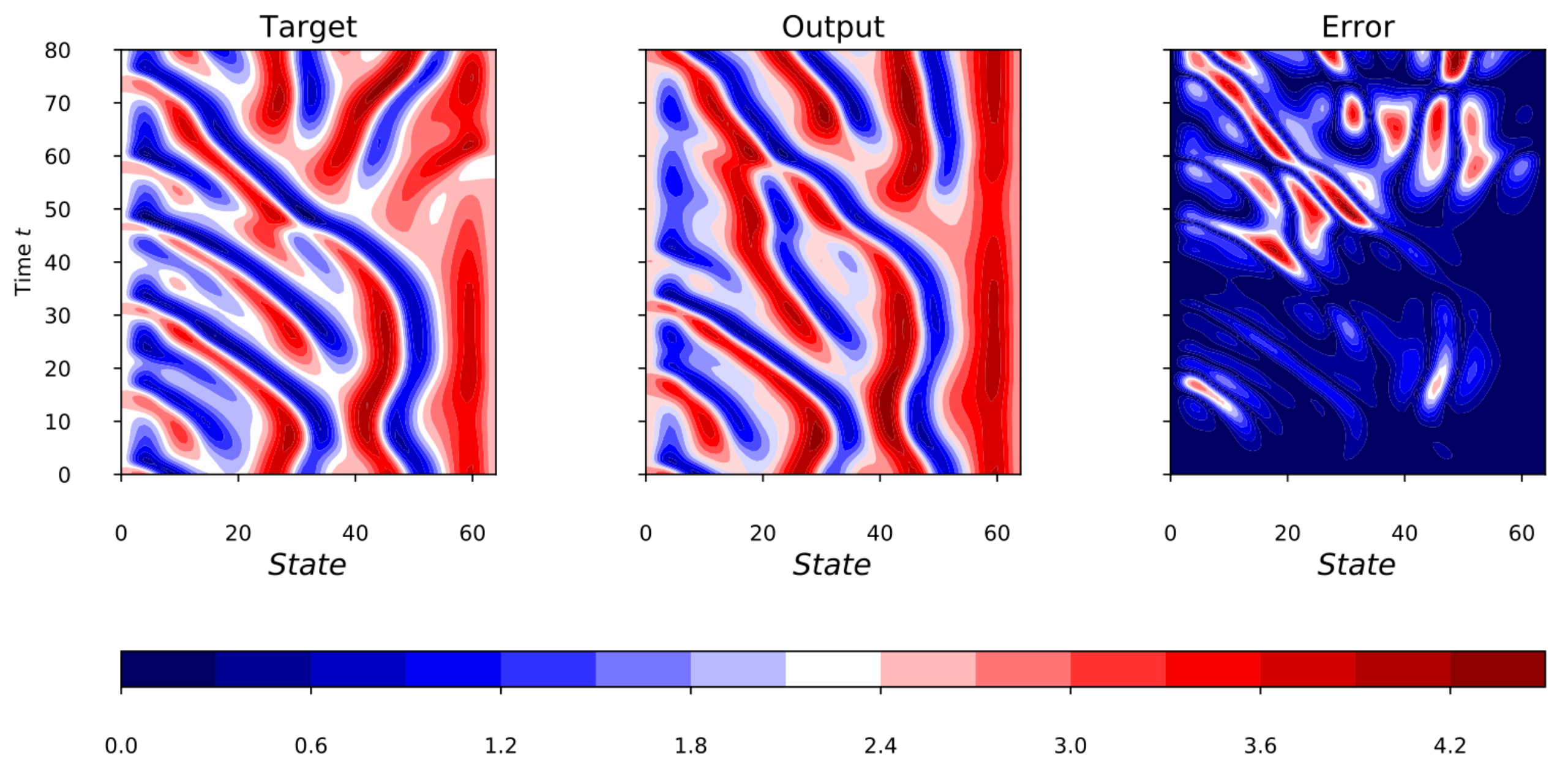}
\vspace{-0.75cm}
\caption{}
\label{ks_example_traj1:c}
\end{subfigure} 
\caption{\textbf{(a)}-\textbf{(d)} True evolution of the state for the KS equation with $\nu=1$, the predicted evolution and the associated RMSE error for three different initial conditions.
}
\label{fig:ks_example_traj1}
\end{figure}

Another interesting question is how the proposed method performs when no dimensionality reduction method is used in a chaotic system with a lower Lyapunov exponent. The dynamics of this system are much easier to capture compared to the applications considered in the main paper, as the chaotic effects are less prominent and there is no missing state information. In the following, we assume that the complete system state information is available and the LSTM forecasts the evolution of the state directly. The KS equation with $\nu=1$ and $L=35$ is simulated with a time-step of $dt=0.25$ and a coarser grid with $D=65$ points instead of $D=513$ of the original paper. The LSTM model used has $h=4096$ hidden units and a truncated back-propagation horizon of $d=32$ was used. The results of three predicted trajectories along with the ground-truth and the RMSE error are illustrated in Figure \ref{fig:ks_example_traj1}. Note that the LSTM can forecast the evolution of the state with high accuracy for a much longer horizon compared to the results shown before where the LSTM was applied in the reduced order dimension of systems with higher Lyapunov exponents. 

\clearpage
\clearpage
\clearpage



\begin{thebibliography}{10}




\bibitem{Kingma2017}
Kingma DP, Ba J. 2017 Adam: A Method for Stochastic Optimization, \textit{ArXiv preprint}.

\bibitem{Williams2015}
Williams MO, Kevrekidis IG, Rowley CW. 2015 A data-driven approximation of the Koopman operator: extending dynamic mode decomposition, \textit{Journal of Nonlinear Science}. \textbf{25}~(6), 1307--1346.

\bibitem{Tu2014}
Tu JH, Rowley CW, Luchtenburg DM, Brunton SL, Kutz JN. 2014 On dynamic mode decomposition: theory and applications, \textit{Journal of Computational Dynamics}. \textbf{1}~(2), 391--421. 

\bibitem{Kerschen2005}
Kerschen G, Golinval JC. Vakakis AF, Bergman LA. 2005 The Method of Proper Orthogonal Decomposition for Dynamical Characterization and Order Reduction of Mechanical Systems: An Overview, \textit{Nonlinear Dynamics}. \textbf{41}~(1), 147--169.


\bibitem{Kutz2016}
Kutz JN, Fu X, Brunton SL. 2016 Multiresolution dynamic mode decomposition, \textit{SIAM Journal on Applied Dynamical Systems}. \textbf{15}~(2), 713--735. 

\bibitem{Arbabi2016}
Arbabi H, Mezic I. 2017 Ergodic theory, Dynamic Mode Decomposition and computation of spectral properties of the Koopman operator, \textit{SIAM Journal on Applied Dynamical Systems}. 

\bibitem{Rowley2005}
Rowley CW. 2005 Model reduction for fluids, using balanced proper orthogonal decomposition, 
\textit{Int. J. Bifurcat. Chaos}. \textbf{15}~(3), 997--1013.

\bibitem{Sapsis2013}
Sapsis TP, Majda AJ. 2013 Statistically accurate low-order models for uncertainty quantification in turbulent dynamical systems, \textit{Proc. Natl Acad. Sci.}. \textbf{110}, 13705--13710.

\bibitem{Bongard2007}
Bongard J, Lipson H. 2007 Automated reverse engineering of nonlinear dynamical systems, \textit{National Academy of Sciences}. \textbf{104}~(24), 9943--9948.

\bibitem{Milano2002}
Milano M, Koumoutsakos P. 2002 Neural Network Modeling for Near Wall Turbulent Flow. \textit{Journal of Computational Physics
} \textbf{182}~(1), 1--26.

\bibitem{Krischer1993}
Krischer K, Rico-Martinez R, Kevrekidis IG, Rotermund HH, Ertl G, Hudson JL. 1992 Model identification of a spatiotemporally varying catalytic reactio. \textit{AIChE Journal} \textbf{39}~(1), 89--98.



\bibitem{Brunton2016}
Brunton SL, Proctor  JL,  Kutz JN. 2016 Discovering governing equations from data by sparse identification of nonlinear dynamical systems, \textit{Proceedings of the National Academy of Sciences}. \textbf{113}~(15), 3932--3937. 

\bibitem{Duriez2016}
Duriez T, Brunton SL, Noack BR. 2016 Machine Learning Control: Taming Nonlinear Dynamics and Turbulence, \textit{Springer}.

\bibitem{Majda2014}
Majda AJ, Lee Y. 2014 Conceptual dynamical models for turbulence, \textit{Proc. Natl Acad. Sci}. \textbf{111}, 6548--6553.

\bibitem{Schaeffer2017}
Schaeffer H. 2017 Learning partial differential equations via data discovery and sparse optimization, \textit{Proc. R. Soc. A}. \textbf{473}, 20160446.

\bibitem{Farazmand2016}
Farazmand M, Sapsis TP. 2016 Dynamical indicators for the prediction of bursting phenomena in high-dimensional systems, \textit{Physical Review E}. \textbf{94}, 032212.





\bibitem{Einicke1999}
Einicke GA, White LB. 1999 Robust Extended Kalman Filtering, \textit{IEEE Trans. Signal Processing}. \textbf{47}~(9), 2596--2599.

\bibitem{Julier1999}
Julier SJ, Uhlmann JK. 1997 A New Extension of the Kalman Filter to Nonlinear Systems, \textit{Proc. SPIE}. \textbf{3068}, 182--193

\bibitem{Lee2016}
Lee Y, Majda AJ. 2016 State estimation and prediction using clustered particle filters, \textit{PNAS}. \textbf{113}~(51), 14609--14614. 

\bibitem{Comeau2017}
Comeau D, Zhao Z, Giannakis D, Majda AJ. 2017 "Data-driven prediction strategies for low-frequency patterns of North Pacific climate variability, \textit{Climate Dynamics}. \textbf{48}~(5-6), 1855--1872. 

\bibitem{Tatsis2017}
Tatsis K, Dertimanis V, Abdallah I, Chatzi E. 2017 A substructure approach for fatigue assessment on wind turbine support structures using output-only measurements, \textit{X International Conference on Structural Dynamics, EURODYN 2017}. \textbf{199}, 1044--1049.

\bibitem{Quade2016}
Quade M, Abel M, Shafi K, Niven RK, Noack BR. 2016 Prediction of dynamical systems by symbolic regression, \textit{Phys. Rev. E.}. \textbf{94}, 012214.



\bibitem{Mirmomeni2010}
Mirmomeni M, Lucas C, Moshiri B, Araabi NB. 2010 Introducing adaptive neurofuzzy modeling with online learning method for prediction of time-varying solar and geomagnetic activity indices, \textit{Expert Systems with Applications}. \textbf{37}~(12), 8267--8277

\bibitem{Gholipour2007}
Gholipour A, Lucas C, Araabi NB, Mirmomeni M, Shafiee M. 2007 Extracting the main patterns of natural time series for long-term neurofuzzy prediction, \textit{Neural Computing and Applications}. \textbf{16}~(4-5), 383--393

\bibitem{Mirmomeni2011a}
Mirmomeni M, Lucas C, Araabi NB, Moshiri B, Bidar MR. 2011 Recursive spectral analysis of natural time series based on eigenvector matrix perturbation for online applications, \textit{IET Signal Processing}. \textbf{5}~(6), 512--526

\bibitem{Mirmomeni2011b}
Mirmomeni M, Lucas C, Araabi NB, Moshiri B, Bidar MR. 2011 Online Multi-step Ahead Prediction of Time-Varying Solar and Geomagnetic Activity Indices via Adaptive Neurofuzzy Modeling and Recursive Spectral Analysis, \textit{Solar Physics}. \textbf{272}, 189--213

\bibitem{Mirmomeni2006}
Marques CAF, Ferreira JA, Rocha A, Castanheira JM, Melo-Goncalves P, Vaz N, Dias JM. 2006 Singular spectrum analysis and forecasting of hydrological time series, \textit{Physics and Chemistry of the Earth, Parts A/B/C}. \textbf{31}~(18), 1172--1179

\bibitem{Abdollahzade2015}
Abdollahzade M, Miranian A, Hassani H, Iranmanesh H. 2015 A new hybrid enhanced local linear neuro-fuzzy model based on the optimized singular spectrum analysis and its application for nonlinear and chaotic time series forecasting, \textit{Information Sciences}. \textbf{295}, 107--125

\bibitem{Ye2011}
Ye L, Liu P. 2011 Combined model based on EMD-SVM for short-term wind power prediction, \textit{Communications in Nonlinear Science and Numerical Simulation}. \textbf{31}, 102--108




\bibitem{Cousins2014}
Cousins W, Sapsis TP. 2014 Quantification and prediction of extreme events in a one-dimensional nonlinear dispersive wave model, \textit{Physica D}. \textbf{280-281}, 48--58.

\bibitem{Cousins2016}
Cousins W, Sapsis TP. 2016 Reduced order precursors of rare events in unidirectional nonlinear water waves, \textit{Journal of Fluid Mechanics}. \textbf{790}, 368--388.



\bibitem{Lorenz1969}
Lorenz EN. 1969 {Atmospheric predictability as revealed by naturally occurring  analogues}, \textit{J. Atmos. Sci.} \textbf{26}, 636--646.

\bibitem{Prince2007}
Xavier PK, Goswami BN. 2007 An analog method for real-time forecasting of summer monsoon subseasonal variability, \textit{Monthly Weather Review} \textbf{135}~(12), 4149--4160.


\bibitem{Zhao2016}
Zhao Z, Giannakis D. 2016 Analog forecasting with dynamics-adapted kernels, \textit{Nonlinearity} \textbf{29}, 2888--2939.

\bibitem{Chiavazzo2014}
Chiavazzo E, Gear CW, Dsilva CJ, Rabin N, Kevrekidis IG. 2014  Reduced models in chemical kinetics via nonlinear data-mining, \textit{Processes} \textbf{2}~(1), 112--140.


\bibitem{Zhong2017}
Wan ZY, Sapsis TP. 2017 Reduced-space Gaussian Process Regression for data-driven probabilistic forecast of chaotic dynamical systems, \textit{Physica D: Nonlinear Phenomena}.

\bibitem{Rasmussen2006}
Rasmussen CE, Williams CKI. 2006 Gaussian Processes for Machine Learning, The MIT Press.


\bibitem{Hochreiter1997}
Hochreiter S, Schmidhuber J. 1997 Long short-term memory, \textit{Neural Computation}. \textbf{9}, 1735--1780.

\bibitem{Graves2013}
Graves A, Mohamed AR, Hinton G. 2013 Speech Recognition with Deep Recurrent Neural Networks, \textit{IEEE ICASSP}. 6645--6649.

\bibitem{Graves2007}
Graves A, Fern\'andez S,  Liwicki M, Bunke H, Schmidhuber, J. 2007 {Unconstrained Online Handwriting Recognition with Recurrent Neural Networks}, \textit{Proc. 20th NIPS.} 577--584

\bibitem{Wierstra2005}
Wierstra D, Schmidhuber J, Gomez FJ. 2005 {Evolino: Hybrid Neuroevolution/Optimal Linear Search for Sequence Learning}, \textit{Proc. 19th IJCAI}, 853--858.

\bibitem{Wu2016}
Wu Y, Schuster M, Chen Z, Le QV, Norouzi M, Macherey W, Krikun M, Cao Y, Gao Q. 2016 {Google's Neural Machine Translation System: Bridging the Gap between Human and Machine Translation}, \textit{arXiv:1609.08144}.

\bibitem{Gers2002}
Gers FA, Eck D, Schmidhuber J. 2012 Applying LSTM to Time Series Predictable Through Time-Window Approaches, \textit{Neural Nets WIRN Vietri-01: Proceedings of the 12th Italian Workshop on Neural Nets.} Springer, London, 193--200.

\bibitem{Rico1992}
Rico-Martinez R, Krischer K, Kevrekidis IG, Kube MC, Hudson JL. 1992 Discrete- vs. Continuous-Time Nonlinear Signal Processing of Cu Electrodissolution Data. \textit{Chemical Engineering Communications} \textbf{118}, 25--48.

\bibitem{Jaeger2004}
Jaeger M, Haas H. 2004 Harnessing Nonlinearity: Predicting Chaotic Systems and Saving Energy in Wireless Communication, \textit{Science} \textbf{304}~(5667), 78--80. 

\bibitem{Chatzis2011}
Chatzis SP, Demiris Y. 2011 {Echo State Gaussian Process}, \textit{IEEE Transactions on Neural Networks}, \textbf{22}~(9), 1435--1445.

\bibitem{Broomhead1988}
Broomhead DS, Lowe D. 1988 Multivariable functional interpolation and adaptive networks, \textit{Complex Systems} \textbf{2}, 321--355.

\bibitem{Kim2000ControlOC}
Kim KB, Park JB, Choi YH, Chen G. 2000 Control of chaotic dynamical systems using radial basis function network approximators, \textit{Inf. Sci.} \textbf{130}, 165--183.

\bibitem{Pathak2018}
Pathak J, Hunt BR, Girvan M, Lu Z, Ott E. 2018 Model-Free Prediction of Large Spatiotemporally Chaotic Systems from Data: A Reservoir Computing Approach, \textit{Phys. Rev. Lett.} \textbf{120}~(2), 024102.

\bibitem{Pathak2017}
Pathak J, Lu Z, Hunt BR, Girvan M, Ott E. 2017 Using machine learning to replicate chaotic attractors and calculate Lyapunov exponents from data, \textit{Chaos.} \textbf{27}~(12), 121102.



\bibitem{Takens1981}
Takens F. 1981 Detecting strange attractors in fluid turbulence, \textit{Symposium on dynamical systems and turbulence, Springer}.366--381.

\bibitem{Hochreiter1991}
Hochreiter J. 1991 {Untersuchungen zu dynamischen neuronalen Netzen}, \textit{Master thesis Institut fur Informatik}, Technische Universitat, Munchen.

\bibitem{Bengio1994}
Bengio Y, Simard P, Frasconi P. 1994 Long short-term memory, \textit{IEEE Transactions on Neural Networks}, \textbf{5}~(2), 157--166.

\bibitem{Majda2012}
Majda A, Harlim J. 2012 \textit{Filtering Complex Turbulent Systems}, Cambridge University Press.

\bibitem{Majda2005}
Majda A, Grote MJ, Abramov RV. 2005 \textit{Information Theory and Stochastics for Multiscale Nonlinear Systems} \textbf{25}, AMS and Centre de Recherches Mathematiques.

\bibitem{Lorenz96}
Lorenz NE. 1996 Predictability - A problem partly solved, \textit{Proc. Seminar on Predictability}. Reading, Berkshire, 1--18.

\bibitem{Basnarkov2012}
Basnarkov L, Kocarev L. 2012 Forecast improvement in Lorenz96 system, \textit{Nonlin. Processes Geophy.} \textbf{19}, 569--575.

\bibitem{Allgaier2012}
Allgaier NA, Harris KD, Danforth CM.: 2012 Empirical correction of a toy climate model, \textit{ Phys. Rev. E} \textbf{85}~(2), 026201

\bibitem{Crommelin2004}
Crommelin DT, Majda AJ. 2004 Strategies for model reduction: Comparing different optimal bases. \textit{J. Atmos. Sci.}, \textbf{61}~(17), 2206--2217.

\bibitem{Kuramoto1976}
Kuramoto Y, Tsuzuki T. 1976 Persistent Propagation of Concentration Waves in Dissipative Media Far from Thermal Equilibrium, \textit{Progress of Theoretical Physics} \textbf{55}~(2), 356--369.

\bibitem{Kuramoto1978}
Kuramoto Y. 1978 Diffusion-induced chaos in reaction systems, \textit{Progress of Theoretical Physics Supplement} \textbf{64}, 346--367.

\bibitem{Sivashinsky1977}
Sivashinsky G, 1977. Nonlinear analysis of hydrodynamic instability in laminar flames--I. Derivation of basic equations, \textit{Acta Astronautica} \textbf{4}, 1177--1206.

\bibitem{Sivashinsky1980}
Sivashinsky G, Michelson DM. 1980 On irregular wavy flow of a liquid film down a vertical plane, \textit{Progress of Theoretical Physics} \textbf{63}~(6), 2112--2114.

\bibitem{Blonigan2014}
Blonigan PJ, Wang Q. 2014 Least squares shadowing sensitivity analysis of a modified Kuramoto-Sivashinsky equation, \textit{Chaos, Solitons and Fractals}. \textbf{64}, 16--25.

\bibitem{Kevrekidis1990}
Kevrekidis IG, Nicolaenko B, Scovel JC. 1990 Back in the saddle again: A computer assisted study of the kuramoto-sivashinsky equation.  \textit{SIAM J. Appl. Math.} \textbf{50}~(3), 760--790.

\bibitem{Selten1995}
Selten FM. 1995 An efficient description of the dynamics of barotropic flow,  \textit{Journal of the Atmospheric Sciences} \textbf{52}~(7), 915--936.

\bibitem{Thompson2000}
Thompson DWJ, Wallace JM. 2000 Annular modes in the extratropical circulation: Part i: Month-to-month variability. \textit{J. Climate} \textbf{13}, 1000--1016.

\bibitem{Mo1986}
Mo KC, Livezey RE. 1986 Tropical-extratropical geopotential height teleconnections during the northern hemisphere winter. \textit{Mon. Weather Rev.} \textbf{114}, 2488--2512.






\bibitem{Xavier2010}
Glorot X, Bengio Y. 2010 Understanding the difficulty of training deep feedforward neural networks. \textit{Proc. 13th AISTATS} \textbf{114}~(9), 249--256.

\end{thebibliography}
\end{document}